\documentclass{bmcart}

\usepackage{amsthm,amsmath}
\usepackage{listings}
\RequirePackage[]{natbib}
\usepackage[utf8]{inputenc}

\usepackage{xcolor}
\usepackage{graphicx}
\usepackage{abbreviations}
\graphicspath{{figures/}}

\startlocaldefs
\newcommand{\ie}{i.e.,~}
\newcommand{\eg}{e.g.,~}
\newcommand{\del}[1]{\partial_{#1}}
\newcommand{\vt}{\mathcal{V}}
\newcommand{\bhac}{\texttt{BHAC}}
\newcommand{\harm}{\texttt{HARM}~}
\newcommand{\harmthreed}{\texttt{HARM3D}~}
\newcommand{\amrvac}{\texttt{MPI-AMRVAC}~}
\newcommand{\paramesh}{\texttt{PARAMESH}~}
\newcommand{\athenapp}{\texttt{Athena++}~}

\newcommand{\bhoss}{\texttt{BHOSS}}

\endlocaldefs

\usepackage{amssymb,amsmath}
\usepackage{hyperref}
\hypersetup{
  colorlinks=true,        
  linkcolor=blue,         
  citecolor=cyan,         
}
\usepackage{float}

\begin{document}

\begin{frontmatter}

\begin{fmbox}
\dochead{Research}

\title{The Black Hole Accretion Code}

\author[
   addressref={aff1},                   
   corref={aff1},                       
   noteref={},                        
   email={porth@itp.uni-frankfurt.de}   
]{\inits{OP}\fnm{Oliver} \snm{Porth}}
\author[
   addressref={aff1},
   email={}
]{\inits{HO}\fnm{Hector} \snm{Olivares}}
\author[
   addressref={aff1},
   email={}
]{\inits{YM}\fnm{Yosuke} \snm{Mizuno}}
\author[
   addressref={aff1},
   email={}
]{\inits{ZY}\fnm{Ziri} \snm{Younsi}}
\author[
   addressref={aff1,aff2},
   noteref={},                        
      email={rezzolla@itp.uni-frankfurt.de}   
]{\inits{LR}\fnm{Luciano} \snm{Rezzolla}}
\author[
   addressref={aff3},
   email={}
]{\inits{MM}\fnm{Monika} \snm{Moscibrodzka}}
\author[
   addressref={aff3},
   email={}
]{\inits{HF}\fnm{Heino} \snm{Falcke}}
\author[
   addressref={aff4},
   email={}
]{\inits{MK}\fnm{Michael} \snm{Kramer}}

\address[id=aff1]{
  \orgname{Institute for Theoretical Physics}, 
  \street{Max-von-Laue-Str. 1},                     
  \postcode{60438}                                
  \city{Frankfurt am Main},                              
  \cny{Germany}                                    
}

\address[id=aff2]{
  \orgname{Frankfurt Institute for Advanced Studies}, 
  \street{Ruth-Moufang-Stra\ss e 1},                     
  \postcode{D-60438}                                
  \city{Frankfurt am Main},                              
  \cny{Germany}                                    
}

\address[id=aff3]{
  \orgname{Department of Astrophysics/IMAPP, Radboud University Nijmegen}, 
  \street{P.O. Box 9010},                     
  \postcode{65008}                                
  \city{Nijmegen},                              
  \cny{The Netherlands}                                    
}

\address[id=aff4]{
  \orgname{Max-Planck-Institut f\"ur Radioastronomie}, 
  \street{Auf dem H\"ugel 69},                     
  \postcode{D-53121}                                
  \city{Bonn},                              
  \cny{Germany}                                    
}

\begin{artnotes}
\end{artnotes}

\end{fmbox}

\begin{abstractbox}

\begin{abstract} 
We present the black hole accretion code (\bhac), a new
multidimensional general-relativistic magnetohydrodynamics module for the
\amrvac framework
. \bhac~has been designed
to solve the equations of ideal general-relativistic magnetohydrodynamics
in arbitrary spacetimes and exploits adaptive mesh refinement techniques
with an efficient block-based approach. Several spacetimes have already
been implemented and tested. We demonstrate the validity of \bhac~by
means of various one-, two-, and three-dimensional test problems, as well
as through a close comparison with the \harmthreed code in the case of a
torus accreting onto a black hole. The convergence of a turbulent
accretion scenario is investigated with several diagnostics and we find
accretion rates and horizon-penetrating fluxes to be convergent to within
a few percent when the problem is run in three dimensions. Our analysis
also involves the study of the corresponding thermal synchrotron
emission, which is performed by means of a new general-relativistic
radiative transfer code, \bhoss\!\!. The resulting synthetic intensity
maps of accretion onto black holes are found to be convergent with
increasing resolution and are anticipated to play a crucial role in the
interpretation of horizon-scale images resulting from upcoming radio
observations of the source at the Galactic Center.
\end{abstract}

\end{abstractbox}

\end{frontmatter}

\section{Introduction}

Accreting black holes (BHs) are amongst the most powerful astrophysical
objects in the Universe. A substantial fraction of the gravitational
binding energy of the accreting gas is released within tens of
gravitational radii from the BH, and this energy supplies the power for a
rich phenomenology of astrophysical systems including active galactic
nuclei, X-ray binaries and gamma-ray bursts. Since the radiated energy
originates from the vicinity of the BH, a fully general-relativistic
treatment is essential for the modelling of these objects and the flows
of plasma in their vicinity.

Depending on the mass accretion rate, a given system can be found in
various spectral states, with different radiation mechanisms dominating
and varying degrees of coupling between radiation and gas
\citep{Fender2004,Markoff2005}. Some supermassive BHs, including the
primary targets of observations by the Event-Horizon-Telescope
Collaboration (EHTC\footnote{http://www.eventhorizontelescope.org}), \ie
Sgr A* and M87, are accreting well below the Eddington accretion rate
\cite{Marrone2007,Ho2009}. In this regime, the accretion flow advects
most of the viscously released energy into the BH rather than radiating
it to infinity. Such optically thin, radiatively inefficient and
geometrically thick flows are termed advection-dominated accretion flows
(ADAFs, see \cite{Narayan1994, Narayan1995, Abramowicz1995, Yuan2014})
and can be modelled without radiation feedback.  Next to the ADAF,
  two additional radiatively inefficient accretion flows (RIAFs) exist:
  The advection-dominated inflow-outflow solution (ADIOS)
  \cite{Blandford1999, Begelman2012} and the convection-dominated
  accretion flow (CDAF) \cite{Narayan2000, Quataert2000}, which include
  respectively, the physical effects of outflows and convection.
Analytical and semi-analytical approaches are reasonably successful in
reproducing the main features in the spectra of ADAFs \citep[see,
  \eg][]{Yuan2003}. However, numerical general-relativistic
magnetohydrodynamic (GRMHD) simulations are essential to gain an
understanding of the detailed physical processes at play in the Galactic
Centre and other low-luminosity compact objects.

Modern BH accretion-disk theory suggests that angular momentum transport
is due to MHD turbulence driven by the magnetorotational instability
(MRI) within a differentially rotating disk \citep{Balbus1991,
  BalbusHawley1998}. Recent non-radiative GRMHD simulations of BH
accretion systems in an ADAF regime have resolved these processes and
reveal a flow structure that can be decomposed into a disk, a corona, a
disk-wind and a highly magnetized polar funnel \cite[see,
  \eg][]{DeVilliers03, McKinney2004, McKinney2006c, McKinney2009}. The
simulations show complex time-dependent behaviour in the disk, corona and
wind. Depending on BH spin, the polar regions of the flow contain a
nearly force-free, Poynting--flux-dominated jet \cite[see,
  \eg][]{Blandford1977, McKinney2004, Hawley2006, McKinney2006c}.

In addition to having to deal with highly nonlinear dynamics that spans a
large range in plasma parameters, the numerical simulations also need to
follow phenomena that occur across multiple physical scales. For example,
in the MHD paradigm, jet acceleration is an intrinsically inefficient
process that requires a few thousand gravitational radii to reach
equipartition of the energy fluxes \cite{Komissarov2007b, Barkov2008}
(purely hydrodynamical mechanisms can however be far more efficient
\cite{Aloy:2006rd}). Jet-environment interactions like the prominent
HST-1 feature of the radio-galaxy M87 \cite{Biretta1989, Stawarz2006,
  Asada2012} occur on scales of $\sim 5\times10^5$ gravitational radii.
Hence, for a self-consistent picture of accretion and ejection, jet
formation and recollimation due to interaction with the environment
\cite[see, \eg][]{Mizuno2015}, numerical simulations must capture
horizon-scale processes, as well as parsec-scale interactions with an
overall spatial dynamic range of $\sim10^5$.  The computational cost
of such large-scale grid-based simulations quickly becomes
prohibitive. Adaptive mesh refinement (AMR) techniques promise an
effective solution for problems where it is necessary to resolve small
and large scale dynamics simultaneously.

Another challenging scenario is presented by radiatively efficient
geometrically thin accretion disks that mandate extreme resolution in the
equatorial plane in order to resolve the growth of MRI instabilities.
Typically this is dealt with by means of stretched grids that concentrate
resolution where needed \citep{Avara2016,Sadowski2016a}. However, when
the disk is additionally tilted with respect to the spin axis of the BH
\citep{Fragile2007b,McKinney2013}, lack of symmetry forbids such an
approach. Here an adaptive mesh that follows the warping dynamics of the
disk can be of great value. The list of scenarios where AMR can have
transformative qualities due to the lack of symmetries goes on, the
modelling of star-disk interactions \cite{Giannios2013}, star-jet
interactions \cite{BarkovAharonian2010}, tidal disruption events
\citep{Tchekhovskoy2014}, complex shock geometries
\cite{Nagakura2008,Meliani2017}, and intermittency in driven-turbulence
phenomena \cite{Radice2012b,Zrake2013}, will benefit greatly from
adaptive mesh refinement.

Over the past few years, the development of general-relativistic
numerical codes employing the $3+1$ decomposition of spacetime and
conservative ``Godunov'' schemes based on approximate Riemann solvers
\cite{Rezzolla_book:2013, Font03, Marti2015} have led to great advances
in numerical relativity. Many general-relativistic hydrodynamic (HD) and
MHD codes have been developed \citep{Hawley84a, Koide00, DeVilliers03a,
  Gammie03, Baiotti04, Duez05MHD0, Anninos05c, Anton05, Mizuno06,
  DelZanna2007, Giacomazzo:2007ti, Radice2012a, Radice2013b,
  McKinney2014, Etienne2015, White2015, Zanotti2015, Meliani2017} and
applied to study a variety of problems in high-energy astrophysics. Some
of these implementations provide additional capabilities that incorporate
approximate radiation transfer \citep[see, \eg][]{Sadowski2013,
  McKinney2013b, Takahashi2016} and/or non-ideal MHD processes
\citep[see, \eg][]{Dionysopoulou:2012pp, Foucart2015b}. Although these
codes have been applied to many astrophysical scenarios involving compact
objects and matter \cite[for recent reviews see,
  \eg][]{Marti2015,Baiotti2016}, full AMR is still not commonly utilised
and exploited \citep[with the exception
  of][]{Anninos05c,Zanotti2015b,White2015}. \bhac~attempts to fill this
gap by providing a fully-adaptive multidimensional GRMHD framework that
features state-of-the-art numerical schemes.

Qualitative aspects of BH accretion simulations are code-independent
\cite[see, \eg][]{DeVilliers03, Gammie03, Anninos05c}, but quantitative
variations raise questions regarding numerical convergence of the
observables \citep{Shiokawa2012,White2015}. In preparation for the
upcoming EHTC observations, a large international effort, whose European
contribution is represented in part by the BlackHoleCam
project\footnote{http://www.blackholecam.org} \citep{Goddi2016}, is
concerned with forward modelling of the future event horizon-scale
interferometric observations of Sgr A* and M87 at submillimeter (EHTC;
\cite{Doeleman2009a}) and near-infrared wavelengths (VLTI GRAVITY;
\cite{Eisenhauer2008}). To this end, GRMHD simulations have been coupled
to general-relativistic radiative transfer (GRRT) calculations \cite[see,
  \eg][]{Moscibrodzka2009, Dexter2009, Chan2015, Gold2016, Dexter2012,
  Moscibrodzka2016}. In order to assess the credibility of these
radiative models, it is necessary to assess the quantitative convergence
of the underlying GRMHD simulations. In order to demonstrate the utility
of \bhac~for the EHTC science-case, we therefore validate the results
obtained with \bhac~against the \harmthreed code
\cite{Gammie03,Noble2009} and investigate the convergence of the GRMHD
simulations and resulting observables obtained with the GRRT
post-processing code \bhoss~\cite{Younsi2017}.

The structure of the paper is as follows. In
Sect.~\ref{sec:numerics} we describe the governing equations and
numerical methods. In Sect.~\ref{sec:tests} we show numerical tests in
special-relativistic and general-relativistic MHD. In
Sect.~\ref{sec:tori} the results of 2D and 3D GRMHD simulations of
magnetised accreting tori are presented. In Sect.~\ref{sec:radiation} we
briefly describe the GRRT post-processing calculation and the resulting
image maps from the magnetised torus simulation shown in
Sect.~\ref{sec:tori}. In Sect.~\ref{sec:outlook} we present our
conclusions and outlook.

\smallskip
Throughout this paper, we adopt units where the speed of light, $c=1$,
the gravitational constant, $G=1$, and the gas mass is normalised to the
central compact object mass. Greek indices run over space and time, i.e.,
$(0,1,2,3),$ and Roman indices run over space only, i.e., $(1,2,3)$. We
assume a $(-,+,+,+)$ signature for the spacetime metric. Self-gravity
arising from the gas is neglected.

\section{GRMHD formulation and numerical methods}\label{sec:numerics}

In this section we briefly describe the covariant GRMHD equations,
introduce the notation used throughout this paper, and present the
numerical approach taken in our solution of the GRMHD system. The
computational infrastructure underlying \bhac~is the versatile
open-source \amrvac toolkit \cite{Keppens2012,Porth2014}.

In-depth derivations of the covariant fluid- and magneto-fluid dynamical
equations can be found in the textbooks by
\cite{Landau-Lifshitz2,Weinberg72,Rezzolla_book:2013}. We follow closely
the derivation of the GRMHD equations by \cite{DelZanna2007}. This is
very similar to the ``Valencia formulation'',
cf. \cite{Rezzolla_book:2013} and \cite{Anton05}. The general
considerations of the ``3+1'' split of spacetime are discussed in greater
detail in \cite{MTW1973,Gourgoulhon2007,Alcubierre:2008}.

We start from the usual set of MHD equations in covariant notation
\begin{align}
\begin{split}
\nabla_{\mu} (\rho u^{\mu}) = 0 \,, \\
\nabla_{\mu} T^{{\mu \nu}} = 0 \,, \\
\nabla_{\mu}\, ^{*}\!F^{\mu\nu} = 0 \,, \label{eq:grmhd}
\end{split}
\end{align}
which respectively constitute mass conservation, conservation of the
energy-momentum tensor $T^{\mu\nu}$, and the homogeneous Faraday's law.
The Faraday tensor $F^{\mu\nu}$ may be constructed from the electric and
magnetic fields $E^\alpha$, $B^\alpha$ as measured in a generic frame
$U^\alpha$ as
\begin{align}
F^{\mu\nu} = U^\mu E^\nu - U^\nu E^\mu - (-g)^{-1/2}\eta^{\mu\nu\lambda\delta} U_\lambda B_\delta \,, \label{eq:genericFaraday}
\end{align}
where $\eta^{\mu\nu\lambda\delta}$ is the fully-antisymmetric symbol
(see, \eg \cite{Rezzolla_book:2013}) and $g$ the determinant of the
spacetime four-metric. The dual Faraday tensor $^{*}\!F^{\mu\nu} :=
\frac{1}{2} (-g)^{-1/2}\eta^{\mu\nu\lambda\delta} F_{\lambda\delta}$ is
then
\begin{align}
^{*}\!F^{\mu\nu} = U^\mu B^\nu - U^\nu B^\mu - (-g)^{-1/2}\eta^{\mu\nu\lambda\delta} U_\lambda E_\delta\,.
\end{align}
We are interested only in the ideal MHD limit of vanishing electric fields in the fluid frame $u^\mu$, hence
\begin{align}
F^{\mu \nu} u_{\nu} = 0 \,, \label{eq:ideal}
\end{align}
such that the inhomogeneous Faraday's law is not required and electric
fields are dependent functions of velocities and magnetic fields. To
eliminate the electric fields from the equations it is convenient to
introduce vectors in the fluid frame and therefore we define the
corresponding electric and magnetic field four-vectors as
\begin{align}
e^{\mu} := F^{\mu\nu}u_{\nu}\,,\qquad b^{\mu} :=\, ^{*}\!F^{\mu\nu} u_{\nu} \,,
\end{align}
where $e^{\mu}=0$ and we obtain the constraint $u_\mu b^\mu = 0$. The Faraday tensor is then 
\begin{align}
  F^{\mu\nu}  = -(-g)^{-1/2}\eta^{\mu\nu\lambda\delta} u_{\lambda}b_{\delta}\,, \qquad 
  ^{*}\!F^{\mu\nu} = b^\mu u^\nu - b^\nu u^\mu \,, \label{eq:fmunu}
\end{align}
and we can write the total energy-momentum tensor in terms of the vectors $u^{\mu}$ and $b^{\mu}$ alone \citep[][]{Anile1990} as
\begin{align}
T^{\mu\nu} = \rho h_{\rm tot}u^{\mu}u^{\nu} + p_{\rm tot} g^{\mu \nu} -b^{\mu}b^{\nu}\,. \label{eq:tmunub}
\end{align}
Here the total pressure $p_{\rm tot}=p+b^{2}/2$ was introduced, as well
as the total specific enthalpy $h_{\rm tot} = h + b^{2}/\rho$. In
addition, we define the scalar $b^{2}:= b^{\nu}b_{\nu}$, denoting the
square of the fluid frame magnetic field strength as $b^{2}=B^{2}-E^{2}$.

\subsection{3+1 split of spacetime}
\label{sub:3+1split}

We proceed to split spacetime into 3+1 components by introducing a
foliation into space-like hyper-surfaces $\Sigma_{t}$ defined as
iso-surfaces of a scalar time function $t$. This leads to the timelike
unit vector normal to the slices $\Sigma_{t}$
\cite{Alcubierre:2008,Rezzolla_book:2013}

\begin{align}
n_{\mu} := -\alpha \nabla_{\mu}t \,, \label{eq:proj-n}
\end{align}
where $\alpha$ is the so-called \textit{lapse}-function. The
four-velocity $n^{\mu}$ defines the frame of the \textit{Eulerian
  observer}. If $g_{\mu\nu}$ is the metric associated with the
four-dimensional manifold, we can define the metric associated with each
timelike slice as
\begin{align}
\gamma_{\mu\nu} := g_{\mu\nu} + n_{\mu} n_{\nu} \,.
\end{align}
This also allows us to introduce the spatial projection operator
\begin{align}
\gamma^\mu_\nu:=\delta_{\nu}^{\mu} + n^{\mu}n_{\nu}  \label{eq:proj-perp}
\end{align}
such that $\gamma^\mu_\nu n_\mu = 0$ 
and through which we can
project any four-vector $V^{\mu}$ (or tensor) into its temporal and
spatial components.

Introducing a coordinate system adapted to the foliation $\Sigma_{t}$,
the line element is given in 3+1 form \cite{Arnowitt2008} as

\begin{align}
ds^{2} = -\alpha^{2} dt^{2} + \gamma_{ij} (dx^{i}+\beta^{i} dt) (dx^{j} + \beta^{j} dt) \,, \label{eq:metric}
\end{align}
where the spatial vector $\beta^{\mu}$ is called the \textit{shift} vector. Written in terms of coordinates, it describes the motion of coordinate lines as seen by an Eulerian observer
\begin{align}
x^{i}_{t+dt} = x^{i}_{t} - \beta^{i}(t,x^{j}) \, dt \,.
\end{align}
More explicitly, we write the metric $g_{\mu\nu}$ and its inverse $g^{\mu\nu}$ as

\begin{align}
g_{\mu\nu} &= 
\left(
\begin{array}{cc}
-\alpha^{2}+\beta_{k}\beta^{k}  &  \beta_{i}   \\
\beta_{j}  &   \gamma_{ij}
\end{array}
\right)\,, 
\qquad g^{\mu\nu} \hspace*{-2.5mm} &= 
\left(
\begin{array}{cc}
-1/\alpha^{2} &  \beta^{i}/\alpha^{2}   \\
\beta^{j}/\alpha^{2}  &   \gamma^{ij} - \beta^{i}\beta^{j}/\alpha^{2} \label{eq:gmunu}
\end{array}
\right) \,.
\end{align}
From (\ref{eq:gmunu}) we find the following useful relation between the determinants of the 3-metric and 4-metric
\begin{align}
(-g)^{1/2} = \alpha \gamma^{1/2} \,.
\end{align}
In a coordinate system specified by (\ref{eq:metric}), the four-velocity of the Eulerian observer reads
\begin{align}
n_{\mu} = (-\alpha,0_{i}),\qquad n^{\mu} = (1/\alpha,-\beta^{i}/\alpha) \,.
\end{align}
It is easy to verify that this normalised vector is indeed orthogonal to
any space-like vector on the foliation $\Sigma_{t}$. Given a fluid
element with four-velocity $u^\mu$, the Lorentz factor with respect to
the Eulerian observer is\footnote{This quantity is often
  indicated as $W$ \cite{Anton05,Rezzolla_book:2013}} $\Gamma:=-u^\mu
n_\mu=\alpha u^0$ and we introduce the three-vectors
\begin{align}
  v^i := \frac{\gamma^i_\mu u^\mu}{\Gamma}=\frac{u^{i}}{\Gamma}
  +\frac{\beta^{i}}{\alpha} \,, \qquad v_i := \gamma_{ij} v^j = \frac{u_i}{\Gamma} \,,
\end{align}
which denote the fluid three-velocity. 

In the following, purely spatial vectors (\eg $v^0=0$) are denoted by Roman indices. 
Note that $\Gamma=(1-v^2)^{-1/2}$ with $v^2=v_i v^i$ just as in special relativity.

Further useful three-vectors are the electric and magnetic fields in the Eulerian frame
\begin{align}
  E^i := F^{i\nu}n_\nu =\alpha F^{i0} \,,\qquad B^i :=\, ^{*}F^{i \nu}n_{\nu} =\, \alpha ^{*}F^{i 0} \,, \label{eq:Beuler}
\end{align}
which differ by a factor $\alpha$ from the definitions used in
\cite{Komissarov1999,Gammie03}. Writing the general Faraday tensor
(\ref{eq:genericFaraday}) in terms of quantities in the Eulerian frame,
the ideal MHD condition (\ref{eq:ideal}) leads to the well known relation
\begin{align}
E^i = \gamma^{-1/2} \eta^{ijk} B_j v_k \,,
\end{align}
or put simply: $\boldsymbol{E}=\boldsymbol{B}\times \boldsymbol{v}$ (here
$\eta_{ijk}$ is the standard Levi-Civita antisymmetric symbol).
Combining (\ref{eq:fmunu}) with (\ref{eq:Beuler}), one obtains the
transformation between $b^\mu$ and $B^\mu$ as
\begin{align}
b^i = \frac{B^\mu+\alpha b^0 u^i}{\Gamma} \,, \qquad b^0 = \frac{\Gamma (B^i v_i)}{\alpha} \label{eq:littleb}
\end{align}
which enables the dual Faraday tensor (\ref{eq:fmunu}) to be expressed in terms of the Eulerian fields
\begin{align}
	^{*}\!F^{\mu\nu} = \frac{B^\mu u^\nu - B^\nu u^\mu}{\Gamma} \,. \label{eq:faradayBeuler}
\end{align}
Equation (\ref{eq:grmhd}) with the Faraday tensor in the form (\ref{eq:faradayBeuler}) yields the final evolution equation for $B^\mu$. The time component of this leads to the constraint $\partial_i \sqrt{\gamma} B^i=0$ or put more simply: $\boldsymbol{\nabla\cdot B}=0$. 
Following \eqref{eq:littleb} we obtain the scalar $b^2$ as
\begin{align}
b^{2} = \frac{B^{2}+\alpha^{2} (b^{0})^{2}}{\Gamma^{2}} = \frac{B^{2}}{\Gamma^{2}}+(B^{i}v_{i})^{2} \,,
\end{align}
where $B^{2}:= B^{i}B_{i}$. 

Using the spatial projection operator, the GRMHD equations
(\ref{eq:grmhd}) can be decomposed into spatial and temporal components.
We skip ahead over the involved algebra \citep[see \eg][]{DelZanna2007}
and directly state the final conservation laws
\begin{align}
\partial_{t} (\sqrt{\gamma} \, \boldsymbol{U}) + \partial_{i} (\sqrt{\gamma} \, \boldsymbol{F}^{i}) = \sqrt{\gamma} \, \boldsymbol{S} \,, \label{eq:conservationlaw}
\end{align}
with the conserved variables $\boldsymbol{U}$ and fluxes $\boldsymbol{F}^{i}$ defined as

\begin{align}
\boldsymbol{U} = 
\left[
\begin{array}{c}
D  \\
S_{j}  \\
\tau \\
B^{j}
\end{array}
\right] \,, \ \qquad 
\boldsymbol{F}^{i} = 
\left[
\begin{array}{c}
\mathcal{V}^{i} D \\
\alpha W^{i}_{j} - \beta^{i} S_{j} \\
\alpha (S^{i}-v^{i} D) - \beta^{i} \tau \\
\mathcal{V}^{i}B^{j} - B^{i}\mathcal{V}^{j}
\end{array}
\right] \,, \label{eq:uandf}
\end{align}
where we define the \textit{transport velocity} $\mathcal{V}^{i} :=
\alpha v^{i} - \beta^{i}$. Hence we solve for conservation of quantities
in the Eulerian frame: the density $D:=-\rho u^\nu n_\nu$, the covariant
three-momentum $S_j$, the rescaled energy density
$\tau=U-D$ \footnote{Using $\tau=U-D$ instead of $U$ improves accuracy in
  regions of low energy and enables one to consistently recover the
  Newtonian limit.} (where $U$ denotes the total energy density as seen
by the Eulerian observer), and the Eulerian magnetic three-fields $B^j$.
The conserved energy density $U$ is given by
\begin{align}
U:= T^{\mu\nu}n_{\mu}n_{\nu} &= \rho h \Gamma^{2} - p +\frac{1}{2} \left(E^{2}+B^{2}\right) \\
&= \rho h \Gamma^{2} - p +\frac{1}{2} \left[B^{2}(1+v^{2}) - (B^{j}v_{j})^{2} \right] \,. \label{eq:uideal}
\end{align}
The purely spatial variant of the stress-energy tensor $W^{ij}$ was
introduced for example in (\ref{eq:uandf}). It reads just as in special relativity
\begin{align}
W^{ij} := \gamma^i_\mu \gamma^j_\nu T^{\mu\nu} 
&= \rho h \Gamma^{2} v^{i} v^{j} - E^{i} E^{j} - B^{i} B^{j} 
+ \left[p+\frac{1}{2} (E^{2}+B^{2})\right] \gamma^{ij} \\
&= S^{i} v^{j} +p_{\rm tot}\gamma^{ij} -\frac{B^{i}B^{j}}{\Gamma^2} - (B^{k}v_{k})v^{i}B^j \,.
\end{align}
Correspondingly, the covariant three-momentum density in the Eulerian frame is
\begin{align}
  S_i:= \gamma^\mu_i n^{\alpha}T_{\alpha\mu}
  &= \rho h\Gamma^2 v_i + \eta_{ijk} \gamma^{1/2} E^j B^k \\
  &= \rho h \Gamma^2 v_i + B^{2} v_{i} - (B^{j}v_{j})B_{i} \,, \label{eq:Si}
\end{align}
as usual. For the sources $\boldsymbol{S}$ we employ the convenient Valencia formulation without Christoffel symbols, yielding 
\begin{align}
\boldsymbol{S} = 
\left[
\begin{array}{c}
0  \\
\frac{1}{2}\alpha W^{ik}\partial_{j}\gamma_{ik} + S_{i}\partial_{j}\beta^{i} - U\partial_{j}\alpha \\
\frac{1}{2} W^{ik} \beta^{j} \partial_{j} \gamma_{ik} + W_{i}^{j}\partial_{j}\beta^{i} - S^{j} \partial_{j} \alpha \\ 
0
\end{array}
\right] \, \label{eq:ss}
\end{align}
which is valid for stationary spacetimes that are considered for the
  remainder of this work (Cowling approximation).  Following the
definitions (\ref{eq:uandf}) and (\ref{eq:ss}), all vectors and tensors
are now specified through their purely spatial variants and thus apart
from the occurrence of the lapse function $\alpha$ and the shift vector
$\beta^i$, the equations take on a form identical to the
special-relativistic MHD (SRMHD) equations. This fact allows for a
straightforward transformation from the SRMHD physics module of \amrvac
into a full GRMHD code.

In addition to the set of conserved variables $\boldsymbol{U}$, knowledge of the primitive variables $\boldsymbol{P}(\boldsymbol{U})$ is required for the calculation of fluxes and source terms. They are given by 
\begin{align}
\boldsymbol{P} = [\rho,\Gamma v^i, p, B^i] \,. 
\end{align}
While the transformation $\boldsymbol{U}(\boldsymbol{P})$ is straightforward, the inversion $\boldsymbol{P}(\boldsymbol{U})$ is a non-trivial matter which will be discussed further in Sect~\ref{sec:con2prim}. 
Note that just like in \amrvac, we do not store the primitive variables $\boldsymbol{P}$ but extend the conserved variables by the set of \emph{auxiliary} variables 
\begin{align}
\boldsymbol{A}=[\Gamma,\xi] \,,
\end{align}
where $\xi:=\Gamma^2\rho h$. Knowledge of $\boldsymbol{A}$ allows for quick transformation of $\boldsymbol{P}(\boldsymbol{U})$. The issue of inversion then becomes a matter of finding an $\boldsymbol{A}$ consistent with both $\boldsymbol{P}$ and $\boldsymbol{U}$. 

\subsection{Finite volume formulation}
\label{sub:finite-volume}
Since \bhac~solves the equations in a finite volume formulation, we take the integral of equation (\ref{eq:conservationlaw}) over the spatial element of each cell $\int dx^{1}dx^{2}dx^{3}$

\begin{align}
\int \partial_{t} (\gamma^{1/2} \boldsymbol{U^{}}) dx^{1}dx^{2}dx^{3} + \int \partial_{i}(\gamma^{1/2} \boldsymbol{F^{i}}) dx^{1}dx^{2}dx^{3} 
= \int \gamma^{1/2} \boldsymbol{S} dx^{1}dx^{2}dx^{3}\,.
\end{align}
This can be written (cf. \cite{Banyuls97}) as 

\begin{align}
\begin{split}
\partial_{t} (\boldsymbol{\bar{U}} \Delta V) + 
\int_{\partial V(x^{1}+{\Delta x^{1}}/{2})} \gamma^{1/2} \boldsymbol{F^{1}} dx^{2} dx^{3} 
- \int_{\partial V(x^{1}-{\Delta x^{1}}/{2})} \gamma^{1/2} \boldsymbol{F^{1}} dx^{2} dx^{3} \\
+ \int_{\partial V(x^{2}+{\Delta x^{2}}/{2})} \gamma^{1/2} \boldsymbol{F^{2}} dx^{1} dx^{3} 
- \int_{\partial V(x^{2}-{\Delta x^{2}}/{2})} \gamma^{1/2} \boldsymbol{F^{2}} dx^{1} dx^{3} \\
+ \int_{\partial V(x^{3}+{\Delta x^{3}}/{2})} \gamma^{1/2} \boldsymbol{F^{3}} dx^{1} dx^{2} 
- \int_{\partial V(x^{3}-{\Delta x^{3}}/{2})} \gamma^{1/2} \boldsymbol{F^{3}} dx^{1} dx^{2} \\
= \bar{\boldsymbol{S}} \Delta V \,,
\end{split}
\end{align}
with the volume averages defined as 
\begin{align}
\boldsymbol{\bar{U}} := \frac{\int \gamma^{1/2}\boldsymbol{U}dx^{1}dx^{2}dx^{3}}{\Delta V} \,, \qquad 
\boldsymbol{\bar{S}} := \frac{\int \gamma^{1/2}\boldsymbol{S}dx^{1}dx^{2}dx^{3}}{\Delta V} \,,
\end{align}
and
\begin{align}
\Delta V = \int \gamma^{1/2}dx^{1}dx^{2}dx^{3} \,.
\end{align}
We next define also the ``surfaces'' $\Delta S^{i}$ and corresponding
surface-averaged fluxes
\begin{align}
\Delta S^{i}_{\partial V(x^{i}+{\Delta x^{i}}/{2})} = \int_{\partial V(x^{i}+{\Delta x^{i}}/{2})} \gamma^{{1/2}} dx^{j,j\ne i} \,,
\end{align}
and
\begin{align}
\label{eq:averagedFluxes}
\boldsymbol{\bar{F^{i}}}_{\partial V(x^{i}+{\Delta x^{i}}/{2})} = \frac{\int_{\partial V(x^{i}+{\Delta x^{i}}/{2})} \gamma^{{1/2}} \boldsymbol{F^i} dx^{j,j\ne i}
}{\Delta S^{i}} \,.
\end{align}
Considering that $\Delta V$ is assumed constant in time, this leads
to the evolution equation
\begin{align}
\begin{split}
\partial_{t} \boldsymbol{\bar{U}} = 
- \frac{1}{\Delta V}
\Biggl[
&\boldsymbol{\bar{F}^{1}}\Delta S^{1}\bigr|_{\partial V(x^{1}+{\Delta x^{1}}/{2})} - \boldsymbol{\bar{F}^{1}}\Delta S^{1}\bigr|_{\partial V(x^{1}-{\Delta x^{1}}/{2})} + \\
&\boldsymbol{\bar{F}^{2}}\Delta S^{2}\bigr|_{\partial V(x^{2}+{\Delta x^{2}}/{2})} - \boldsymbol{\bar{F}^{2}}\Delta S^{2}\bigr|_{\partial V(x^{2}-{\Delta x^{2}}/{2})} + \\
&\boldsymbol{\bar{F}^{3}}\Delta S^{3}\bigr|_{\partial V(x^{3}+{\Delta x^{3}}/{2})} - \boldsymbol{\bar{F}^{3}}\Delta S^{3}\bigr|_{\partial V(x^{3}-{\Delta x^{3}}/{2})} \bigr] + \boldsymbol{\bar{S}} \,.
\end{split}
\end{align}
We aim to achieve second-order accuracy and represent the
interface-averaged flux, \eg $\boldsymbol{\bar{F}^{1}}_{\partial
  V(x^{1}+{\Delta x^{1}}/{2})}$, with the value at the midpoint, change
to an intuitive index notation $\boldsymbol{F^{1}}_{i+1/2,j,k}$, and then
arrive at a semi-discrete equation for the average state in the cell
$(i,j,k)$ as
\begin{align}
\begin{split}
\frac{d\boldsymbol{\bar{U}}_{i,j,k}}{dt} = 
- \frac{1}{\Delta V_{i,j,k}}
\Biggl[
&\boldsymbol{F^{1}}\Delta S^{1}\bigr|_{i+1/2,j,k} - \boldsymbol{F^{1}}\Delta S^{1}\bigr|_{i-1/2,j,k} + \\
&\boldsymbol{F^{2}}\Delta S^{2}\bigr|_{i,j+1/2,k} - \boldsymbol{F^{2}}\Delta S^{2}\bigr|_{i,j-1/2,k} + \\
&\boldsymbol{F^{3}}\Delta S^{3}\bigr|_{i,j,k+1/2} - \boldsymbol{F^{3}}\Delta S^{3}\bigr|_{i,j,k-1/2} \bigr]+ \boldsymbol{S}_{i,j,k} \,.
\label{eq:fvolume}
\end{split}
\end{align}
Here the source term $\boldsymbol{S}_{i,j,k}$ is also evaluated at the cell barycenter to second-order accuracy \citep{Mignone2014}. 
Barycenter coordinates $\bar{x}^{i}$ are straightforwardly defined as
\begin{align}
\bar{x}^{i} = \frac{\int \gamma^{1/2} x^{i} dx^{1}dx^{2}dx^{3}} {\Delta V}\,.
\end{align}
This finite volume form is readily solved with the {\tt MPI-AMRVAC} toolkit. 
For ease of implementation, we pre-compute all static integrals yielding cell volumes $\Delta V$, Surfaces $\Delta S^{i}$  and barycenter coordinates. The integrations are performed numerically at the phase of initialisation using a fourth-order Simpson's rule. 

For the temporal update, we interpret the semi-discrete form
(\ref{eq:fvolume}) as an ordinary differential equation in time for each
cell and employ a multi-step Runge-Kutta scheme to evolve the average
state in the cell $\boldsymbol{\bar{U}}_{{i,j,k}}$, a procedure also
known as ``method of lines''.  At each sub-step, the point-wise interface
fluxes $\boldsymbol{F^{i}}$ are obtained by performing a limited
  reconstruction operation of the cell-averaged state
$\boldsymbol{\bar{U}}$ to the interfaces (see
Sect.~\ref{sec:reconstructions}) and employing approximate Riemann
solvers, \eg \textit{HLL} or \textit{TVDLF} (Sect.~\ref{sec:riemann}).

Several temporal update schemes are available: simple predictor-corrector, third-order Runge-Kutta (RK) RK3 \citep{Gottlieb98} and the strong-stability preserving $s$-step, $p$th-order RK schemes SSPRK($s$,$p$) schemes: SSPRK(4,3), SSPRK(5,4) due to \cite{Spiteri2002}.\footnote{For implementation details, see \cite{Porth2014}.}

\subsection{Metric data-structure} \label{sec:datastructure}

The metric data-structure is built to be optimal in terms of storage
while remaining convenient to use. Since the metric and its derivatives
are often sparsely populated, the data is ultimately stored using index
lists.  For example, each element in the index list for the four-metric
$g_{\mu\nu}$ holds the indices of the non-zero element together with a
\texttt{Fortran90} array of the corresponding metric coefficient for the
grid block.  A summation over indices, \eg ``lowering'' can then be cast
as a loop over entries in the index-list only.  For convenience, all
elements can also be accessed directly over intuitive identifiers which
point to the storage in the index list, \eg {\tt m\%g(mu,nu)\%elem}
yields the grid array of the $g_{\mu\nu}$ metric coefficients as
expected.  Similarly, the lower-triangular indices point to the
transposed indices in the presence of symmetries.  In addition, one block
of zeros is allocated in the metric data-structure and all zero elements
are set to point towards it. An overview of the available identifiers is
given in Table~\ref{tab:metricdatastructure}.
\begin{table}[htp]
\caption{Elements of the metric data-structure}
\begin{center}
\begin{tabular}{l|l|l}
Symbol & Identifier & Index list\\
\hline
$g_{\mu\nu}$ & {\tt m\%g(mu,nu)} & {\tt m\%nnonzero, m\%nonzero(inonzero)} \\
$\alpha$ & {\tt m\%alpha} & - \\
$\beta^{i}$ & {\tt m\%beta(i)} & {\tt m\%nnonzeroBeta, m\%nonzeroBeta(inonzero)} \\
$\sqrt{\gamma}$ & {\tt m\%sqrtgamma} & -\\
$\gamma^{ij}$ & {\tt m\%gammainv(i,j)} & -\\
$\beta_{i}$ & {\tt m\%betaD(i)} &  - \\
$\del{k}\gamma_{ij}$ & {\tt m\%dgdk(i,j,k)} & {\tt m\%nnonzeroDgDk, m\%nonzeroDgDk(inonzero)}\\
$\del{j}\beta^{i}$ & {\tt m\%DbetaiDj(i,j)} & {\tt m\%nnonzeroDbetaiDj, m\%nonzeroDbetaiDj(inonzero)} \\
$\del{j}\alpha$ & {\tt m\%DalphaDj(j)} & {\tt m\%nnonzeroDalphaDj, m\%nonzeroDalphaDj(inonzero)}  \\
$0$ & {\tt m\%zero} & - 
\end{tabular}
\end{center}
\label{tab:metricdatastructure}
\end{table}

As a consequence, only 14 grid functions are required for the
Schwarzschild coordinates and 29 grid functions need to be allocated
in the Kerr-Schild (KS) case. This is still less than half of the $68$
grid functions which a brute-force approach would yield.  The need for
efficient storage management becomes apparent when we consider that the
metric is required in the barycenter as well as on the interfaces, thus
multiplying the required grid functions by a factor of four for
three-dimensional simulations (yielding $116$ grid functions in the KS
case).

In order to eliminate the error-prone process of implementing complicated
functions for metric derivatives, \bhac~can obtain derivatives by means
of an accurate complex-step numerical differentiation \cite{Squire1998}.
This elegant method takes advantage of the Cauchy-Riemann differential
equations for complex derivatives and achieves full double-precision
accuracy, thereby avoiding the stepsize dilemma of common
finite-differencing formulae \cite{Martins2003}. The small price to pay
is that at the initialisation stage, metric elements are provided via
functions of the complexified coordinates. However, the intrinsic complex
arithmetic of \texttt{Fortran90} allows for seamless implementation.

To promote full flexibility in the spacetime, we always calculate the
inverse metric $\gamma^{ij}$ using the standard LU decomposition
technique \cite{Press2007}. As a result, GRMHD simulations on any metric
can be performed after providing only the non-zero elements of the
three-metric $\gamma_{ij}(x^1,x^2,x^3)$, the lapse function
$\alpha(x^1,x^2,x^3)$ and the shift vector $\beta^i(x^1,x^2,x^3)$. As an
additional convenience, \bhac~can calculate the required elements and
their derivatives entirely from the four-metric
$g_{\mu\nu}(x^0,x^1,x^2,x^3)$.

\subsection{Equations of state} \label{sec:eos}

For closure of the system (\ref{eq:grmhd})-(\ref{eq:ideal}), an equation
of state (EOS) connecting the specific enthalpy $h$ with the remaining
thermodynamic variables $h(\rho,p)$ is required
\cite{Rezzolla_book:2013}. The currently implemented closures are

\begin{itemize}
\item
\emph{Ideal gas}: $h(\rho,p)= 1 + \dfrac{\hat{\gamma}}{\hat{\gamma}-1}
\dfrac{p}{\rho}$ with adiabatic index $\hat{\gamma}$.
\item
\emph{Synge gas}: $h(\Theta) =
\dfrac{K_{3}(\Theta^{-1})}{K_{2}(\Theta^{-1})}$, where the relativistic
temperature is given by $\Theta=p/\rho$ and $K_n$ denotes the modified
Bessel function of the second kind. In fact, we use an approximation to
the previous expression that does not contain Bessel functions
\cite[see][]{Meliani2004,Keppens2012}.
\item
\emph{Isentropic flow}: Assumes an ideal gas with the additional
constraint $p=\kappa \rho^{\hat{\gamma}}$, where the pseudo-entropy
$\kappa$ may be chosen arbitrarily. This allows one to omit the energy
equation entirely and only the reduced set
$\boldsymbol{P}=\{\rho,v^{j},B^{j}\}$ is solved.
\end{itemize}
As long as $h(\rho,p)$ is analytic, its implementation in \bhac~is
straightforward.

\subsection{Divergence cleaning and augmented Faraday's law}

To control the $\boldsymbol{\nabla \cdot B}=0$ constraint on AMR grids,
we have adopted a constraint dampening approach customarily used in
Newtonian MHD \cite{Dedner:2002}. In this approach, which is usually
referred as Generalized Lagrangian Multiplier (GLM) of the Maxwell
equations (but is also known as the ``divergence-cleaning'' approach), we
extend the usual Faraday tensor by the scalar $\phi$, such that the
homogeneous Maxwell equation reads
\begin{align}
\nabla_{\nu} (^{*}\!F^{\mu\nu} - \phi g^{\mu\nu}) = -\kappa n^{\mu} \phi \,, \label{eq:maxglm}
\end{align}
and the scalar $\phi$ follows from contraction $\phi=(^{*}\!F^{\mu\nu}-
\phi g^{\mu\nu})n_{\mu}n_{\nu}$. Naturally, for $\phi\to0$, the usual
set of Maxwell equations is recovered. It is straightforward to show
\citep[see, \eg][]{Palenzuela:2008sf} that (\ref{eq:maxglm}) leads to a
telegraph equation for the constraint violation parameter $\phi$ which
becomes advected at the speed of light and decays on a timescale
$1/\kappa$. With the modification (\ref{eq:maxglm}), the time-component
of Maxwell's equation now becomes an evolution equation for $\phi$.
After some algebra (see Appendix~\ref{sec:phi}), we obtain
\begin{align}
\begin{split}
\partial_{t} \sqrt{\gamma} \phi 
 + \partial_{i}[\sqrt{\gamma}(\alpha B^{i}-\phi \beta^{i})] = 
& - \sqrt{\gamma} \alpha  \kappa \phi - \sqrt{\gamma} \phi \partial_{i}\beta^{i} \\
& - \frac{1}{2}\sqrt{\gamma} \phi \gamma^{ij}\beta^{k}\partial_{k}\gamma_{ij} 
 + \sqrt{\gamma} B^i\partial_i \alpha
\, . \label{eq:phievol}
\end{split}
\end{align}
Equivalently, the modified evolution equations for $B^i$ (see Appendix~\ref{sec:modfaraday}) read 
\begin{align}
\begin{split}
\del{t}(\sqrt{\gamma} B^{j}) 
+ \del{i}\left(\sqrt{\gamma} (\vt^{i}B^{j}-\vt^{j}B^{i}-B^{i}\beta^{j})\right) 
= 
- \sqrt{\gamma}B^{i} \del{i}\beta^{j} 
- \sqrt{\gamma}\alpha \gamma^{ij}\del{i}\phi\, .
\end{split}
\label{eq:fglm2}
\end{align}
Now equation (\ref{eq:fglm2}) replaces the usual Faraday's law and
(\ref{eq:phievol}) is evolved alongside the modified MHD system. Due to
the term $\partial_i\phi$ on the right hand side of equation
(\ref{eq:fglm2}), the new equation is non-hyperbolic.  Hence,
  numerical stability can be a more involved issue than for hyperbolic
  equations.  We find that the numerical stability of the system is
enhanced when using an upwinded discretisation for $\partial_i\phi$.
Note that Equations (\ref{eq:phievol}) and (\ref{eq:fglm2}) are in
  agreement with \cite{Dionysopoulou:2012pp} when accounting for
  $\frac{\partial_i\sqrt{\gamma}}{\sqrt{\gamma}}=\frac{1}{2}\gamma^{lm}\partial_i\gamma_{lm}$
  and taking the ideal MHD limit.  

\subsection{Flux-interpolated Constrained Transport}
As an alternative to the GLM approach, the $\boldsymbol{\nabla \cdot
  B}=0$ constraint can be enforced using a cell-centred version of
Flux-interpolated Constrained Transport (FCT) consistent with the finite
volume scheme used to evolve the hydrodynamic variables. Constrained
Transport (CT) schemes aim to keep to zero at machine precision the sum
of the magnetic fluxes through all surfaces bounding a cell, and
therefore (in the continuous limit) the divergence of the magnetic field
inside the cell. In the original version \cite{Evans1988} this is
achieved by evolving the magnetic flux through the cell faces and
computing the circulation of the electric field along the edges bounding
each face. Since each edge appears with opposite signs in the time
update of two faces belonging to the same cell, the total magnetic flux
leaving each cell is conserved during evolution. The magnetic field
components at cell centers, necessary for performing the transformation
from primitive to conserved variables and vice-versa, are then found
using interpolation from the cell faces. \cite{Toth2000} showed that it
is possible to find cell centred variants of CT schemes that go from the
average field components at the cell center at a given time to those one
(partial) time step ahead in a single step, without the need to compute
magnetic fluxes at cell faces. The CT variant known as FCT is
particularly well suited for finite volume conservative schemes as that
employed by \bhac, as it calculates the electric fields necessary for the
update as an average of the fluxes given by the Riemann solver. In this
way, the time update for its cell centred version can be written using a
form similar to \eqref{eq:fvolume}. For example, for the update of the
$\bar{B}^1$ component, we obtain

\begin{align}
\begin{split}
\frac{d\bar{B}_{i,j,k}^1}{dt} = -\frac{1}{\Delta V_{i,j,k}}
\Biggl[
&F^{*\; 2} \Delta S^2 \bigr|_{i,j+1/2,k} -  F^{*\; 2} \Delta S^2\bigr|_{i,j-1/2,k} +\\
&F^{*\; 3} \Delta S^3 \bigr|_{i,j,k+1/2} -  F^{*\; 3} \Delta S^3\bigr|_{i,j,k-1/2}\Biggr] \,,
\end{split}
\label{eq:Bbar_evol}
\end{align}

\noindent
where the modified fluxes in the $x^1$-direction are zero and the remaining fluxes are calculated as

\begin{align}
\begin{split}
F^{* 2} \Delta S^2 \bigr|_{i,j-1/2,k} =
\frac{\Delta x^1_i}{8}
\Biggl(&2\frac{ \bar{F}^2 \Delta S^2 \bigr|_{i,j-1/2,k}}{\Delta x^1_i}+  \\  
&\phantom{2} \frac{\bar{F}^2 \Delta S^2 \bigr|_{i+1,j-1/2,k}}{\Delta x^1_{i+1}} 
            +\frac{\bar{F}^2 \Delta S^2 \bigr|_{i-1,j-1/2,k}}{\Delta x^1_{i-1}} - \\ 
&\phantom{2} \frac{\bar{F}^1 \Delta S^1 \bigr|_{i-1/2,j,k}}{\Delta y_{j}}
            -\frac{\bar{F}^1 \Delta S^1 \bigr|_{i-1/2,j-1,k}}{\Delta x^2_{j-1}} - \\
&\phantom{2} \frac{\bar{F}^1 \Delta S^1 \bigr|_{i+1/2,j,k}}{\Delta x^2_{j}}
            -\frac{\bar{F}^1 \Delta S^1 \bigr|_{i+1/2,j-1,k}}{\Delta x^2_{j-1}}\Biggr) \,.
\end{split}
\label{eq:average_fluxes}
\end{align}
The derivation of equations \eqref{eq:Bbar_evol} and
\eqref{eq:average_fluxes} from the staggered version with magnetic fields
located at cell faces is given in Appendix~\ref{sec:fvolumeFCT}. Since
magnetic fields are stored at the cell center and not at the faces, the
divergence conserved by the FCT method corresponds to a particular
discretisation
\begin{align}
\begin{split}
&\left. \frac{1}{2}\Delta V^* (\boldsymbol{\nabla \cdot B}) \right|_{i+1/2,j+1/2,k+1/2} = \\
&\sum_{l_1,l_2,l_3=0,1}\left[(-1)^{1+l_1}\frac{\bar{B}^1\Delta V}{\Delta x^1} +(-1)^{1+l_2}\frac{\bar{B}^2\Delta V}{\Delta x^2}
                            +(-1)^{1+l_3}\frac{\bar{B}^3\Delta V}{\Delta x^3}\right]_{i+l_1,j+l_2,k+l_3} \,,
\end{split}
\label{eq:div}
\end{align}
\noindent
where
\begin{equation}
\left. \Delta V^* \right|_{i+1/2,j+1/2,k+1/2} = \sum_{l_1,l_2,l_3=0,1}\left.\Delta V \right|_{i+l_1,j+l_2,k+l_3} \,.
\end{equation}

\noindent
Equation \eqref{eq:div} is closely related to the integral over the surface of a volume containing eight cells in 3D (see Appendix~\ref{sec:initial_div} for the derivation), and it reduces to equation (27) from \cite{Toth2000} in the special case of Cartesian coordinates.
As mentioned before, this scheme can maintain $\boldsymbol{\nabla \cdot B}=0$ to machine precision only if it was already zero at the initial condition.
The corresponding curl operator used to setup initial conditions is derived in Appendix~\ref{sec:initial_div}. 

In its current form, \bhac~cannot handle both constrained transport and AMR. The reason is that special prolongation and restriction operators are required
in order to avoid the creation of divergence when refining or coarsening.
Due to the lack of information about the magnetic flux on cell faces, the problem of finding such divergence-preserving prolongation operators
becomes underdetermined.
However, storing the face-allocated (staggered) magnetic fluxes and applying the appropriate prolongation and restriction operators
requires a large change in the code infrastructure on which we will report in an accompanying work. 
 
\subsection{Coordinates}
Since one of the main motivations for the development of the \bhac~code is to simulate BH accretion in arbitrary metric theories of gravity, the coordinates and metric data-structures have been designed to allow for maximum flexibility and can easily be extended. 
A list of the currently available coordinate systems is given in Table~\ref{tab:coordinates}. In addition to the identifiers used in the code, the table lists whether numerical derivatives are used and whether the coordinates are initialised from the three-metric or the four-metric. 
The less well-known spacetimes and coordinates are described in the following subsection.  

\begin{table}[htp]
\caption{Coordinates available in \bhac}
\begin{center}
\begin{tabular}{l|l|ll}
Coordinates & Identifier & Num. derivatives & Init. $g_{\mu\nu}$ \\
\hline
Cartesian & {\tt cart} & No & No\\
Boyer-Lindquist & {\tt bl} & No & No \\
Kerr-Schild & {\tt ks} & No & No \\
Modified Kerr-Schild & {\tt mks} & No & No \\
Cartesian Kerr-Schild & {\tt cks} & Yes & Yes \\
Rezzolla \& Zhidenko parametrization \cite{Rezzolla2014} & {\tt rz} & Yes  & No\\
Horizon penetrating Rezzolla \& Zhidenko coordinates & {\tt rzks} & Yes  & Yes\\
Hartle-Thorne \cite{Hartle68} & {\tt ht} & Yes  & Yes
\end{tabular}
\end{center}
\label{tab:coordinates}
\end{table}

\subsubsection{Modified Kerr-Schild coordinates}\label{sec:modkscoord}

Modified KS coordinates were introduced by \eg \cite{McKinney2004} with the purpose of stretching the grid radially and being able to concentrate resolution in the equatorial region. 

The original coordinate transformation is equivalent to:
\begin{align}
 r_{\rm KS}(s) &= R_0 + e^s \,, \\
 \theta_{\rm KS}(\vartheta) &= \vartheta + \frac{h}{2} \sin(2\vartheta) \,, \label{eq:varthMK}
\end{align}
where $R_0$ and $h$ are parameters which control, respectively, how much resolution is concentrated near the horizon and near the equator. 

\begin{figure}[htbp]
\begin{center}
\includegraphics[width=.6\textwidth]{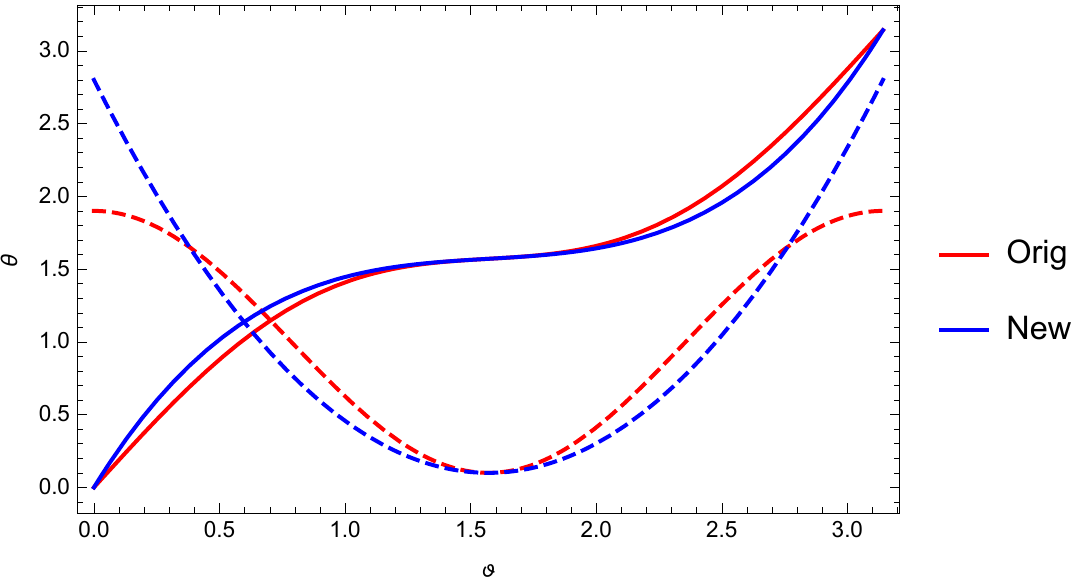}
\caption{$\theta$-grid stretching functions comparing the transcendental
  function $\vartheta(\theta_{\rm KS})$ (solid red curves) with the cubic
  approach (solid blue curves) for $h=0.9$. We also give the respective
  derivatives $d\theta/d\vartheta$ (dashed). }
\label{fig:modKS}
\end{center}
\end{figure}

Unfortunately, the inverse of $\vartheta(\theta)$ is a transcendental equation that has to be solved numerically. To avoid this complication and still capture the functionality of the modified coordinates, we instead use the following $\theta-$ transformation
\begin{align}
 \theta_{\rm KS}(\vartheta) &= \vartheta + \frac{2 h \vartheta}{\pi^2}(\pi-2\vartheta)(\pi-\vartheta) \,.
\end{align}
Now the solution to the cubic equation can be expressed in closed-form, and the only real root reads
\begin{align}
\vartheta(\theta_{\rm KS})=\frac{1}{12} \pi ^{2/3} \left(-\frac{2 \sqrt[3]{2} (3 \pi )^{2/3} (h-1)}{R(\theta_{\rm KS})}-\frac{2^{2/3} \sqrt[3]{3} R(\theta_{\rm KS})}{h}+6 \sqrt[3]{\pi }\right) \,,
\end{align}
where
\begin{align}
R(\theta_{\rm KS}) = \biggl[&
h \sqrt{-3h \left[-108 h \theta_{\rm KS} ^2+108 \pi  h \theta_{\rm KS} +(h-4) (2 \pi  h+\pi )^2\right]}\nonumber \\
& 9 (\pi -2 \theta_{\rm KS} ) h^2+\biggr]^{1/3} \,.
\end{align}
This is compared with the original version (\ref{eq:varthMK}) in Fig.~\ref{fig:modKS} and shows a good match between the two versions of modified Kerr-Schild coordinates. 
The radial back-transformation follows trivially as
\begin{align}
s(r_{\rm KS})=\ln(r_{\rm KS}-R_0) \,,
\end{align}
and the derivatives for the diagonal Jacobian are
\begin{align}
\partial_s r_{\rm KS} &= e^s \\
\partial_\vartheta \theta_{\rm KS} &= 1 + 2h +12h((\vartheta/\pi)^2-\vartheta/\pi) \,.
\end{align}

With these transformations, we obtain the new metric $g_{\rm{MKS}} = J^\mathrm{T} g_{\rm{KS}} J$. Note that whenever the parameters $h=0$ and $R_{0}=0$ are set, our MKS coordinates reduce to the standard \emph{logarithmic} Kerr-Schild coordinates.

\subsubsection{Rezzolla \& Zhidenko parametrization}\label{sec:RZ}
The Rezzolla-Zhidenko parameterisation \cite{Rezzolla2014} has been
proposed to describe spherically-symmetric BH geometries in metric
theories of gravity. In particular, using a continued-fraction expansion
(Pad\'e expansion) along the radial coordinate, deviations from general
relativity can be expressed using a small number of coefficients. The
line element reads
\begin{equation}
\label{ssbh}
ds^2=-N^2(r)dt^2+\frac{B^2(r)}{N^2(r)}dr^2+r^2 d\theta^2+r^2\sin^2\theta d\phi^2\,,
\end{equation}
with $N(r)$ and $B(r)$ being functions of the radial coordinate $r$.
The radial position of the event horizon is fixed at $r = r_0 > 0$ which implies that $N(r_0)=0$.
Furthermore, the radial coordinate is compactified by means of the dimensionless coordinate 
\begin{equation}
x := 1-\frac{r_0}{r} \,,
\end{equation}
in which $x=0$ corresponds to the position of the event horizon, while
$x=1$ corresponds to spatial infinity. Through this dimensionless coordinate, the function $N$ can be written as
\begin{equation}
N^2=x \, A(x) \,, \label{N2}
\end{equation}
where $A(x)>0 \quad\mbox{for}\quad 0\leq x\leq1$. Introducing additional
coefficients $\epsilon$, $a_n$, and $b_n$, the metric functions $A$ and
$B$ are then expressed as follows
\begin{eqnarray}
\label{asympfix_1}
A(x) &=& 1-\epsilon (1-x)+(a_0-\epsilon)(1-x)^2+{\widetilde A}(x)(1-x)^3 \,, \label{asympfix_2} \\
B(x) &=& 1+b_0(1-x)+{\widetilde B}(x)(1-x)^2 \,.
\end{eqnarray}
Here ${\widetilde A}$ and ${\widetilde B}$ are functions
describing the metric near the event horizon and at spatial infinity. In particular, ${\widetilde A}$ and ${\widetilde B}$ have rapid convergence properties, that is by Pad\'e approximants
\begin{align}
\label{eq:contfrac}
{\widetilde A}(x)=\frac{a_1}{\displaystyle 1+\frac{\displaystyle
    a_2x}{\displaystyle 1+\frac{\displaystyle a_3x}{\displaystyle
      1+\ldots}}}
\,, \qquad \ 
{\widetilde B}(x)=\frac{b_1}{\displaystyle 1+\frac{\displaystyle
    b_2x}{\displaystyle 1+\frac{\displaystyle b_3x}{\displaystyle
      1+\ldots}}} \,,
\end{align}
where $a_1, a_2, a_3,\ldots$ and $b_1, b_2, b_3,\ldots$ are dimensionless
coefficients that can, in principle, be constrained from observations.
The dimensionless parameter $\epsilon$ is fixed by the ADM mass $M$ and the coordinate of the horizon $r_0$. It measures the deviation from the Schwarzschild case as
\begin{equation}
\epsilon=\frac{2M-r_0}{r_0} = - \left(1 - \frac{2M}{r_0}\right)\,.
\end{equation}
It is easy to see that at spatial infinity ($x=1$), all coefficients contribute to \eqref{eq:contfrac}, while at event horizon only the first two terms remain, {\it i.e.}
\begin{align}
{\tilde A}(0)={a_1}\,,
\qquad 
{\tilde B}(0)={b_1}\,.
\end{align}
Given a number of coefficients, any spherical spacetime can hence
directly be simulated in \bhac. For example, the coefficients in the
Rezzolla-Zhidenko parametrization for the Johannsen-Psaltis
\citep{Johannsen2011} metric and for Einstein-Dilaton BHs
\citep{Garcia1995} have already been provided in
\cite{Rezzolla2014}. Typically, expansion up to $a_2, b_2$ yields
sufficient numerical accuracy for the GRMHD simulations. The first
simulations in the related horizon penetrating form of the
Rezzolla-Zhidenko parametrization are discussed in \cite{Mizuno2017}.

\subsection{Available reconstruction schemes} \label{sec:reconstructions}

The second-order finite volume algorithm \eqref{eq:fvolume} requires
numerical fluxes centered on the interface mid-point. As in any
Godunov-type scheme \citep[see \eg][]{Toro99,Komissarov1999}, the fluxes
are in fact computed by solving (approximate) Riemann problems at the
interfaces (see Sect.~\ref{sec:riemann}). Hence for higher than
first-order accuracy, the fluid variables need to be \emph{reconstructed}
at the interface by means of an appropriate spatial interpolation.  Our
reconstruction strategy is as follows. \emph{1)} Compute primitive
variables $\boldsymbol{\bar{P}}$ from the averages of the conserved
variables $\boldsymbol{\bar{U}}$ located at the cell
barycenter. \emph{2)} Use the reconstruction formulae to obtain two
representations for the state at the interface, one with a left-biased
reconstruction stencil $\boldsymbol{P}^\mathrm{L}$ and the other with a
right-biased stencil $\boldsymbol{P}^\mathrm{R}$. \emph{3)} Convert the
now point-wise values back to their conserved states
$\boldsymbol{U}^\mathrm{L}$ and $\boldsymbol{U}^\mathrm{R}$. The latter
two states then serve as input for the approximate Riemann solver.

A large variety of reconstruction schemes are available, which can be
grouped into standard second-order total variation diminishing (TVD)
schemes like ``minmod'', ``vanLeer'', ``monotonized-central'',
``woodward'' and ``koren'' \cite[see][for details]{Keppens2012} and
higher order methods like the third-order methods ``PPM''
\citep{Colella84}, ``LIMO3'' \citep{Cada2009} and the fifth-order
monotonicity preserving scheme ``MP5'' due to \citep{suresh_1997_amp}.
While the overall order of the scheme will remain second-order, the
higher accuracy of the spatial discretisation usually reduces the
diffusion of the scheme and improves accuracy of the solution
\citep[see, \eg][]{Porth2014}.  For typical GRMHD simulations with
near-evacuated funnel/atmosphere regions, we find the PPM reconstruction
scheme to be a good compromise between high accuracy and robustness. For
simple flows, \eg the stationary toroidal field torus discussed in
Sect.~\ref{sec:magtor}, the compact stencil LIMO3 method is recommended.

\subsection{Characteristic speed and approximate Riemann solvers} \label{sec:riemann}

The time-update of \bhac~proceeds in a dimensionally unsplit manner, thus
at each Runge-Kutta substep the interface-fluxes in all directions are
computed based on the previous substep. The state is then advanced to the
next substep with the combined fluxes of the cell.  To compute these
fluxes from the reconstructed conserved variables at the interface
$\boldsymbol{U}^L$ and $\boldsymbol{U}^R$, we provide two approximate
Riemann solvers: \emph{1)} the Rusanov flux, also known as Total
variation diminishing Lax-Friedrichs scheme (TVDLF) which is based on the
largest absolute value of the characteristic waves normal to the
interface $c^i$, and \emph{2)} the HLL solver \citep{Harten83}, which is
based on the leftmost ($c_{-}^i$) and rightmost ($c_{+}^i$) waves of the
characteristic fan with respect to the interface. The HLL upwind flux
function for the conserved variable $u\in\boldsymbol{U}$ is calculated as
\begin{equation}
F^{i}(u)=\left\{
\begin{array}{lcc}
F^i(\boldsymbol{U}^\mathrm{L}) &; & c_{-}^{i} > 0 \\
F^i(\boldsymbol{U}^\mathrm{R}) &; & c_{+}^{i} < 0 \\
\tilde{F}^i(\boldsymbol{U}^\mathrm{L},\boldsymbol{U}^\mathrm{R}) &; & {\rm otherwise} \\
\end{array}
\right.
\end{equation}
where
\begin{equation}
\tilde{F}^i(\boldsymbol{U}^\mathrm{L},\boldsymbol{U}^\mathrm{R}) := \frac{c_{+}^{i}\,F^{i}
  \left(\boldsymbol{U}^\mathrm{L}\right)-c_{-}^{i}\,F^{i}\left(\boldsymbol{U}^\mathrm{R}\right)+c_{+}^{i}\,c_{-}^{i}\,
  \left(u^{\mathrm{R}} - u^{\mathrm{L}}\right)}{c_{+}^{i}\,-\,c_{-}^{i}} \,,
\end{equation}
and we set in accordance with \cite{Davis1988}: $c_{-}^i={\rm min}\left(\lambda_{i,-}^{\mathrm{L}},\lambda_{i,-}^{\mathrm{R}}\right)$, $c_{+}^i={\rm max}\left(\lambda_{+}^{\mathrm{L}},\lambda_{+}^{\mathrm{R}}\right)$. 

The TVDLF flux is simply
\begin{equation}
F^{i}(u) = \frac{1}{2}\left[F^i(\boldsymbol{U}^\mathrm{L}) + F^i(\boldsymbol{U}^\mathrm{R})\right] -
\frac{1}{2} c^i \left(u^\mathrm{R}-u^\mathrm{L}\right)\,.
\end{equation}
with $c^i={\rm  max}\left( |c_-^i|, |c_+^i|\right) \,$.

In addition to these two standard approximate Riemann solvers, we also
provide a modified TVDLF solver that preserves positivity of the
conserved density $D$. The algorithm was first described in the context
of Newtonian hydrodynamics by \cite{Hu2013} and was successfully applied
in GRHD simulations by \cite{Radice2013c}. It takes advantage of the
fact that the first-order Lax-Friedrichs flux $F^{i,\mathrm{LO}}(u)$ is
positivity preserving under a CFL condition ${\rm CFL}\le1/2$. Hence the
fluxes can be constructed by combining the high order flux
$F^{i,\mathrm{HO}}(u)$ (obtained \eg by PPM reconstruction) and
$F^{i,\mathrm{LO}}(u)$ such that the updated density does not fall below
a certain threshold.\footnote{In the general-relativistic hydrodynamic
  WhiskyTHC code \citep{Radice2012a,Radice2013b}, this desirable property
  allows to set floors on density close to the limit of floating point precision $\sim10^{-16}\rho_{\rm ref}$.}
Specifically, the modified fluxes read
\begin{align}
F^i(u) = \theta F^{i,\mathrm{HO}}(u) + (1-\theta)F^{i,\mathrm{LO}}(u) \,,  \label{eq:positivity}
\end{align}
where $\theta\in[0,1]$ is chosen as a maximum value which ensures positivity of the cells adjacent to the interface (see \cite{Hu2013} for details of its construction). Note that although we only stipulate the density be positive, the formula \eqref{eq:positivity} must be applied to all conserved variables $u\in\boldsymbol{U}$. 

In relativistic MHD, the exact form of the characteristic wave speeds
$\lambda_\pm$ involves solution of a quartic equation
\citep[see, \eg][]{Anile1990} which can add to the computational overhead.
For simplicity, instead of calculating the exact characteristic
velocities, we follow the strategy of \cite{Gammie03} who propose a
simplified dispersion relation for the fast MHD wave $\omega^2 = a^2
k^2$. As a trade-off, the simplification can overestimate the wavespeed
in the fluid frame by up to a factor of $2$, yielding a slightly more
diffusive behaviour. The upper bound $a$ for the fast wavespeed is given
by
\begin{align}
a^2 = c_s^2+c_a^2 - c_s^2c_a^2 \,,
\end{align}
which depends on the usual sound speed and Alfv\'en speed
\begin{align}
c_s^2 = \hat{\gamma} \frac{p}{\rho h} \,, \qquad c_a^2 = \frac{b^2}{\rho h +b^2} \,,
\end{align}
here given for an ideal EOS with adiabatic index $\hat{\gamma}$. As pointed out by \cite{DelZanna2007}, the 3+1 structure of the fluxes leads to characteristic waves of the form 
\begin{equation}\label{Eq_lambda}
\lambda_{\pm}^{i} =\alpha\,\lambda_{\pm}^{' i}-\beta^{i}\,,
\end{equation}
where $\lambda_{\pm}^{' i}$ is the characteristic velocity in the corresponding special relativistic system ($\alpha\to 1$, $\beta^i\to 0$). 

For the simplified isotropic dispersion relation, the characteristics can then be obtained just like in special relativistic hydrodynamics \citep[see, \eg][]{Font1994,Banyuls97,Keppens2008}
\begin{align}
\label{Eq_lambdaprim}
\lambda_{\pm}^{' i}  = &\frac{\left(1-a^2\right)\,v^{i}
  \pm\sqrt{a^2\,\left(1-v^2\right)\,\left[\left(1-v^2 a^2\right)\gamma^{ii}-\left(1-a^2\right){\left(v^{i}\right)}^2\right]}}
{(1-v^2\,a^2)} .
\end{align}

\subsection{Primitive variable recovery} \label{sec:con2prim}

It is well-known that the nonlinear inversion
$\boldsymbol{P}(\boldsymbol{U})$ is the Achilles heel of any relativistic
(M)HD code and sophisticated schemes with multiple backup strategies have
been developed over the years as a consequence (\eg \cite{Noble2006},
\cite{Faber2007}, \cite[][]{Noble2009}, \cite{Etienne2012},
\cite{Galeazzi2013}, \cite{Hamlin2013}). Here we briefly describe the
methods used throughout this work and refer to the previously mentioned
references for a more detailed discussion.

\subsubsection{Primary inversions}
Two primary inversion strategies are available in \bhac. The first
strategy, which we denote by ``1D'', is a straightforward generalisation
of the one-dimensional strategy described in \citep{vanderHolst2008}. It
involves a non-linear root finding algorithm which is implemented by
means of a Newton-Raphson scheme on the auxiliary variable $\xi$. Once
$\xi$ is found, the velocity follows from (\ref{eq:Si})
\begin{align}
v^i = \frac{S^i}{(\xi+B^2)} +\frac{B^i (B^jS_j)}{\xi(\xi+B^2)} \,,
\end{align}
and we calculate the second auxiliary variable $\Gamma=(1-v^2)^{-1/2}$ so
that $\rho=D/\Gamma$. The thermal pressure $p$ then follows from the
particular EOS in use (Sect.~\ref{sec:eos}). For example, for an ideal
EOS we have
\begin{align}
p = \frac{\hat{\gamma}}{\hat{\gamma}-1}\left( \frac{\xi}{\Gamma^2}-\rho\right)\, .
\end{align}
For details of the consistency checks and bracketing, we refer the
interested reader to \cite{vanderHolst2008}.

In addition to the 1D scheme, we have implemented the ``2DW'' method of
\cite{Noble2006, DelZanna2007}. The 2DW inversion simultaneously solves
the non-linear equations \eqref{eq:uideal} and the square of the
three-momentum $S^2$, following \eqref{eq:Si} by means of a
Newton-Raphson scheme on the two variables $\xi$ and $v^2$. Among all
inversions tested by \cite{Noble2006}, the 2DW method was reported as the
one with the smallest failure rate. We find the same trend, but also
find that the lead of 2DW over 1D is rather minor in our tests.

With two distinct inversions that might fail under different circumstances, one can act as a backup strategy for the other. Typically we first attempt a 2DW inversion and switch to the 1D method when no convergence is found. The next layer of backup can be provided by the entropy method as described in the next section. 

\subsubsection{Entropy switch}\label{sec:entropy}
To deal with highly magnetised regions, \cite[][]{Noble2009,Sadowski2013}
introduced the advection of entropy to provide a backup strategy for the
primitive variable recovery. Similar to
\cite[][]{Noble2009,Sadowski2013}, alongside the usual fluid equations,
\bhac~can be configured to solve an advection equation for the entropy
$S$
\begin{align}
\nabla_{\mu} S u^{\mu} = 0 \,,
\end{align}
where we define
\begin{align}
S:= p/\rho^{\hat{\gamma}-1} \,,
\end{align}
given the adiabatic index $\hat{\gamma}$. This leads to the evolution equation
\begin{align}
\partial_{t} \sqrt{\gamma} \Gamma S + \partial_{i} \sqrt{\gamma} (\alpha v^{i} - \beta^{i}) \Gamma S = 0 \,,
\end{align}
for the conserved quantity $\Gamma S$. The primitive counterpart is the
actual entropy $\kappa = p/\rho^{\hat{\gamma}}$, which can be recovered
via $\kappa = \Gamma S / D$. In case of failure of the primary inversion
scheme, using the advected entropy $\kappa$, we can attempt a recovery of
primitive variables which does not depend on the conserved energy. Note
that after the primitive variables are recovered from the entropy, we
need to discard the conserved energy and set it to the value consistent
with the entropy. On the other hand, after each successful recovery of
primitive variables, the entropy is updated to $\kappa=
p/\rho^{\hat{\gamma}}$, which is then advected to the next step. In
addition, entropy-based inversion can be activated whenever
$\beta=2p/b^{2}\le 10^{-2}$ since the primary inversion scheme is likely
to fail in these highly magnetised regions. Tests of the dynamic
switching of the evolutionary equations are described in
Sect.~\ref{sec:bondi}. In GRMHD simulations of BH accretion, the
``entropy region'' is typically located in the BH magnetosphere, which is
strongly magnetised and the error due to missing shock dissipation is
thus expected to be small.

In the rare instances where the entropy inversion also fails to converge
to a physical solution, the code is normally stopped. To force a
continuation of the simulation, last resort measures that depend on the
physical scenario can be employed. Often the simulation can be continued
when the faulty cell is replaced with averages of the primitive variables
of the neighbouring healthy cells as described in \cite{Keppens2012}. In
the GRMHD accretion simulations described below, failures could happen
occasionally in the highly magnetised evacuated ``funnel'' region close
to the outer horizon where the floors are frequently applied. We found
that the best strategy is then to replace the faulty density and pressure
values with the floor values and set the Eulerian velocity to zero. Note
that in order to avoid generating spurious $\boldsymbol{\nabla \cdot B}$,
the last resort measures should never modify the magnetic fields of the
simulation.

\subsection{Adaptive Mesh refinement}\label{sec:AMR}
The computational grid employed in \bhac~is provided by the \amrvac
toolkit and constitutes a fully adaptive block based (oct-) tree with a
fixed refinement factor of two between successive levels. That is, the
domain is first split into a number of blocks with equal amount of cells
(\eg $10^3$ computational cells per block). Each block can be refined
into two (1D), four (2D) or eight (3D) child-blocks with an again
fixed number of cells. This process of refinement can be repeated ad
libitum and the data-structure can be thought of a forest (collection of
trees). All operations on the grid, for example time-update, IO and
problem initialisation are scheduled via a loop over a space-filling
curve. We adopt the Morton Z-order curve for ease of implementation via
a simple recursive algorithm.

Currently, all cells are updated with the same global time-step and hence
load-balancing is achieved by cutting the space-filling curve into equal
sections that are then distributed over the MPI-processes. The AMR
strategy just described is applied in various astrophysical codes, for
example codes employing the \paramesh library
\cite{MacNeice00,Fryxell2000,Zhang2006}, or the recent \athenapp
framework \citep[see, \eg][]{White2016}. Compared to a patch-based
approach \citep[see, \eg][]{Mignone2012}, the block based AMR has several
advantages: \emph{1)} well-defined boundaries between neighbouring grids
on different levels ,\emph{2)} data is uniquely stored and updated, thus
no unnecessary interpolations are performed, and \emph{3)} simple
data-structure, \eg straightforward integer arithmetic can be used to
locate a particular computational block. For in-depth implementation
details such as refinement/prolongation operations, indexing and
ghost-cell exchange, we refer to \cite{Keppens2012}.  Prolongation
  and restriction can be used on conservative variables or primitive
  variables. Typically primitive variables are chosen to avoid unphysical
  states which can otherwise result from the interpolations in conserved
  variables.  The refinement criteria usually adapted is the
  L\"{o}hner's error estimator \cite{Loehner87} on physical variables. It
  is a modified second derivative, normalised by the average of the
  gradient over one computational cell.  The multidimensional
  generalization is given by
\begin{align}
E_{i_1 i_2 i_3} = \sqrt{ \frac{\sum_{p} \sum_{q} \left( {\partial^2 u \over \partial x_p \partial x_q}\Delta x_p \Delta x_q \right)^2}
{\sum_{p} \sum{q} \left[ \left( \left| {\partial u \over \partial x_p} \right|_{i_p + 1/2} + \left|  {\partial u \over \partial x_p} \right|_{i_p -1/2}\right) \Delta x_p + f_{\rm wave} {\partial^2 |u| \over \partial x_p \partial x_q} \Delta x_p \Delta x_q \right]^2}}.
\end{align}
The indices $p, q$ run over all dimensions $p, q =1, ... , N_D$. The
  last term in the denominator acts as a filter to prevent refinement of
  small ripples, where $f_{\rm wave}$ is typically chosen of order $10^{-2}$.
  This method is also used in other AMR codes such as \texttt{FLASH}
  \citep{Calderetal02}, \texttt{RAM} \citep{Zhang2006}, \texttt{PLUTO}
  \citep{Mignone2012} and \texttt{ECHO} \citep{Zanotti2015b}.

\section{Numerical tests}\label{sec:tests}

\subsection{Shock tube test with gauge effect}
 
The first code test is considered in flat spacetime and therefore no
metric source terms are involved.  Herein we perform one-dimensional
MHD shock tube tests with gauge effects by considering gauge
transformations of the spacetime. Shock tube tests are well-known tests
for code validation and emphasise the nonlinear behaviour of the
equations, as well as the ability to resolve discontinuities in the
solutions \citep[see, \eg][]{Anton05,DelZanna2007}.

The initial condition is given as
\begin{align}
\begin{split}
(\rho,p,B^{x},B^{y}) = \left\{
\begin{array}{ll}
(1,1,0.5,1)  &  x<0 \,, \\
(0.125,0.1,0.5,-1)  & x>0 \,,
\end{array}
\right.
\end{split}
\end{align}
and all other quantities are zero. In order to check whether the
covariant fluxes are correctly implemented, we use different settings for
the flat spacetime as detailed in Table~\ref{tab:shock}.
\begin{table}[h!]
\begin{center}
\caption{Shock tube with gaugeeffect setups.}\label{tab:shock}
\begin{tabular}{ccrrrr}
\hline
Case & $\alpha$ & $\beta^{i}$ & $\gamma_{11}$ & $\gamma_{22}$ & $\gamma_{33}$ \\
\hline
A & 1 & (0,0,0) & 1 & 1 &1
\\
B & 2 & (0,0,0) & 1 & 1 & 1
\\
C & 1 & (0.4,0,0) & 1 & 1 & 1
\\

D & 1 & (0,0,0) & 4 & 1 & 1
\\

E & 1 & (0,0,0) & 1 & 4 & 1
\\

F & 2 & (0.4,0,0) & 4 & 9 & 1
\\
\hline
\end{tabular}
\end{center}
\end{table}

\noindent
In the simulations, an ideal gas EOS is employed with an adiabatic index
of $\hat{\gamma}=2$. The 1D problem is run on a uniform grid in
$x-$direction using 1024 cells spanning over $x\in[-1/2,1/2]$. The
simulations are terminated at $t=0.4$.  For the spatial
  reconstruction, we adopt the second order TVD limiter due to Koren
  \citep{Koren1993}.  Furthermore, RK3 timeintegration is used with
  Courant number set to $0.4$.  

Case A is the reference solution without modification of fluxes due to
the three-metric, lapse or shift\footnote{We note that for the reference
  solution we have relied here on the extensive set of tests performed in
  flat spacetime within the \amrvac framework; however, we could also
  have employed as reference solution the ``exact'' solution as derived
  in Ref. \cite{Giacomazzo:2005jy}.}. By means of simple transformations
of flat-spacetime, all other cases can be matched with the reference
solution. Case B will coincide with solution A if B is viewed at
$t/2=0.2$. Case C will agree with case A when it is shifted in positive
$x-$direction by $\delta x = \beta^{x}t=0.16$.  For case D, we rescale
the domain as $x\in[-1/4,1/4]$ and initialise the contravariant vectors
as $B'^{x} = B^{x}/2$. The state at $t=0.4$ should agree with case A when
the domain is multiplied by the scale factor $h_{x} = 2$.  For case E we
initialise $B'^{y}=B^{y}/2$ and case F is initialised similarly as
$B'^{x} = B^{x}/2$, $B'^{y}=B^{y}/3$.

\begin{figure}[htbp]
\begin{center}
\includegraphics[width=0.9\textwidth]{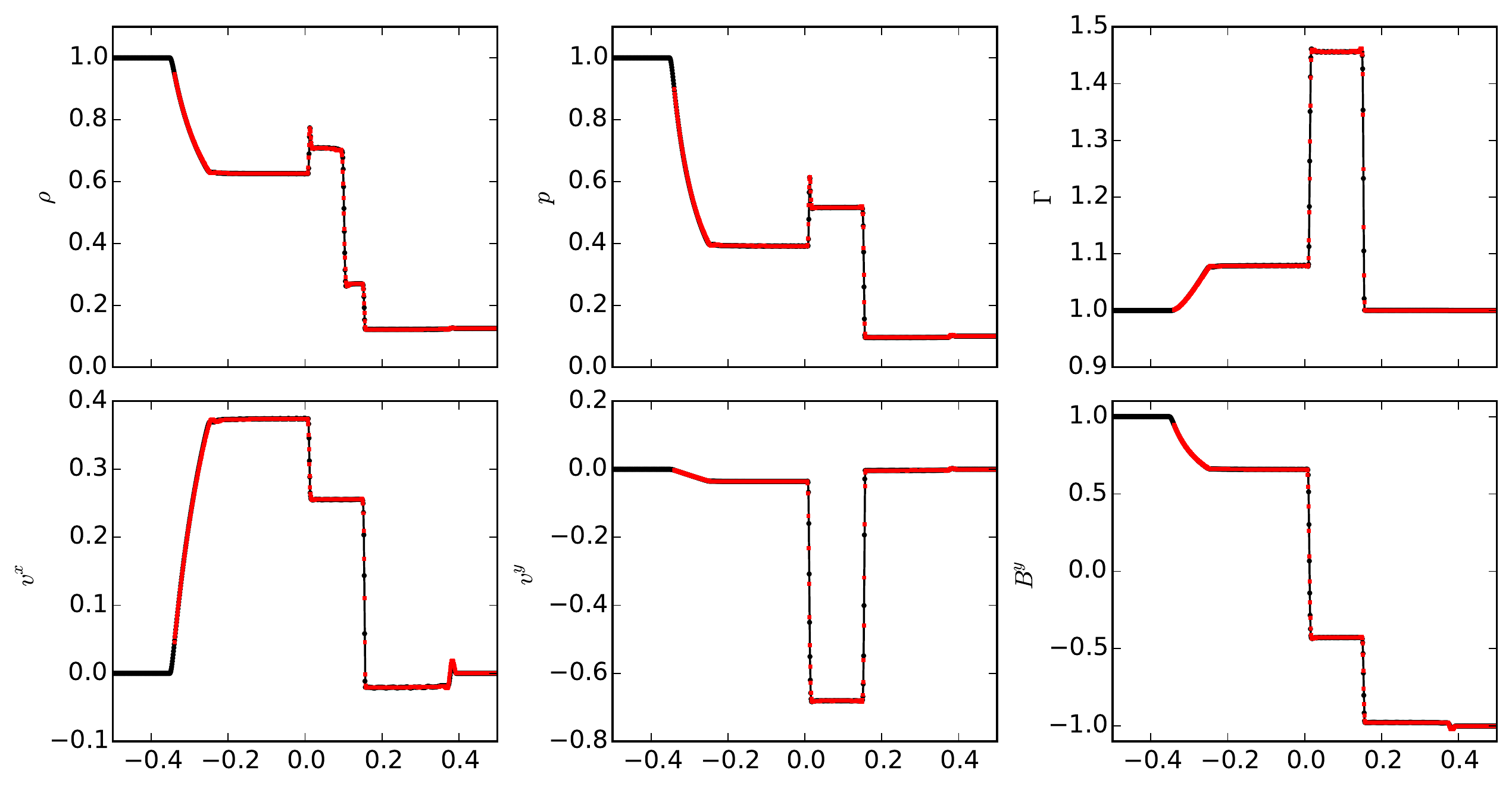}
\caption{1D plots of density $\rho$, gas pressure $p$, Lorentz factor
  $\Gamma$, velocity components $v^x$ and $v^y$, and the $y$-component of
  the magnetic field for the shock tube test at $t=0.4$. The reference
  solution of case A is shown as a solid black line and the rescaled
  solution of case F is overplotted as red squares. }
\label{fig:F}
\end{center}
\end{figure}

In general, all cases agree very well with the rescaled solution. To give
an example, Fig.~\ref{fig:F} shows the rescaled simulation results of
case F compared to the reference solution of case A. This test
demonstrates the shock-capturing ability of the MHD code and enables us
to conclude that the calculation of the covariant fluxes has been
implemented correctly.

 \subsection{Boosted loop advection}

In order to test the implementation of the GLM-GRMHD system, we perform the advection of a force-free flux-tube with poloidal and toroidal components of the magnetic field in a flat spacetime.  

The initial equilibrium configuration of a force-free flux-tube is given by a modified Lundquist tube \citep[see \eg][]{Gourgouliatos2012}, where we avoid sign changes of the vertical field component $B^{z}$ with the additive constant $C=0.01$. Pressure and density are initialized as constant throughout the simulation domain. The initial pressure value is obtained from the central plasma-beta $\beta_{0} = B^{2}(0)/2p$, where $B$ is the magnetic field in the co-moving system. The density is set to $\rho=p/2$ yielding a relativistic hot plasma. Consequently, an adiabatic index $\gamma=4/3$ is used. We set $\beta_{0}=0.01$, which results in a high magnetisation $\sigma_{0}=B^{2}(0)/(\rho c^{2}+4p)\simeq 25$.  
The equations for the magnetic field for $r<1$ read
\begin{align}
B^{\phi}(r) &= J_{1}(\alpha_{t}r) \,, \\
B^{z}(r)    &= \sqrt{J_{0}(\alpha_{t}r)^{2} + C} \,,
\end{align}
and
\begin{align}
B^{\phi}(r) &= 0 \,, \\
B^{z}(r)    &= \sqrt{J_{0}(\alpha_{t})^{2} + C} \,,
\end{align}
otherwise, where $J_{0}$ and $J_{1}$ are Bessel functions of zeroth and
first order respectively and the constant $\alpha_{t}\simeq 3.8317$ is
the first root of $J_{0}$.

This configuration is then boosted to the frame moving at velocity
$\boldsymbol{v}=\sqrt{2}(-v_{c},-v_{c},0)$ and we test values of $v_{c}$
between $0.5$c and $0.99$c.

Standard Lorentz transformation rules result in
\begin{align}
\boldsymbol{r} = \boldsymbol{r}' + (\Gamma-1)(\boldsymbol{r'\cdot n})\boldsymbol{n} -\Gamma t' v_{c}\boldsymbol{n} \,, \ \hspace{1cm}
\boldsymbol{B}' = \Gamma \boldsymbol{B} - \frac{\Gamma^{2}}{\Gamma+1} \boldsymbol{\beta}(\boldsymbol{\beta\cdot B}) \,,
\end{align}
where $t'$ can be set to zero and where we assumed a vanishing electric
field in the co-moving system. Therefore relativistic length contraction
gives the loop a squeezed appearance.  A simulation domain
$(x,y)\in[-1,1]$ at a base resolution of $N_{x}\times N_{y}=64^{2}$ is
initialised with an additional three levels of refinement. We advect the
loop for one period ($P=2\sqrt{2}/v$) across the domain where periodic
boundary conditions are used.

\begin{figure}[htbp]
\begin{center}
\includegraphics[width=.9\textwidth]{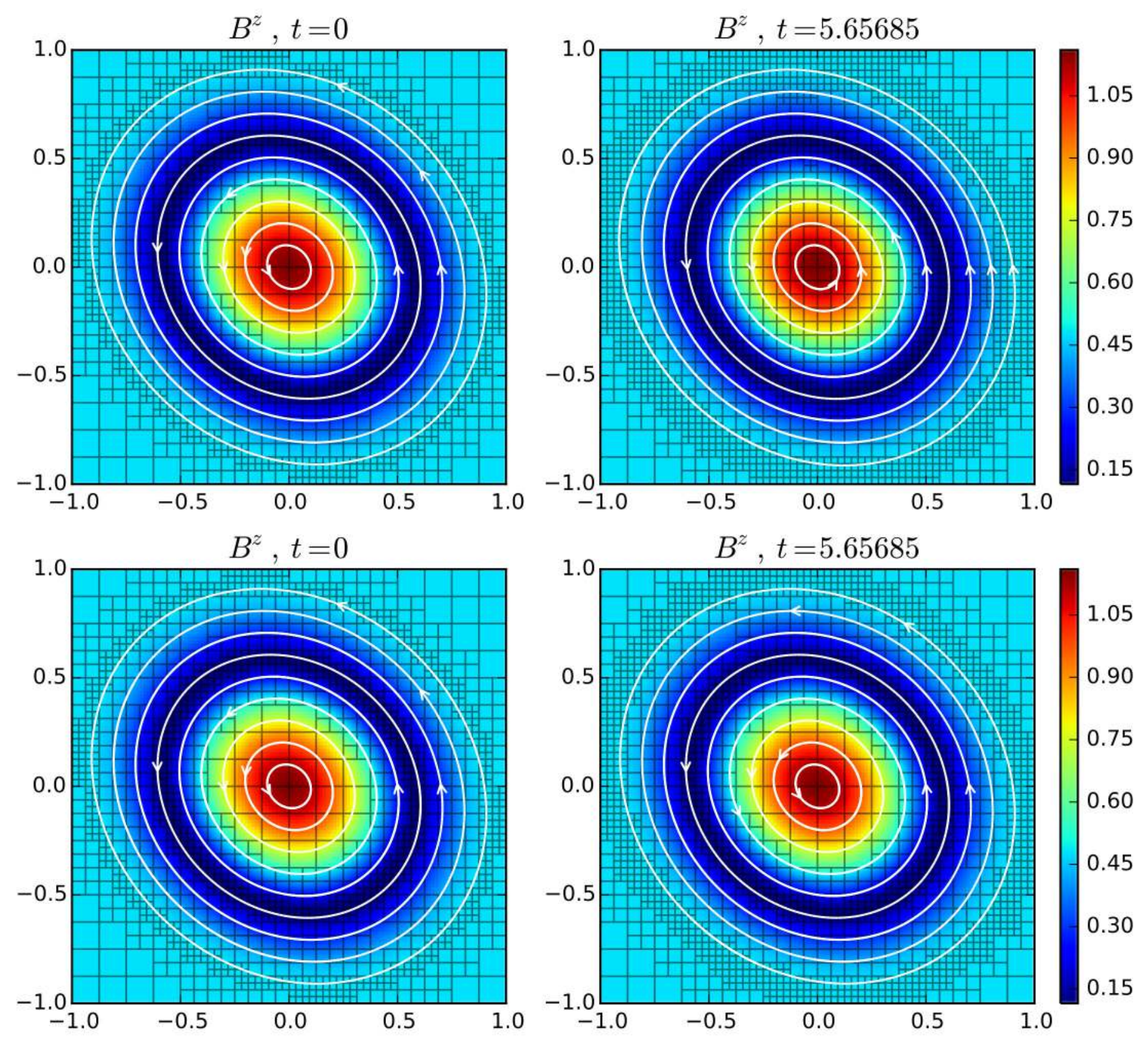}
\caption{Force-free magnetic loop with diagonal boost velocity
  $|v|=0.5c$.  \textit{Top:} No shift, the loop is advected for one
  period. \textit{Bottom:} The shift vector just opposes the (diagonal)
  advection velocity, $|v|=0.5$, hence the loop remains stationary with
  respect to the grid.  Base resolution is $64^{2}$ cells with a total of
  three grid levels.  The color shows the strength of the out-of-plane
  field component $B^{z}$ and white lines are in-plane field lines of
  $(B^{x},B^{z})$.  Blocks containing $8^{2}$ cells are indicated.  }
\label{fig:loopadv1}
\end{center}
\end{figure}

The advection over the coordinates can be counteracted by setting the
shift vector appropriately,
i.e. $\boldsymbol{\beta}=-\boldsymbol{v}$. This is an important
consistency check of the implementation.  Figure~\ref{fig:loopadv1} shows
the initial and final states of the force-free magnetic flux-tube
advected for one period and for the case with spacetime shifted against
the advection velocity. The advected and counter-shifted cases are in
good agreement, with only the truly advected case being slightly more
diffused, the effect of which is reflected in the activation of more
blocks on the third AMR level.

\begin{figure}[htbp]
\begin{center}
\includegraphics[width=.9\textwidth]{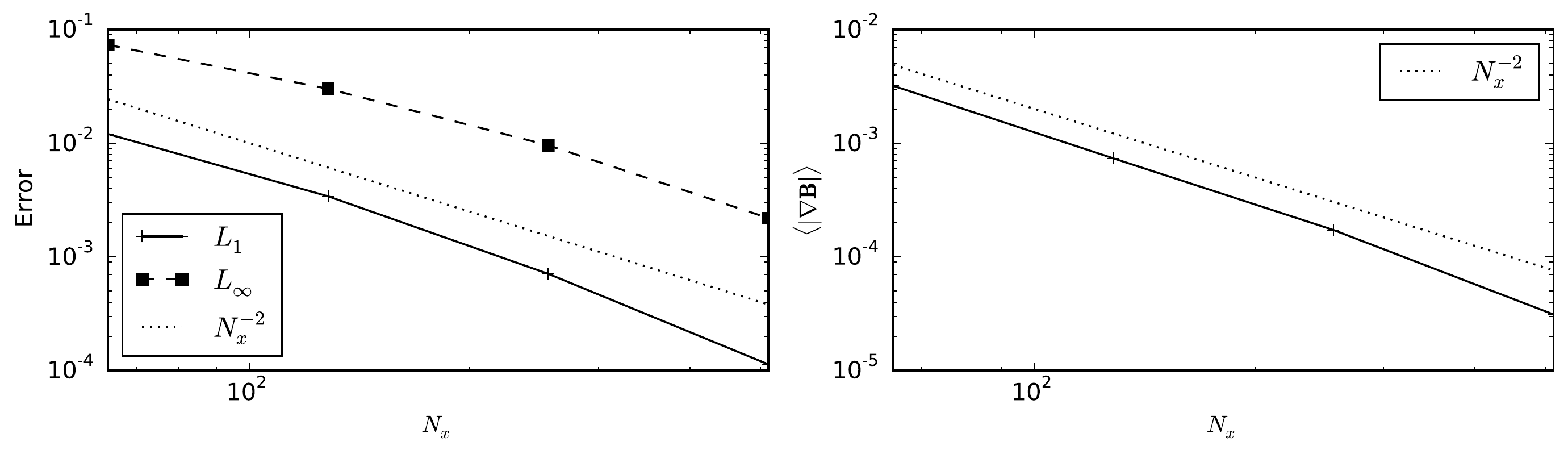}
\caption{Error of the out-of-plane magnetic field component $B^{z}$ \textit{(left)} and divergence of $\boldsymbol{B}$ \textit{(right)}.  For this test, we chose $\boldsymbol{v}=0.5\sqrt{0.5} (1,1,0)$ and $\beta=\sqrt{0.5} (1,0,0)$, resulting in an advection in the direction of the upper-left diagonal.  }
\label{fig:loopadv1-error}
\end{center}
\end{figure}

To investigate the numerical accuracy the $L_1$ and $L_\infty$ norms of
the out-of-plane magnetic field component $B^z$, as well as the
divergence of magnetic field between the initial state and the simulation
at a time after one advection period with different resolutions as seen
in Fig.\ref{fig:loopadv1-error} are checked.  The error norms from
analytically known solutions $u^*$ are defined as
\begin{align}
L_1(u) &= \frac{1}{N_{\rm cells}}\sum_{i,j,k} \left|\bar{u}_{i,j,k}-\frac{1}{\Delta V_{i,j,k}} \int_{V_{i,j,k}} u^* \sqrt{\gamma} dx^1 dx^2 dx^3 \right| \,, \\
L_{\infty}(u) &= {\rm max}_{i,j,k} \left| \bar{u}_{i,j,k} - \frac{1}{\Delta V_{i,j,k}} \int_{V_{i,j,k}} u^* \sqrt{\gamma} dx^1 dx^2 dx^3 \right| \,,
\end{align}
where the summation, respectively maximum operation, includes all cells
in the domain and the integrals are performed over the volume of the cell
$\Delta V_{i,j,k}$.  In this sense, the reported errors correspond to the
mean and maximal error in the computational domain.  We should note that
for the test of convergence, we use a uniform grid and choose
$\boldsymbol{v}=0.5\sqrt{0.5} (1,1,0), \beta=\sqrt{0.5} (1,0,0)$
resulting in an advection in direction of the upper-left diagonal.  A TVD
``Koren" limiter is chosen.  As expected, the convergence is second order
for all cases.

\subsection{Magnetised spherical accretion}\label{sec:bondi}

A useful stress test for the conservative algorithm in a
general-relativistic setting is spherical accretion onto a Schwarzschild
BH with a strong radial magnetic field \cite{Gammie03}. The steady-state
solution is known as the Michel accretion solution \citep{Michel72} and
represents the extension to general relativity of the corresponding
Newtonian solution by \cite{Bondi52}. The steady-state spherical
accretion solution in general relativity is described in a number of
works \citep[see, \eg][]{Hawley84a, Rezzolla_book:2013}. It is easy to
show that the solution is not affected when a radial magnetic field of
the form $B^r\propto \gamma^{-1/2}$ is added \citep{DeVilliers03a}. This
test challenges the robustness of the code and of the inversion procedure
$\boldsymbol{P}(\boldsymbol{U})$ in particular. The calculation of the
initial condition follows that outlined in \cite{Hawley84a}. Here, we
parametrize the field strength via $\sigma=b^{2}/\rho$ at the inner edge
of the domain ($r=1.9\,M$).  The simulation is setup in the equatorial
plane using MKS coordinates corresponding to a domain of $r_{\rm
  KS}\in[1.9,20]\,M$.  The analytic solution remains fixed at the
  inner and outer radial boundaries.   We run two cases, case 1 with
magnetisation up to $\sigma=10^{3}$ and case 2 with a very high
magnetisation reaching up to $\sigma=10^{4}$. Since the problem is
  only 1D, the constraint $\mathbf{\nabla\cdot B}=0$ has a unique
  solution which gets preserved via the FCT algorithm.

\begin{figure}[htbp]
\begin{center}
\includegraphics[width=.9\textwidth]{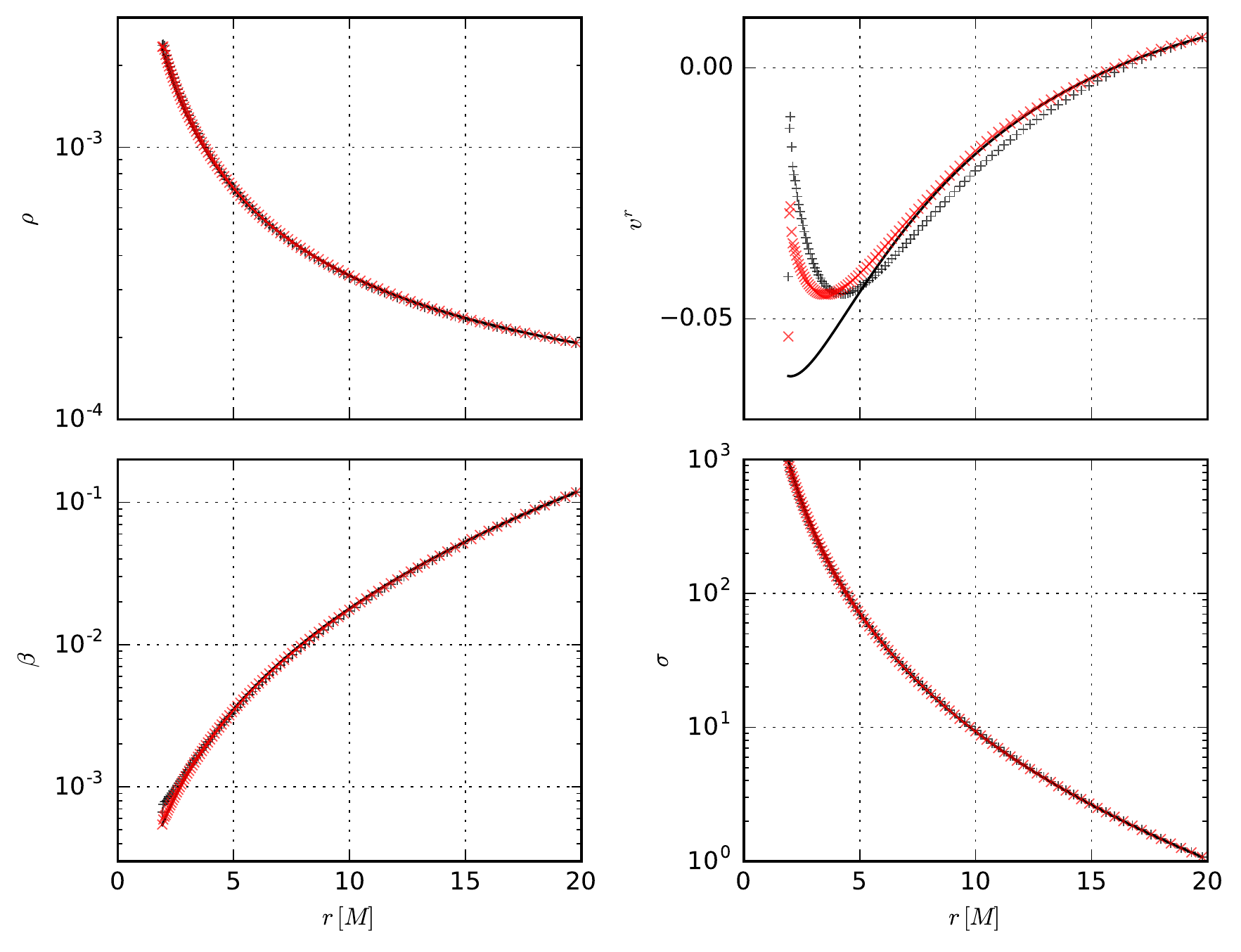}
\caption{Magnetized Bondi flow at $t=100\,M$ in MKS coordinates with
  $\sigma=10^{3}$ at the inner edge of the domain.  The black solid curve
  indicates the initial exact solution.  We show two realisations with
  resolution $N_{r}=100$.  Black crosses are with the standard treatment
  for the inversion.  Red crosses switch to the entropy evolution at
  $\beta\le10^{-2}$ (here in the middle of the domain).  In particular,
  the error in the radial three-velocity $v^{r}$ decreases when switching
  to the entropy evolution.  }
\label{fig:highsigmabondi-profiles}
\end{center}
\end{figure}

Figure~\ref{fig:highsigmabondi-profiles} illustrates the profiles for
$\sigma=10^{3}$ and two inversion strategies: 2DW (black $+$) and 2DW
with entropy switching in regions of high magnetization $b^2/2p>100$ (red
$\times$). With the exception of the radial three-velocity near the BH
horizon ($r \le 5\,M$), in both cases the simulations maintain well the
steady-state solution.\footnote{Note that the discrepancy in $v^r$
  appears less dramatic when viewed in terms of the four-velocity
  $u^{r}$.}  Comparing theses results with and without entropy switching,
the entropy strategy actually keeps the solution closer to the
steady-state solution (solid black curves) even though the change of
inversion strategy occurs in the middle of the domain, $r\simeq10$.

\begin{figure}[htbp]
\begin{center}
\includegraphics[width=.94\textwidth]{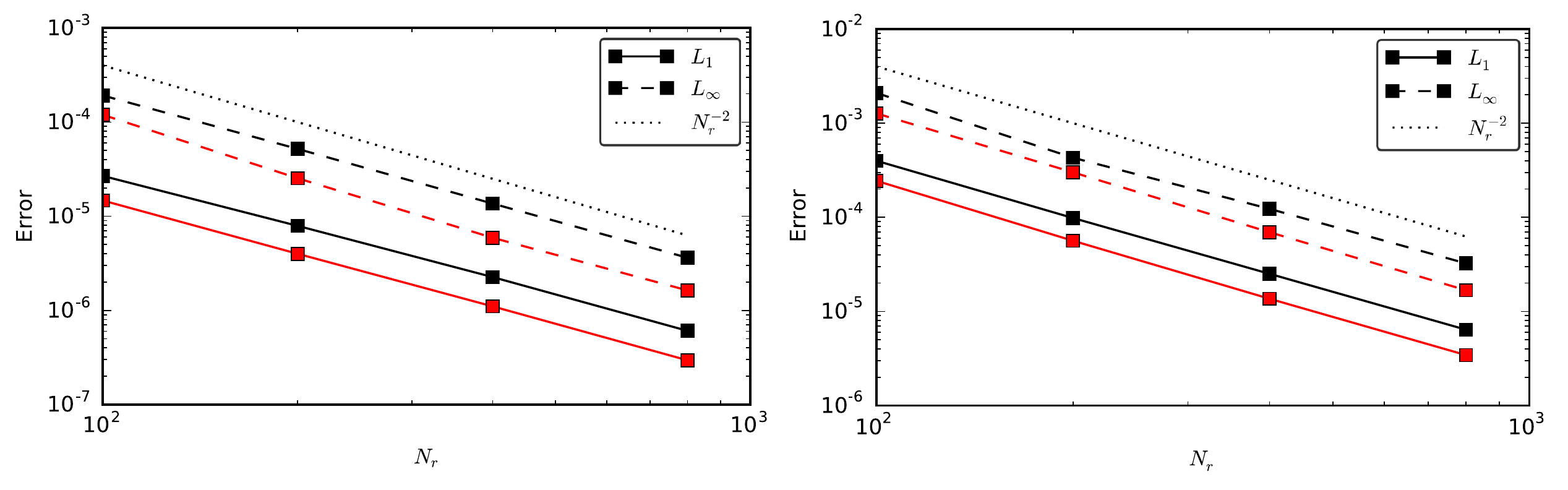}
\caption{Error of density $\rho$ in the highly magnetised Bondi flow:
  $\sigma=10^{3}$ (left) and $\sigma=10^{4}$ (right).  The black data
  points are obtained with the standard 2D inversion and the red
  datapoints switch to the entropy inversion at $\beta\le10^{-2}$.  One
  can see that both recipes are convergent with the expected order and
  that the error in the entropy strategy is decreased by roughly a factor
  of two.  }
\label{fig:highsigmabondi-error}
\end{center}
\end{figure}

The errors ($L_1$ and $L_\infty$ norms) for the four cases are shown in
Fig.~\ref{fig:highsigmabondi-error}. Again, the second-order accuracy of
the algorithm is recovered.  Using the entropy strategy increases the
numerical accuracy by around a factor of two and we suggest its use in
the highly magnetised regime of BH magnetospheres.

\subsection{Magnetised equilibrium torus}\label{sec:magtor}

As a final validation of the code in the GRMHD regime, we perform the
simulation of a magnetised equilibrium torus around a spinning BH. A
generalisation of the steady-state solution of the standard
hydrodynamical equilibrium torus with constant angular momentum
\cite[see, \eg][]{Fishbone76, Hawley84a, Font02a} to MHD equilibria with
toroidal magnetic fields was proposed by \cite{Komissarov2006a}. This
steady-state solution is important since it constitutes a rare case of a
non-trivial analytic solution in GRMHD\footnote{We thank Chris Fragile
  for providing subroutines for this test case.}.

For the initial setup of the equilibrium torus, we adopt a particular
relationship $\omega=\omega(p)$, where $\omega=\rho h$ is the fluid
enthalpy and $\tilde{\omega} = \tilde{\omega}(\tilde{p_m})$, where
$\tilde{p}_m = \mathcal{L}p_m$, $\tilde{\omega}=\mathcal{L}\omega$, $p_m
= b^2/2$ is the magnetic pressure, and $\mathcal{L}=g_{t \phi} g_{t
  \phi}-g_{tt}g_{\phi \phi}$. From these relationships, thermal and
magnetic pressures are described as
\begin{align}
p &= K \omega^\kappa \,, \\
\tilde{p}_m &= K_m \tilde{\omega}^\eta \,.
\end{align}
The analytical solutions can be constructed from 
\begin{align}
W - W_{in} + {\kappa \over \kappa -1} {p \over \omega} + {\eta \over \eta-1} {p_{m} \over \omega} = 0 \,,
\end{align}
for the introduced total potential $W$, where $W = \ln |u_t|$.  The
centre of the torus is located at $(r_\mathrm{c}, \pi/2)$. At this point,
we parametrize the magnetic field strength in terms of the pressure ratio
\begin{align}
\beta_c = p_g (r_c, \pi/2) / p_m (r_c, \pi/2) \,.
\end{align} 
The gas pressure and magnetic pressure at the centre of the torus are
given by
\begin{align}
p_c = \omega_c (W_{in} - W_{c}) \left( {\kappa \over \kappa-1} + {\eta
  \over \eta-1} {1 \over \beta_c} \right)^{-1} \,, 
\qquad p_{m_{c}} = p_c / \beta_c \,.
\end{align}
From these, the constants $K$ and $K_m$ for barotropic fluids are obtained.

The magnetic field distribution is given by the distribution of magnetic
pressure $p_\mathrm{m}$. From the consideration of a purely toroidal
magnetic field one obtains
\begin{align}
b^\phi = \sqrt{2 p_m/\mathcal{A}} \,, \qquad \ b^t = \ell \, b^\phi \,,
\end{align}
where $\mathcal{A}=g_{\phi\phi}+2l g_{t\phi}+l^2 g_{tt}$ and $\ell := -
u_\phi/u_t$ is the specific angular momentum.

We perform 2D simulations using logarithmic KS coordinates with $h=0$ and
$R_0 =0$. The simulation domain is $\theta\in[0,\pi]$, $r_{\rm
  KS}\in[0.95\, r_{\mathrm{h}},\ 50\,M]$, where $r_{\mathrm{h}}$ is the
(outer) event horizon radius of the BH. The BH has the dimensionless spin
parameter $a=0.9$. For simplicity, we set the two indices to the same
value of $\kappa=\eta=4/3$ and also set the adiabatic index of the
adopted ideal EOS to $\gamma=4/3$. The remaining parameters are listed in
the Table~\ref{torus:table1}.

\begin{table}
\begin{center}
\caption{Parameters for the MHD equilibrium torus test}
\begin{tabular}{rcccccc} \hline
Case & $l_0$ & $r_c$ & $W_{in}$ & $W_c$ & $\omega_c$ & $\beta_c$ \\
\hline
A & 2.8 & 4.62 & -0.030 & -0.103 & 1.0 & 0.1 \\
\hline
\end{tabular}
\label{torus:table1}
\end{center}
\end{table}

Initially, the velocity of the atmosphere outside of the torus is set to
zero in the Eulerian frame, with density and gas pressure set to very
small values of $\rho=\rho_{\mathrm{min}} \, r_{\rm BL}^{-3/2}$,
$p=p_{\mathrm{min}} \, r_{\rm BL}^{-5/2}$ with
$\rho_{\mathrm{min}}=10^{-5}$ and $p_{\mathrm{min}}=10^{-7}$.  It is
important to note that the atmosphere is free to evolve and only
densities and pressures are floored according to the initial state.  In
the simulations we use the HLL approximate Riemann solver, third order
LIMO3 reconstruction, two-step time update, and a CFL number of
$0.5$. We impose outflow conditions on the inner and outer boundaries
  of the radial direction and reflecting boundary conditions in the
  $\theta$ direction.  As the magnetic field is purely toroidal, and will
  remain so during the time-evolution of this axisymmetric case, no 
  particular $\mathbf{\nabla\cdot B=0}$ treatment is used.
\begin{figure}[htbp]
\begin{center}
\includegraphics[width=.95\textwidth]{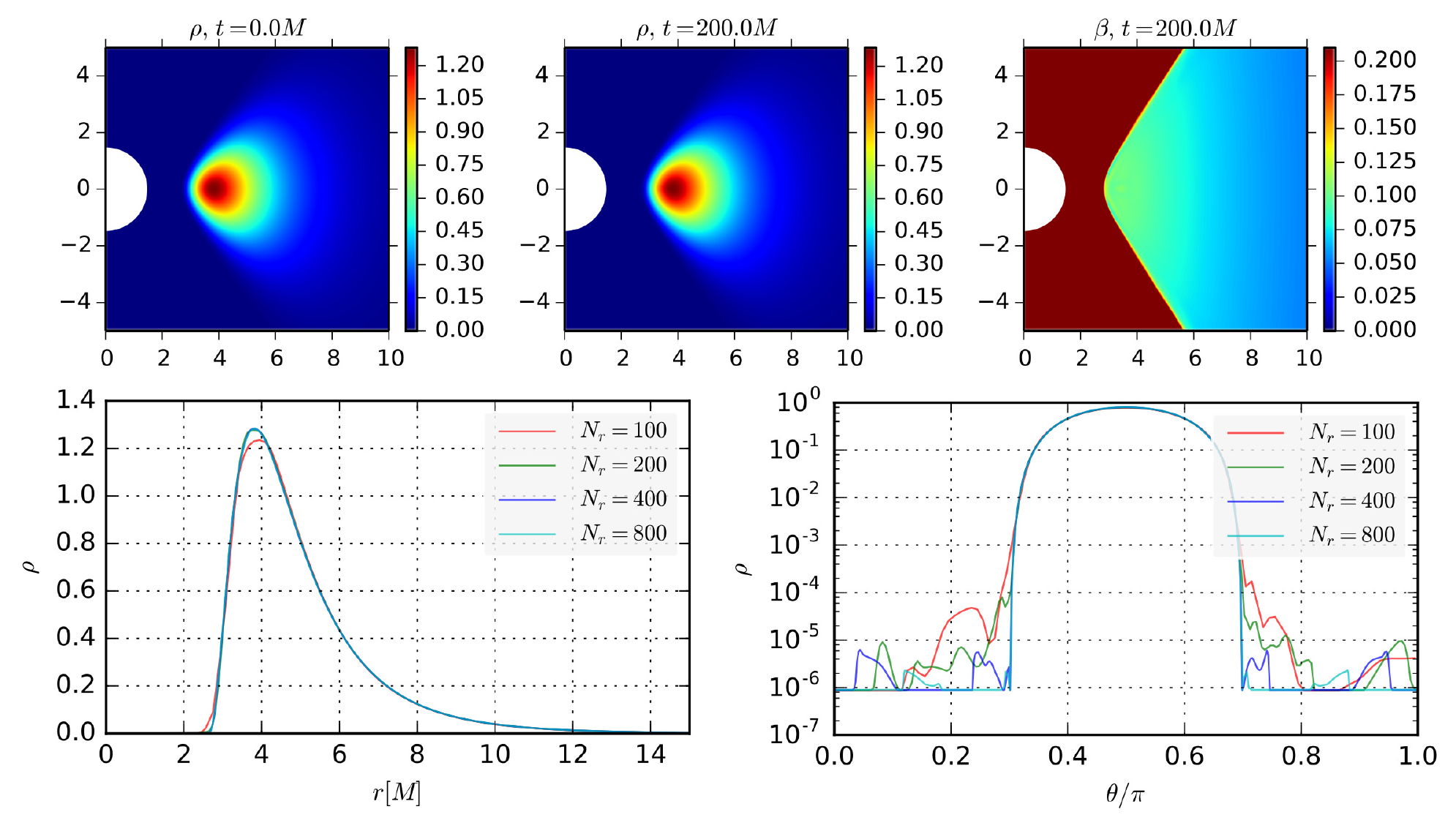}
\caption{\textit{Top}: qualitative view of the torus evolution at a
  resolution of $N_r=N_\theta=400$. The spatial scale is given in units
  of $M$. \textit{Left}: Initial rest-frame density distribution,
  \textit{center}: density at $t=200\,M$, \textit{right}: plasma $\beta$
  parameter at $t=200\,M$.  \textit{Bottom}: density slices through the
  torus at $t=100\,M$ for constant $\theta=\pi/2$ \textit{(left)} and
  $r=5$ \textit{(right)}.  }
\label{fig:komissarov-snapshots}
\end{center}
\end{figure}

\begin{figure}[htbp]
\begin{center}
\includegraphics[width=.65\textwidth]{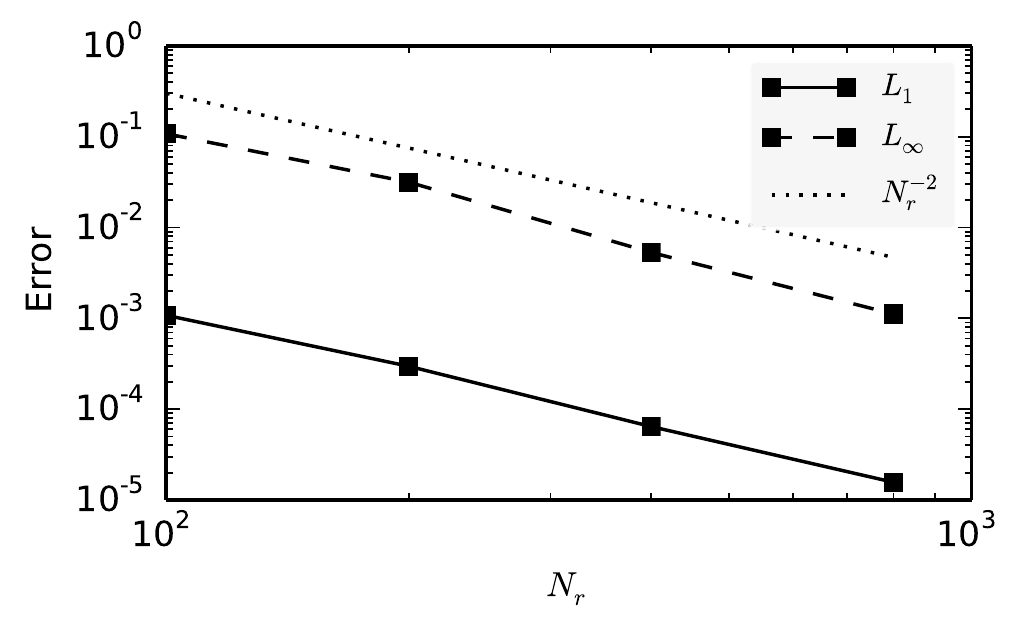}
\caption{Error of the density $\rho$ in the strongly-magnetised
  Komissarov torus, comparing the solution at $t=60\,M$ with the exact
  solution.  The second-order behaviour of the numerical scheme is well
  recovered.}
\label{fig:komissarov-error}
\end{center}
\end{figure}

The top panels of Fig.~\ref{fig:komissarov-snapshots} show the density
distribution at the initial state and at $t=200\,M$, as well as the
plasma beta distribution at $t=200\,M$. The rotational period of the disk
centre is $t_{\mathrm{r}} = 68\,M$. The initial torus configuration is
well maintained after several rotation period. For a qualitative view of
the simulations, the 1D radial and azimuthal distributions of the density
are shown in the lower two panels in Fig.~\ref{fig:komissarov-snapshots}
with different grid resolutions.  All but the low resolution case are
visually indistinguishable from the initial condition in the bottom-left
panel, showing $\rho-r$ with a linear scale.  Since the atmosphere is
evolved freely, small density waves propagate in the ambient medium of
the torus, as seen in the $\rho-\theta$ cut. This does not adversely
affect the equilibrium solution in the bulk of the torus however.  Error
quantification ($L_1$ and $L_\infty$) is provided in
Fig.~\ref{fig:komissarov-error}. The second-order properties of the
numerical scheme are well recovered.

\subsection{Differences between FCT and GLM}

Having implemented two methods for divergence control, we took the
opportunity to compare the results of simulations using both methods.  We
analysed three tests: a relativistic Orszag-Tang vortex, magnetised
Michel accretion, and magnetised accretion from a Fishbone-Moncrief
torus.  Although much less in-depth, this comparison is in the same
spirit as those performed in previous works in non-relativistic MHD
\cite{Toth2000, Balsara2004, Mocz2016}.  The well-known work by
\cite{Toth2000} compares seven divergence-control methods, including an
early non-conservative divergence-cleaning method known as the {\it
  eight-wave method} \cite{Powell1994}, and three CT methods, finding
that FCT is among the three most accurate schemes for the test problems
studied.  In \cite{Balsara2004}, three divergence-cleaning schemes and
one CT scheme were applied to the same test problem of supernova-induced
MHD turbulence in the interstellar medium.  It was found that the three
divergence-cleaning methods studied suffer from, among other problems,
spurious oscillations in the magnetic energy, which is attributed to the
non-locality introduced by the loss of hyperbolicity in the equations.
Finally, in \cite{Mocz2016}, a non-staggered version of CT adapted to a
moving mesh is compared to the divergence-cleaning Powell scheme
\cite{Powell1999}, an improved version of the eight-wave method.  They
observe greater numerical stability and accuracy, and a better
preservation of the magnetic field topology for the CT scheme. In their
tests, the Powell scheme suffers from an artificial growth of the
magnetic field.  This is explained to be a result of the scheme being
non-conservative.

\subsubsection{Orszag-Tang Vortex}
\begin{figure}
\includegraphics[height=7cm]{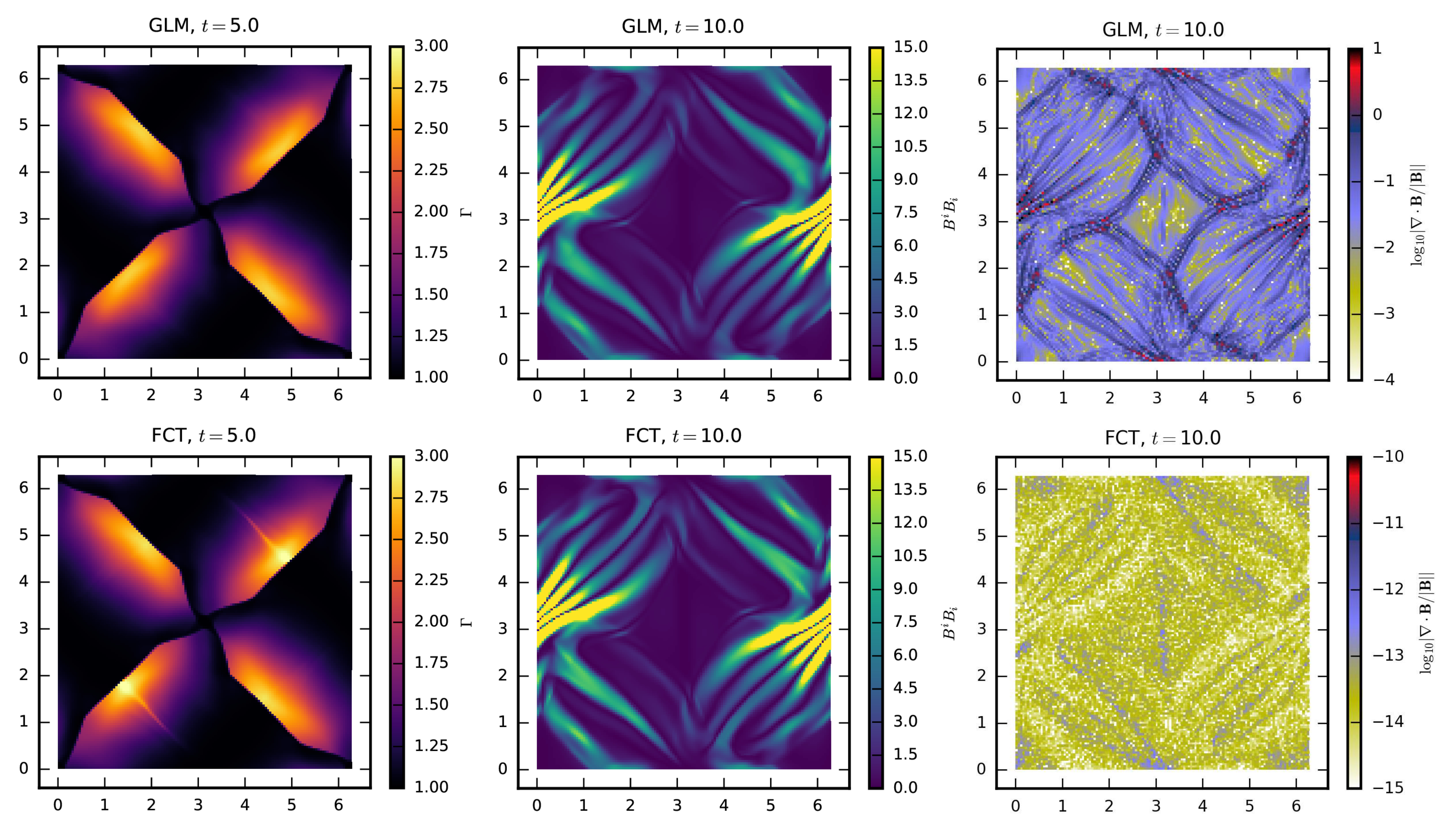}
\caption{Relativistic Orszag-Tang vortex. {\it Left column}: small
  differences can be observed in this snapshot of the Lorentz factor at
  $t=5.0$.  Some features that appear when using CT are flattened when
  using GLM, possibly due to a greater diffusivity of the latter.  {\it
    Middle column}: final snapshot of $B^iB_i$. Good agreement between
  the two methods can be seen, except at some extreme points.  {\it Right
    column}: violation of $\nabla \cdot \boldsymbol{B} = 0 $.}
\label{fig:GLM-FCT-OT}
\end{figure}

The Orszag-Tang vortex \cite{Orszag1979} is a common problem that can be
used to test MHD codes for violations of $\nabla\cdot\boldsymbol{B}$.
The relativistic version presented here was performed in 2D using
Cartesian coordinates in a $128\times128$-resolution domain of
$2\pi\times2\pi$ length units with periodic boundary conditions, and
evolved for $10$ time units ($c=1$).  The equation of state was chosen to
be that of an ideal fluid with $\hat{\gamma}=4/3$ and the initial
conditions were set to $\rho=1.0$, $p=10.0$, $v^x=-0.99\sin y$,
$v^y=0.99\sin x$, $B^x=-\sin y$ and $B^y = \sin 2x$ .  Snapshots of the
evolution are shown in Fig.~\ref{fig:GLM-FCT-OT}.

\begin{figure}
\includegraphics[width=0.9\textwidth]{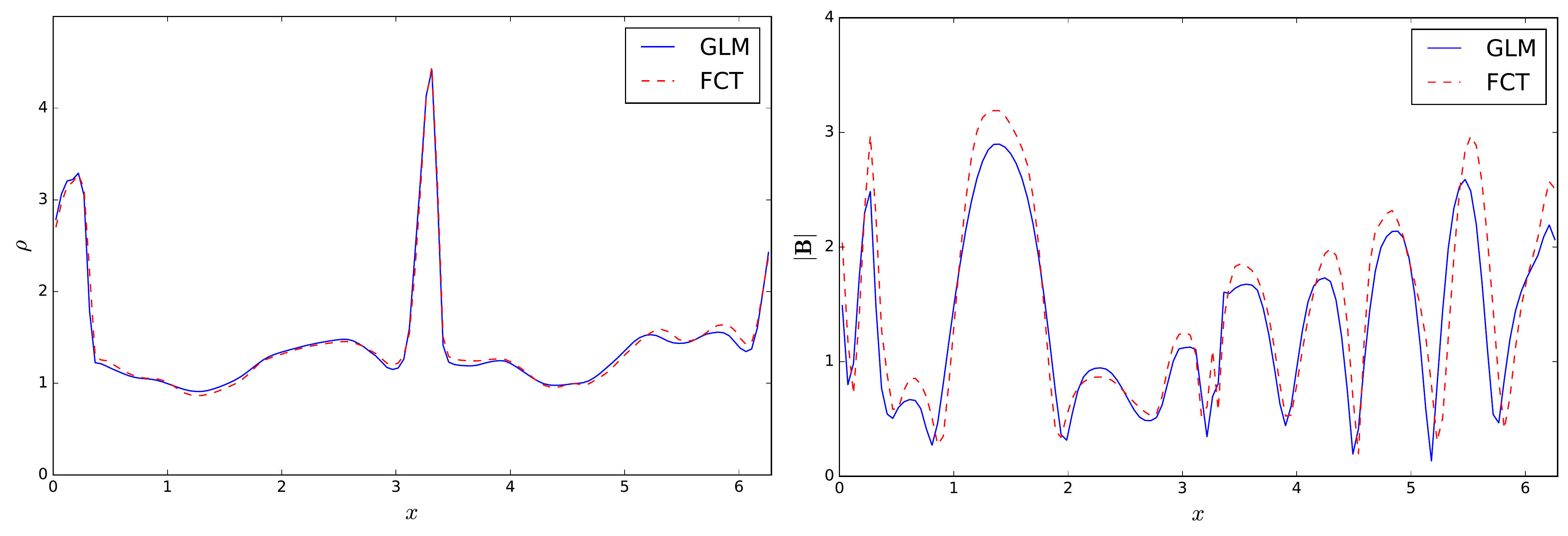}
\caption{Relativistic Orszag-Tang vortex: cuts at $y=\pi/2$ and $t=10.0$
  of the density $\rho$ (left) and the magnitude of the magnetic field
  $|B|$ (right).  While for $\rho$ there is in general a good agreement,
  FCT tends to produce higher maxima for the magnetic field.}
\label{fig:GLM-FCT-OTcuts}
\end{figure}

As can be seen in Figs.~\ref{fig:GLM-FCT-OT} and
~\ref{fig:GLM-FCT-OTcuts}, the general behaviour in both cases is quite
similar qualitatively, with only slight differences at specific
locations.  For instance, when compared to GLM, FCT exhibited higher and
sharper maxima for the magnitude of the magnetic field.  In a similar
fashion, some fine features in the Lorentz factor that can be seen in
Fig. ~\ref{fig:GLM-FCT-OT} for FCT appear to be smeared out when using
GLM, giving a false impression of symmetry under 90$^\circ$ rotations,
while the actual symmetry of the problem is under 180$^\circ$ rotations.
This may be an evidence of FCT being less diffusive than GLM.

\subsubsection{Spherical accretion}

\begin{figure}
\includegraphics[width=0.9\textwidth]{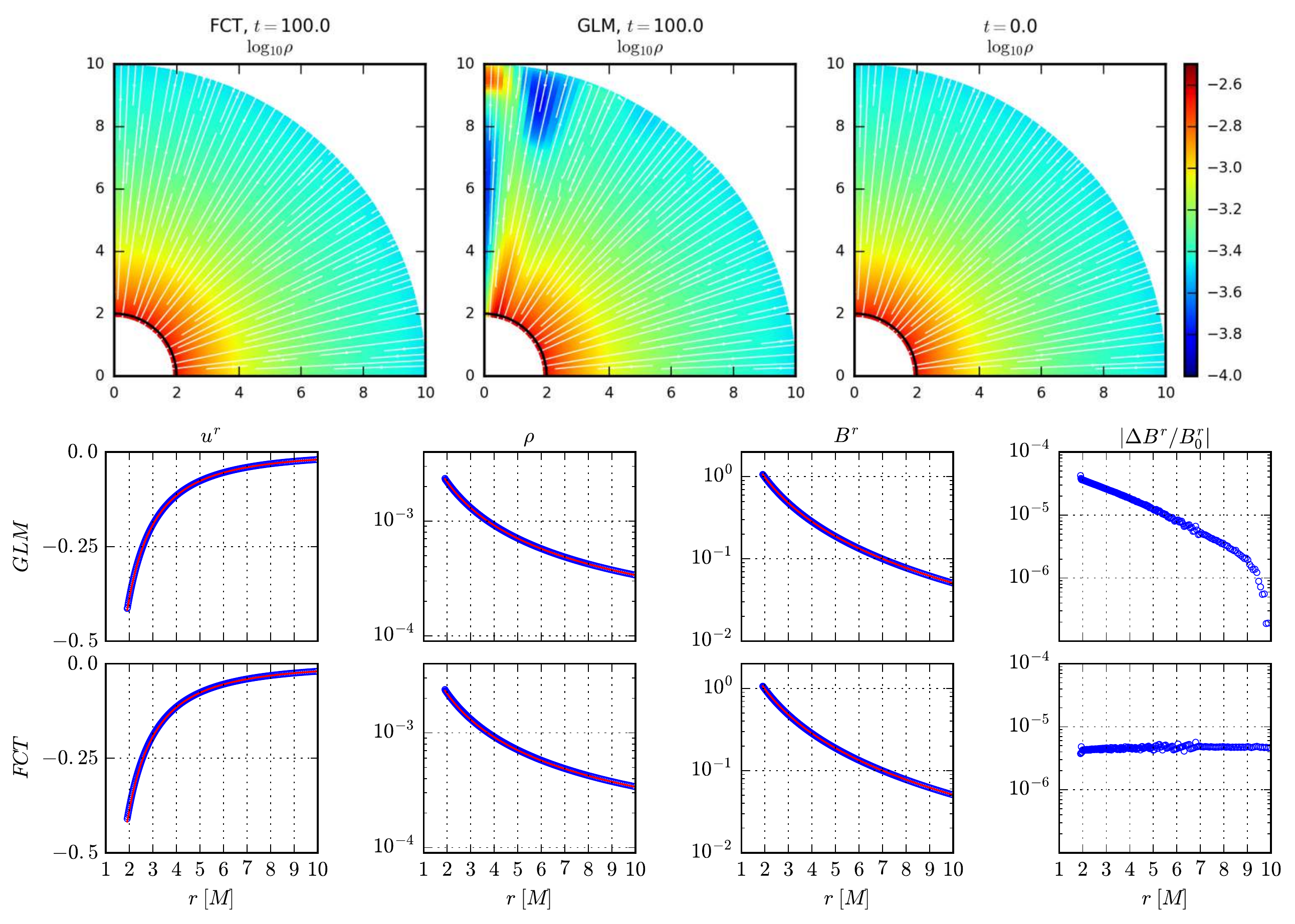}
\caption{{\it Top}: logarithmic density and streamlines in 2D magnetised
  Michel accretion at times $t=0\,M$ (left) and $t=100\,M$ using GLM
  (centre) and FCT (right).  The horizon is marked by the black line at
  $r=2$.  {\it Bottom}: profiles at $\theta=\pi/2$ of, from left to right,
  radial 3-velocity, density and radial magnetic field at $t=0\,M$ (blue
  circles) and $t=100\,M$ (red line) using GLM (upper) and FCT
  (lower). The last column shows the relative difference between the
  magnetic field at $t=100\,M$ and at the initial condition.}
\label{fig:GLM-FCT-Bondi}
\end{figure}

We tested the ability of both methods to preserve a stationary solution
by evolving a magnetised Michel accretion in 2D, as shown in
Fig.~\ref{fig:GLM-FCT-Bondi}.
We employed spherical MKS coordinates (see
  Sect. ~\ref{sec:modkscoord}), $N_r\times N_\theta = 200\times100$
  resolution, and a domain with $r\in [1.9M,10M]$ and
$\theta\in[0,\pi]$.  The fluid obeyed an ideal equation of state with
$\hat{\gamma} = 5/3$ and the sonic radius was located at $r_\mathrm{c} =
8$, and the magnetic field was normalised so that the maximum
  magnetisation was $\sigma=10^3$.  We repeated the numerical experiment
  of section Sect.~\ref{sec:bondi}, now in 2D.  As shown in
Fig.~\ref{fig:GLM-FCT-Bondi}, numerical artefacts start to become
noticeable at these later times.  For instance, with these extreme
  magnetisations, for GLM we observe spurious features near the poles at
  $\theta=0$ and $\theta=\pi$, as well as deviations in the velocity
  field near the outer boundary $r=10\,M$.  The polar region is of
  special interest for jet simulations, where the divergence-control
  method must be robust enough to interplay with the axial boundary
  conditions.  The bottom of Fig.~\ref{fig:GLM-FCT-Bondi} shows the
  profiles of several quantities at $\theta=\pi/2$. Both
  divergence-control methods produce an excellent agreement between the
  solution at different times in the equatorial region.  The rightmost
column in the bottom of Fig.~\ref{fig:GLM-FCT-Bondi} shows the relative
errors in the radial component of the magnetic field for each
method. The errors for FCT are not only one order of magnitude lower
  than for GLM, but also behave differently, remaining at the same level
  near the more-magnetised inner region instead of growing as seen for
  GLM.

\subsubsection{Accreting Torus}\label{sec:GLM-FCT-torus}

To compare both methods in a setting closer to our intended scientific
applications, we simulated accretion from a magnetised perturbed
Fishbone-Moncrief torus around a Kerr BH with $M=1$ and $a=0.9375$.  We
employed modified spherical MKS coordinates as described in
Sect.~\ref{sec:modkscoord} and a domain where $r\in [1.29,2500]$ and
$\theta\in[0,\pi]$ with a resolution of $N_r\times N_\theta = 512 \times
256$, and evolved the system until $t=2000\,M$.
At the radial boundaries, we imposed \textit{noinflow} boundary conditions
while at the boundaries with the polar axis we imposed symmetric
boundary conditions for the scalar variables and the radial vector components
and antisymmetric boundary conditions for the azimuthal and polar components.  
In the BHAC code, \textit{noinflow} boundary conditions are implemented via continuous extrapolation 
of the primitive variables and by replacing the three-velocity with zero in case inflowing velocities are present in the solution.  
The fluid obeyed an ideal
equation of state with $\hat{\gamma} = 4/3$.  The inner edge of the torus
was located at $r_{\mathrm{in}} = 6.0$ and the maximum density was
located at $r_{\mathrm{max}} = 12.0$.  The initial magnetic field
configuration consisted of a single loop with $A_{\phi} \propto
(\rho/\rho_{\mathrm{max}} -\rho_{\mathrm{cut}} )$ and zero for
$\rho < \rho_{\mathrm{cut}} = 0.2$.  To simulate vacuum, the region outside the
torus was filled with a tenuous atmosphere as is customarily done in
these types of simulation.  In this case, the prescription for the
atmosphere was $\rho_{\mathrm{atm}} = \rho_{\mathrm{min}} \, r^{-3/2}$
and $ p_{\mathrm{atm}} = p_{\mathrm{min}} \, r^{-5/2}$, where
$\rho_{\mathrm{min}} = 10^{-5} $ and $p_{\mathrm{min}} = 1/3 \times10^{-7}$.
\begin{figure}
\includegraphics[width=0.8\textwidth]{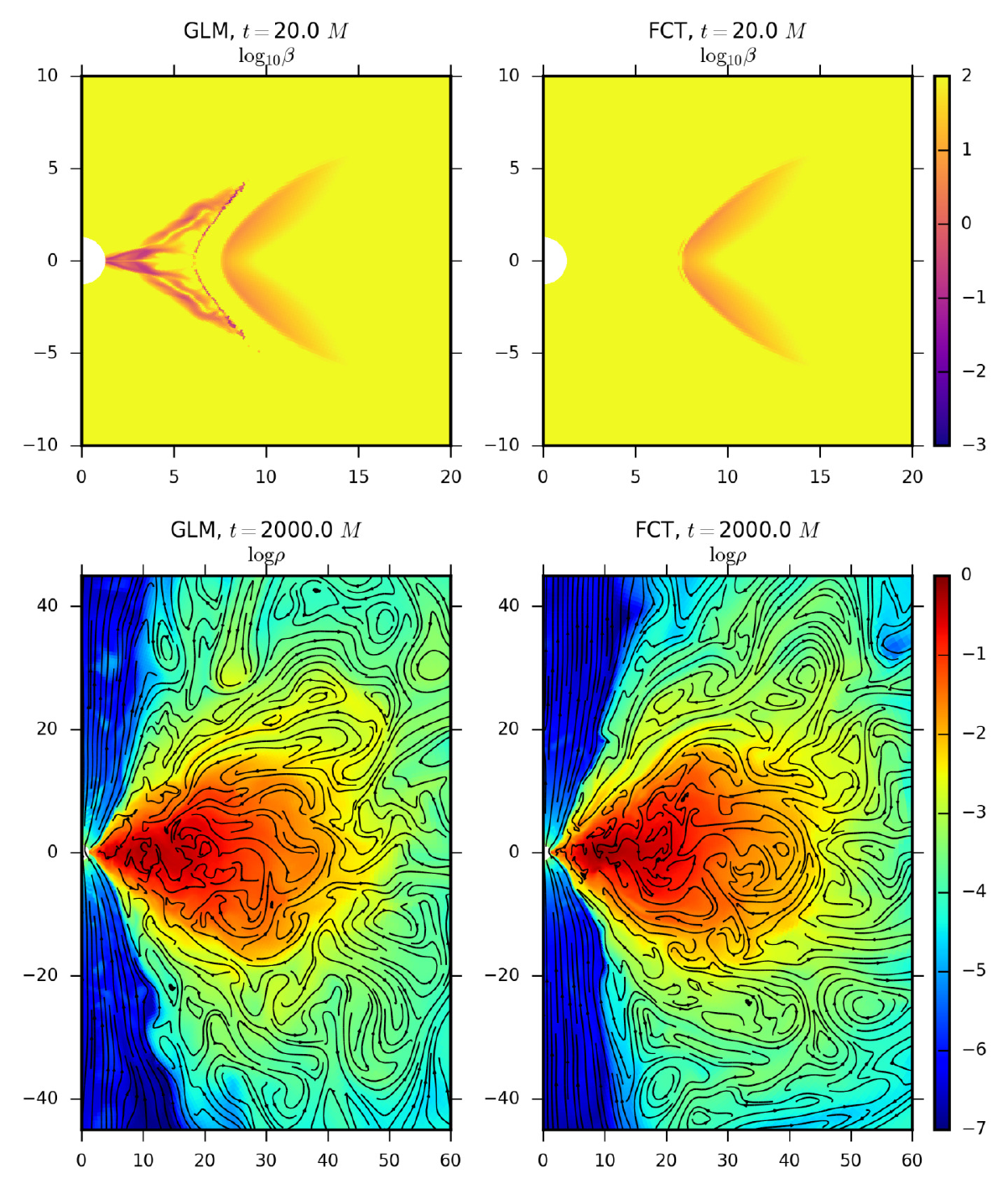}
\caption{Magnetised torus: plasma $\beta$ at $t=20\,M$ (top) and density
  and magnetic field lines $t=2000\,M$ (bottom) using GLM (left) and FCT
  (right).}
\label{GLM-FCT-Torus}
\end{figure}
\noindent
A qualitative difference can be seen even at early times of the
simulation.  The two upper panels of Fig.~\ref{GLM-FCT-Torus} show a
snapshot of the simulation at $t=20\,M$ using both GLM and FCT.
For GLM some of the magnetic field has diffused out of the original
torus, magnetising the atmosphere.  This artefact is visible for GLM from
almost the beginning of the simulation ($t\approx20\,M$), while for FCT
it is minimal.  Even though this particular artefact is not of
crucial importance for the subsequent dynamics of the simulation, this
points to a higher inclination of GLM to produce spurious magnetic field
structures.  At later times (bottom panels of Fig.~\ref{GLM-FCT-Torus}),
the most noticeable difference is the smaller amount of turbulent
magnetic structures and the bigger plasma magnetisation inside the funnel
in FCT, as compared to GLM.  This latter difference indicates that the
choice of technique to control $\nabla\cdot\boldsymbol{B}$ may have an
effect on the possibility of jet formation in GRMHD simulations, although
this specific effect was not extensively studied.

To summarise this small section on the comparison between both
divergence-control techniques, we found from the three tests performed
that FCT seems to be less diffusive than GLM, is able to preserve for a
longer time a stationary solution, and seems to create less spurious
structures in the magnetic field.  However, it still has the inconvenient
property that it is not possible to implement a cell-entered version of
it whilst fully incorporating AMR.  As mentioned previously, we are
currently working on a staggered implementation adapted to AMR, and this
will be described in a separate work.

\section{Torus simulations}\label{sec:tori}

\subsection{Initial conditions}\label{sec:torus_initial}

We consider a hydrodynamic equilibrium torus threaded by a weak magnetic
field loop. The particular equilibrium torus solution with constant
angular momentum was first presented by \cite{Fishbone76} and
\cite{Kozlowski1978} and is now a standard test for GRMHD simulations
\citep[see, \eg][]{Font02a,Zanotti03,Anton05,Rezzolla_book:2013,White2016}. To
facilitate cross-comparison, we set the initial conditions in the torus
close to those adopted by \cite{Shiokawa2012, White2016}. Hence the
spacetime is a Kerr BH with dimensionless spin parameter $a =
0.9375$. The inner radius of the torus is set to $r_{\rm in} = 6\,M$
and the density maximum is located at $r_{\rm max}=12\,M$, where
radial and azimuthal positions refer to Boyer-Lindquist coordinates.
With these choices, the orbital period of the torus at the density
maximum becomes $T=247M$. We adopt an ideal gas EOS with an adiabatic
index of $\hat{\gamma}=4/3$. A weak single magnetic field loop defined
by the vector potential
\begin{align}
A_{\phi} \propto {\rm max} (\rho/\rho_{\rm max} - 0.2, 0) \,,
\end{align}
is added to the stationary solution. The field strength is set such that
$2 p_{\rm max}/b^2_{\rm max}=100$, where global maxima of pressure
$p_{\rm max}$ and magnetic field strength $b^2_{\rm max}$ do not
necessarily coincide. In order to excite the MRI inside the torus, the
thermal pressure is perturbed by $4\%$ white noise.

As with any fluid code, vacuum regions must be avoided and hence we apply
floor values for the rest-mass density ($\rho_{\mathrm{fl}} = 10^{-5}
r^{-3/2}$) and the gas pressure ($p_{\mathrm{fl}} =
1/3\times10^{-7}\ r^{-5/2}$). In practice, for all cells which satisfy
$\rho \le \rho_{\mathrm{fl}} $ we set $\rho=\rho_{\mathrm{fl}}$, in
addition if $p \le p_{\mathrm{fl}}$, we set $p = p_{\mathrm{fl}}$.

The simulations are performed using horizon penetrating logarithmic KS
coordinates (corresponding to our set of modified KS coordinates with
$h=0$ and $R_0=0$). In the 2D cases, the simulation domain covers $r_{\rm
  KS} \in [0.96 r_\mathrm{h}, 2500\,M]$ and $\theta\in[0,\pi]$, where
$r_\mathrm{h} \simeq 1.35\,M$. In the 3D cases, we slightly excise the
axial region $\theta\in[0.02\pi,0.98\pi]$ and adopt $\phi \in [0, 2\pi]$.
We set the boundary conditions in the horizon and at $r=2500\,M$ to zero
gradient in primitive variables. The $\theta$-boundary is handled as
follows: when the domain extends all the way to the poles (as in our 2D
cases), we adopt ``hard'' boundary conditions, thus setting the flux
through the pole manually to zero. For the excised cone in the 3D cases,
we use reflecting ``soft'' boundary conditions on primitive variables.

The time-update is performed with a two-step predictor corrector based on
the TVDLF fluxes and PPM reconstruction. Furthermore, we set the CFL
number to 0.4 and use the FCT algorithm. Typically, the runs are stopped
after an evolution for $t=5000\,M$, ensuring that no signal from the
outflow boundaries can disturb the inner regions. To check convergence,
we adopt the following resolutions: $N_r\times N_\theta \in
\{256\times128,512\times256,1024\times512 \}$ in the 2D cases and
$N_r\times N_\theta \times N_\phi \in \{128\times 64\times64, 192\times
96\times96, 256\times128\times128, 384\times192\times192\}$ in the 3D
runs. In the following, the runs are identified via their resolution in
$\theta$-direction. For the purpose of validation, we ran the 2D cases
also with the \harmthreed code \cite{Noble2009}.\footnote{The results
  were kindly provided by Monika Moscibrodzka.}

To facilitate a quantitative comparison, we report radial profiles of
disk-averaged quantities similar to
\cite[][]{Shiokawa2012,White2016,Beckwith2008}. For a quantity
$q(r,\theta,\phi,t)$, the shell average is defined as
\begin{align}
\langle q(r,t) \rangle := \frac{\int_0^{2\pi}\int_{\theta{\rm min}}^{\theta{\rm max}} q(r,\theta,\phi,t) \sqrt{-g} \, d\phi \, d\theta}{\int_0^{2\pi}\int_{\theta{\rm min}}^{\theta{\rm max}} \sqrt{-g} \, d\phi \, d\theta} \,, \label{eq:averaged}
\end{align}
which is then further averaged over a given time interval to yield
$\langle q(r) \rangle$ (note that we omit the weighting with the density
as done by \cite[][]{Shiokawa2012,White2016}). The limits $\theta_{\rm
  min}=\pi/3$, $\theta_{\rm max}=2\pi/3$ ensure that atmosphere material
is not taken into account in the averaging.  The time-evolution is
monitored with the accretion rate $\dot{M}$ and the magnetic flux
threading the horizon $\phi_B$
\begin{align}
\dot{M} &:= \int_0^{2\pi}\int_{0}^{\pi} \rho u^r\sqrt{-g} \, d\theta \, d\phi \,, \label{eq:mdot} \\ 
\phi_B &:= \frac{1}{2} \int_0^{2\pi}\int_{0}^{\pi} |B^r|\sqrt{-g} \, d\theta \, d\phi \,, \label{eq:phib}
\end{align}
where both quantities are evaluated at the outer horizon $r_{\mathrm{h}}$.

\subsection{2D results}\label{sec:2dtori}

\begin{figure}[htbp]
\begin{center}
\includegraphics[width=.9\textwidth]{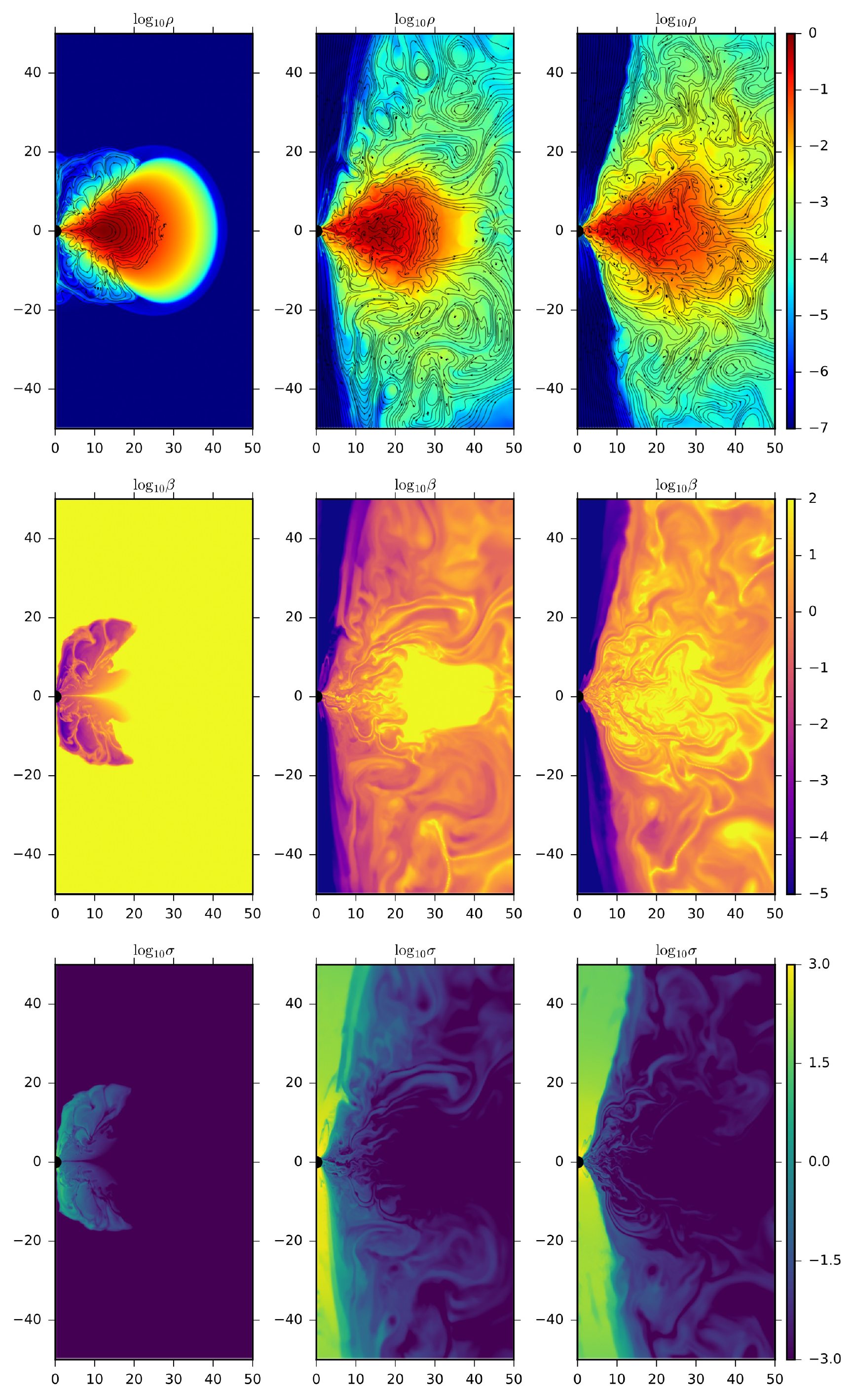}
\caption{Evolution of the 2D magnetised torus with resolution $1024\times 512$ for times $t/M\in\{300,1000,2000\}$. We show logarithmic rest-frame density (top), logarithmic plasma $\beta$ (middle) and the logarithm of the magnetisation parameter $\sigma=b^2/\rho$ (bottom). 
Magnetic field lines are traced out in the first panel using black contour lines. 
One can clearly make out the development of the MRI and evacuation of a strongly magnetised funnel reaching values of $\beta< 10^{-5}$ and $\sigma\approx10^3$. }
\label{fig:magtor-density}
\end{center}
\end{figure}

Figure~\ref{fig:magtor-density} illustrates the qualitative time
evolution of the torus by means of the rest-frame density $\rho$,
plasma-$\beta$ and the magnetisation $\sigma=b^2/\rho$.  After
$t\simeq300\,M$, the MRI-driven turbulence leads to accretion onto the
central BH. The accretion rate and magnetic flux threading the BH then
quickly saturate into a quasi-stationary state (see also
Fig.~\ref{fig:2dharm-MDOT}).  The accreted magnetic flux fills the polar
regions and gives rise to a strongly magnetised funnel with densities and
pressures dropping to their floor values.  For the adopted floor values
we hence obtain values of plasma-$\beta$ as low as $10^{-5}$ and
magnetisations peaking at $\sigma\approx 10^3$ in the inner BH
magnetosphere. These extreme values pose a stringent test for the
robustness of the code and, consequently, the funnel region must be
handled with the auxiliary inversion based on the entropy switch (see
Sect.~\ref{sec:entropy}).

\subsubsection{Comparison to \harmthreed}\label{sec:validation}

For validation purposes we simulated the same initial conditions with the
\harmthreed code. Wherever possible, we have made identical choices for
the algorithm used in both codes, that is: PPM reconstruction, TVDLF
Riemann solver and a two step time update.  It is important to note that
the outer radial boundary differs in both codes: while the \harmthreed
setup implements outflow boundary conditions at $r=50M$, in the
\bhac~runs the domain and radial grid is doubled in the logarithmic
Kerr-Schild coordinates, yielding identical resolution in the region of
interest. This ensures that no boundary effects compromise the
\bhac~simulation.  Next to the boundary conditions, also the initial
random perturbation varies in both codes which can amount to a slightly
different dynamical evolution.

\begin{figure}[htbp]
\begin{center}
\includegraphics[width=0.7\textwidth]{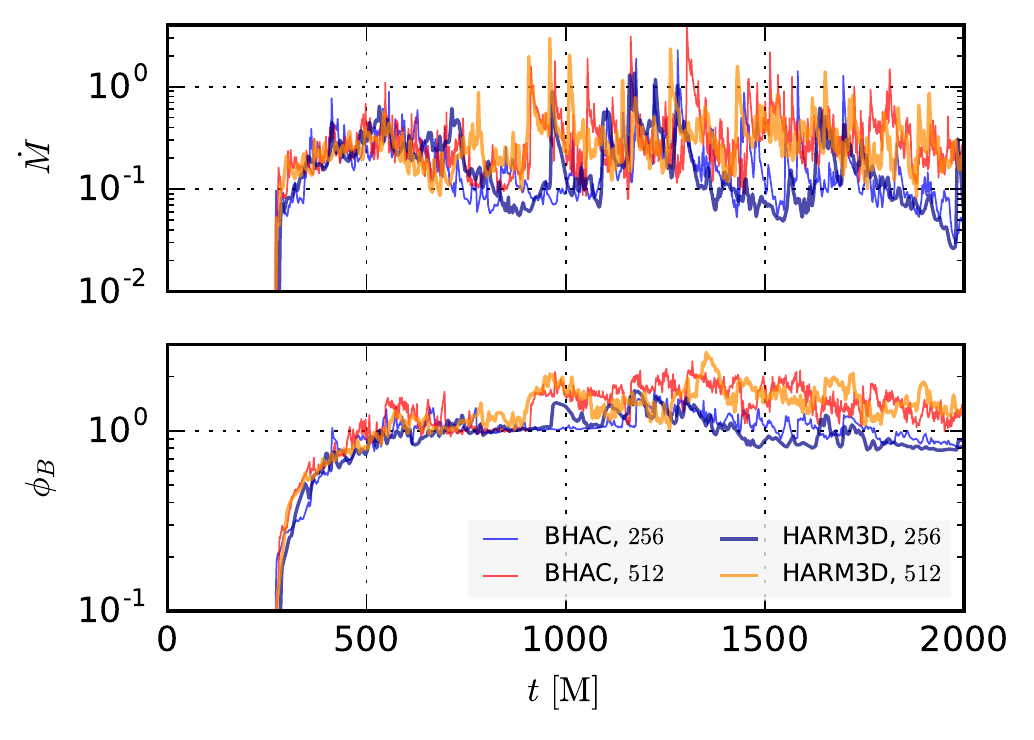}
\caption{Accretion rates and horizon-penetrating magnetic flux in the 2D
  validation runs. We show two resolutions with each code: \bhac~(blue,
  red) and \harmthreed~(dark blue, orange). Despite the chaotic nature of
  the turbulent accretion both codes show very good qualitative and
  quantitative agreement.}
\label{fig:2dharm-MDOT}
\end{center}
\end{figure}

After verifying good agreement in the qualitative evolution, we quantify
with both codes $\dot{M}$ and $\phi_B$ according to equations
\eqref{eq:mdot} and \eqref{eq:phib}. The results are shown in
Fig.~\ref{fig:2dharm-MDOT}. Onset-time of accretion, magnitude and
overall behaviour are in excellent agreement, despite the chaotic nature
of the turbulent flow. We also find the same trend with respect to the
resolution-dependence of the results: upon doubling the resolution, the
accretion rate $\langle \dot{M}\rangle$, averaged over $t\in[1000,2000]$,
increases significantly by a factor of $1.908$ and $1.843$ for \bhac~and
\harm, respectively. For $\langle \phi_B\rangle$, the factors are $1.437$
and $1.484$. At a given resolution, the values for $\langle
\dot{M}\rangle$ and $\langle \phi_B\rangle$ agree between the two codes
within their standard deviations. Furthermore, we have verified that
these same resolution variations are within the run-to-run
deviations due to a different random number seed for the initial
perturbation.

\begin{figure}[htbp]
\begin{center}
\includegraphics[width=0.9\textwidth]{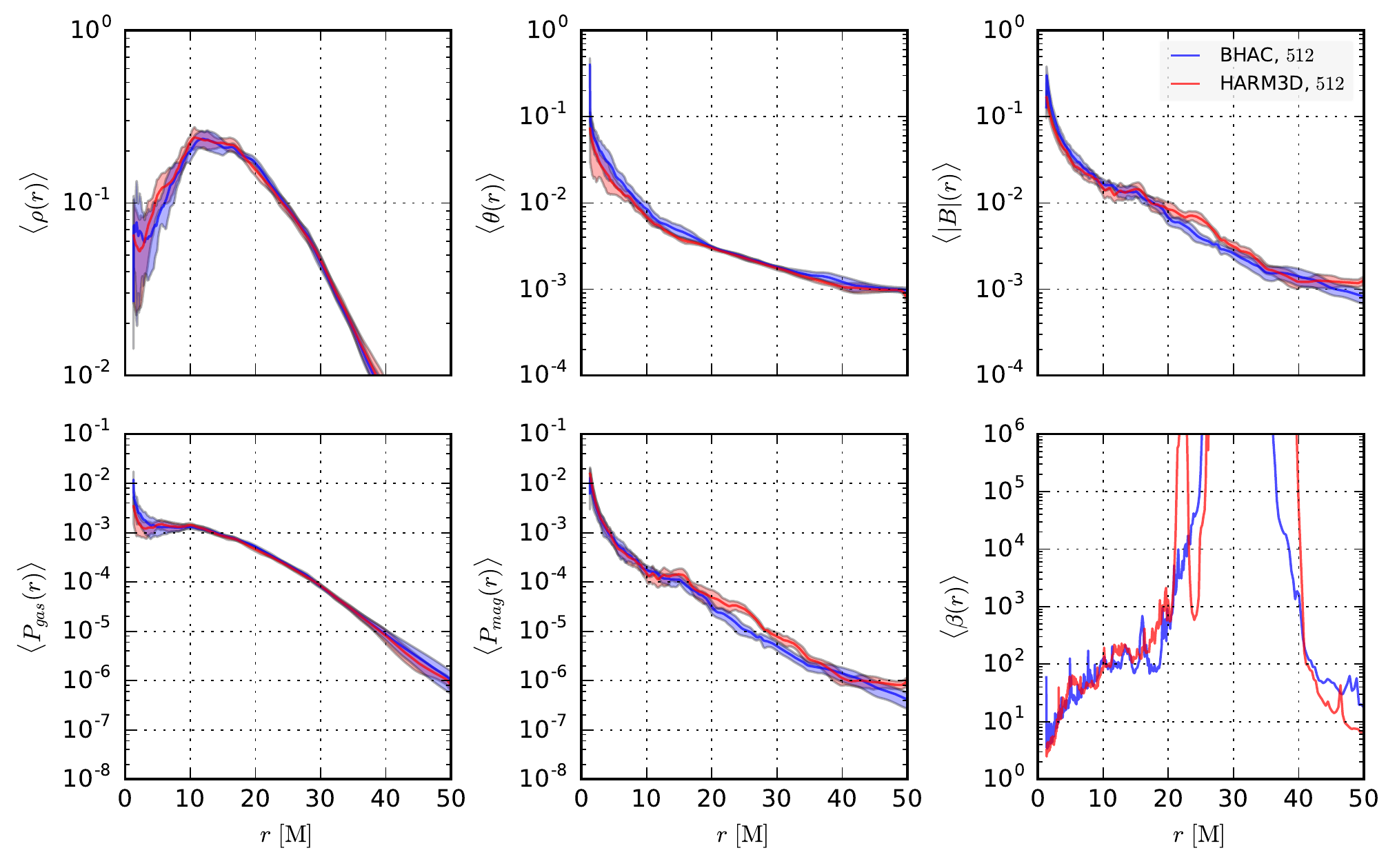}
\caption{Disk-averaged quantities in the 2D validation runs. The blue
  curves are obtained with \bhac~and the red curves with \texttt{HARM3D}
  in a two-dimensional setting. The shaded regions mark the $1\sigma$
  standard deviation of the spatially-averaged snapshots (omitted for the
  highly fluctuating $\langle \beta \rangle$). Apart from a slightly
  higher magnetisation in \harm for $r\in[20,30]$, we find excellent
  agreement between both codes.}
\label{fig:2dharm}
\end{center}
\end{figure}

Further validation is provided in Fig.~\ref{fig:2dharm} where
disk-averaged profiles for the two highest resolution 2D runs are shown
according to equation \eqref{eq:averaged}. The quantities of interest are
the rest-frame density $\rho$, the dimensionless temperature $\Theta:=
p/\rho c^2$, the magnitude of the fluid-frame magnetic field
$|B|=\sqrt{b^2}$, thermal and magnetic pressures $P_{\rm gas}$, $P_{\rm
  mag}$ and the plasma-$\beta$. Again we set the averaging time
$t\in[1000,2000]\,M$ with both codes. The agreement can be considered as
very good, that is apart from a slightly higher magnetisation in \harm
for $r\in[20,30]$, the differences of which are well within the $1\sigma$
standard deviation over the averaging time. Small systematic departures
at the outer edge of the \harm domain are likely attributable to boundary
effects.

\subsection{3D results}

\begin{figure}[htbp]
\begin{center}
\includegraphics[width=.9\textwidth]{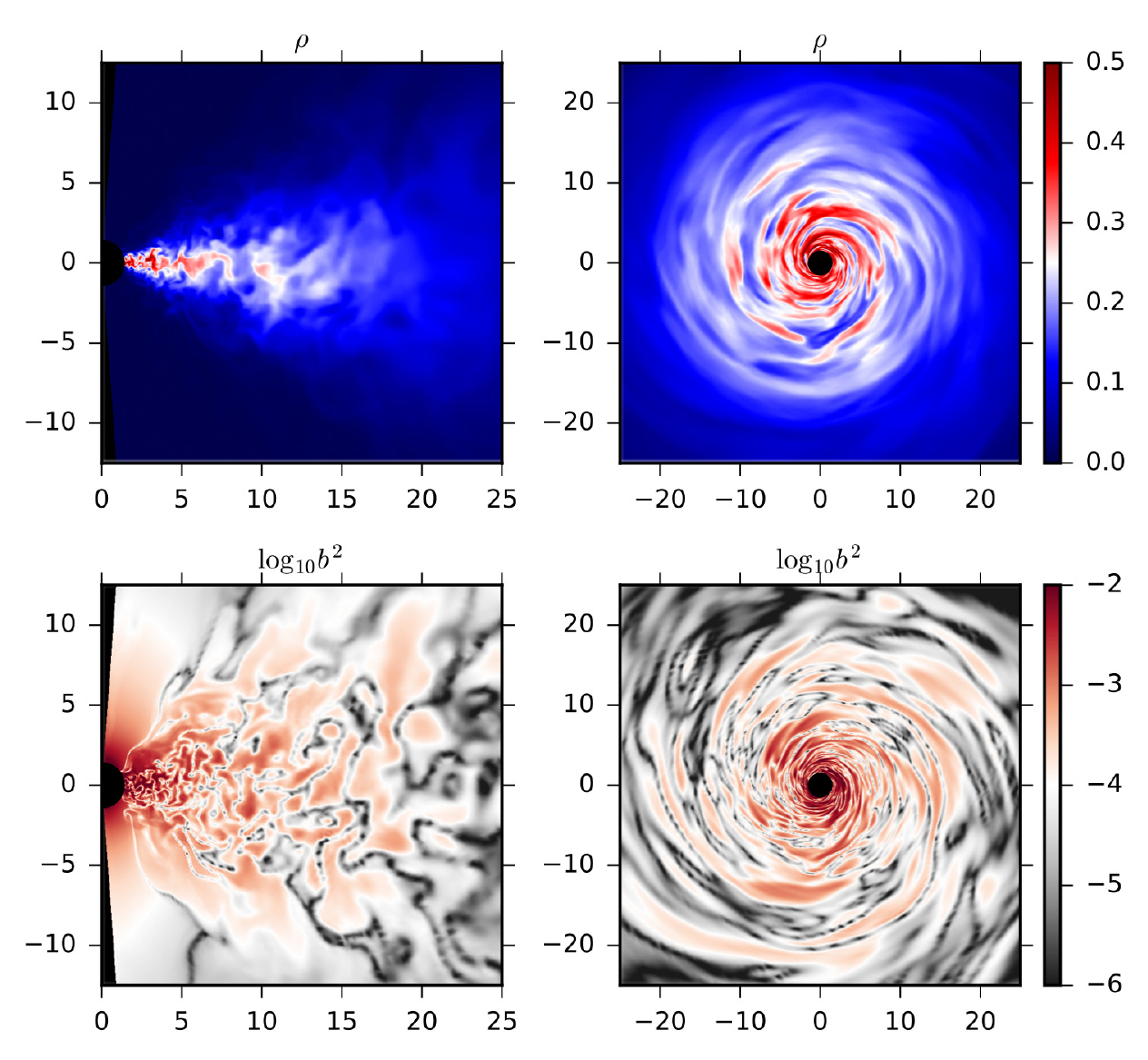}
\caption{Fluid-frame density (top) and $\log_{10}b^2$ (bottom) for $t=3000\,M$ on the $y=0$ plane (left) and the $z=0$ plane (right) in the 3D magnetised torus run with resolution $384\times192\times 192$.}
\label{fig:magtor-density3D}
\end{center}
\end{figure}

We now turn to the 3D runs performed with \bhac. The qualitative
evolution of the high resolution run is illustrated in
Fig.~\ref{fig:magtor-density3D} showing rest-frame density and $b^2$ on
the two slices $z=0$ and $y=0$. Overall, the evolution progresses in a
similar manner to the 2D cases: MRI-driven accretion starts at $t\approx
300\,M$ and enters saturation at around $t\simeq1000\,M$. Similar values
for the magnetisation in the funnel region are also obtained.
However, since the MRI cannot be sustained in axisymmetry as poloidal
  field cannot be re-generated via the ideal MHD induction equation
  \citep{Cowling1933}, we expect to see qualitative differences between the
  2D and 3D cases at late times.

Four different numerical resolutions were run which allows a first
convergence analysis of the magnetised torus accretion scenario.  Based
on the convergence study, we can estimate which numerical resolutions are
required for meaningful observational predictions derived from GRMHD
simulations of this type.

Since we attempt to solve the set of dissipation-free ideal MHD
equations, convergence in the strict sense cannot be achieved in the
presence of a turbulent cascade \citep[see also the discussion
  in][]{SorathiaReynolds2012,HawleyRichers2013}.\footnote{Even when the
  dissipation length is well resolved, high-Reynolds number flows show
  indications for positive Lyapunov exponents and thus non-convergent
  chaotic behaviour \citep[see, \eg][]{Lecoanet2016}.}  Instead, given
sufficient scale separation, one might hope to find convergence in
quantities of interest like the disk averages and accretion rates.  The
convergence of various indicators in similar GRMHD torus simulations was
addressed for example by \cite{Shiokawa2012}. The authors found signs for
convergence in most quantifications when adopting a resolution of
$192\times192\times128$, however no convergence was found in the
correlation length of the magnetic field.  Hence the question as to
whether GRMHD torus simulations can be converged with the available
computational power is still an open one.

\begin{figure}[htbp]
\begin{center}
\includegraphics[width=.7\textwidth]{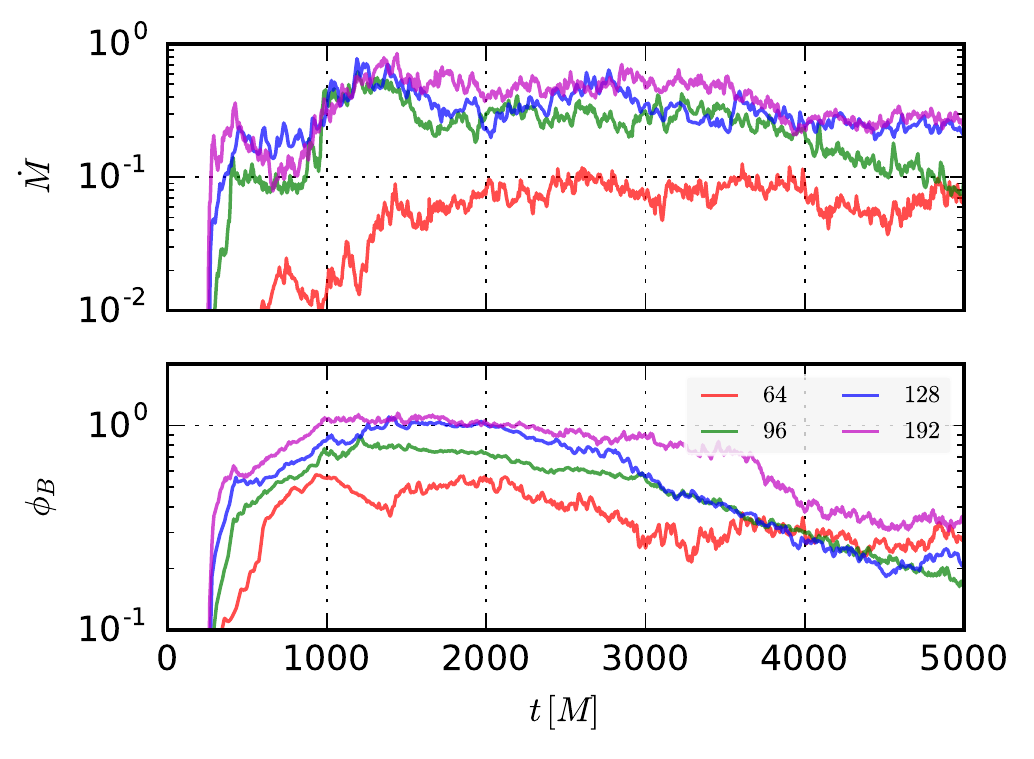}
\caption{Accretion rates and horizon-penetrating magnetic flux in the 3D runs for varying numerical resolution. We show results from four different resolutions labeled according to the number of cells in $\theta$-direction.}
\label{fig:3dbhac-MDOT}
\end{center}
\end{figure}

\begin{figure}[htbp]
\begin{center}
\includegraphics[width=.9\textwidth]{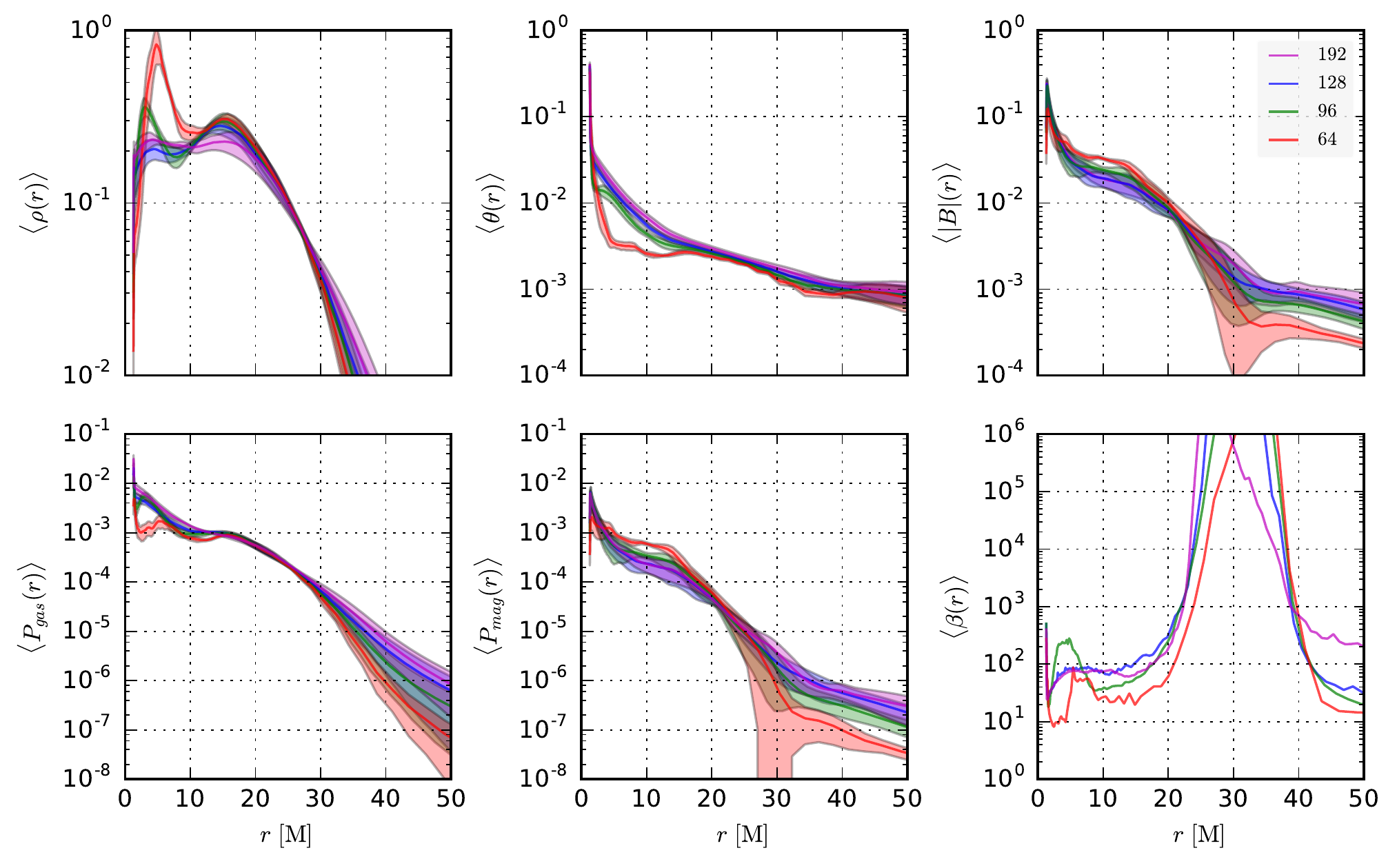}
\caption{Disk-averaged quantities in the 3D runs for varying numerical
  resolution. The shaded regions mark the $1\sigma$ standard deviation of
  the spatially-averaged snapshots as in Fig.~\ref{fig:2dharm}.}
\label{fig:averaged3D}
\end{center}
\end{figure}

From Figs.~\ref{fig:3dbhac-MDOT} and \ref{fig:averaged3D}, it is clear
that the resolution of the $N_\theta=64$ run is insufficient: a peculiar
mini-torus is apparent in the disk-averaged density which diminishes with
increasing resolution. Also the onset-time of accretion and the
saturation values differ significantly between the $N_\theta=64$ run and
its high-resolution counterparts. These differences diminish between the
high-resolution runs and we can see signs of convergence in the accretion
rate: increasing resolution from $N_\theta=128$ to $N_\theta=192$ appears
to not have a strong effect on $\dot{M}$. Also the evolution of $\phi_B$
agrees quite well between $N_\theta=128$ and $N_\theta=192$. Hence the
systematic resolution dependence of $\dot{M}$ and $\phi_B$ in the (even
higher resolution) 2D simulations appears to be an artefact of the
axisymmetry.  It is also noteworthy that the variability amplitude of the
accretion rate is reduced in the 3D cases. It appears that the
superposition of uncorrelated accretion events distributed over the
$\phi$-coordinate tends to smear out the sharp variability that results
in the axisymmetric case.

Although the simulations were run until $t=5000\,M$, in order to enable
comparison with the 2D simulations, we deliberately set the averaging
time to $t\in [1000\,M,2000\,M]$. Figure~\ref{fig:averaged3D} shows that
as the resolution is increased, the disk-averaged 3D data approaches the
much higher resolution 2D results shown in Fig.~\ref{fig:2dharm},
indicating that the dynamics are dominated by the axisymmetric MRI modes
at early times.  When the resolution is increased from $N_\theta=128$ to
$N_\theta=192$, the disk-averaged profiles generally agree within their
standard deviations, although we observe a continuing trend towards
higher gas pressures and magnetic pressures in the outer regions
$r\in[30M,50M]$.  The overall computational cost quickly becomes
significant: for the $N_\theta=128$ simulation we spent $100\,K$ CPU
hours on the Iboga cluster equipped with Intel(R) Xeon(R) E5-2640 v4
processors. As the runtime scales with resolution according to
$N_\theta^4$, doubling resolution would cost a considerable $1.6\,M$ CPU
hours.

\subsection{Effect of AMR} 

In order to investigate the effect of the AMR treatment, we have
performed a 2D AMR-GRMHD simulation of the torus setup.  It is clear that
whether a simulation can benefit from adaptive mesh refinement is very
much dependent on the physical scenario under investigation. For example,
in the hydrodynamic simulations of recoiling BHs due to
\cite{Meliani2017}, refinement on the spiral shock was demonstrated to
yield significant speedups at a comparable quality of solution. This is
understandable as the numerical error is dominated by the shock
hypersurface.  In the turbulent accretion problem, whether automated mesh
refinement yields any benefits is not clear.

The initial conditions for this test are the same as those used in
Sect.~\ref{sec:torus_initial}. However, due to the limitation of current
AMR treatment, we resort to the GLM divergence cleaning method. Three
refinement levels are used and refinement is triggered by the error
estimator due to \cite{Loehner87} with the tolerance set to
$\epsilon_\mathrm{t} =0.1$ (see Sect.~\ref{sec:AMR}).  The numerical
resolution in the base level is set to $N_r \times N_\theta = 128 \times
128$. To test the validity and efficiency, we also perform the same
simulation in a uniform grid with resolution of $N_r \times N_\theta =
512 \times 512$ which corresponds to the resolution on the highest AMR
level.

\begin{figure}[htbp]
\begin{center}
\includegraphics[width=.9\textwidth]{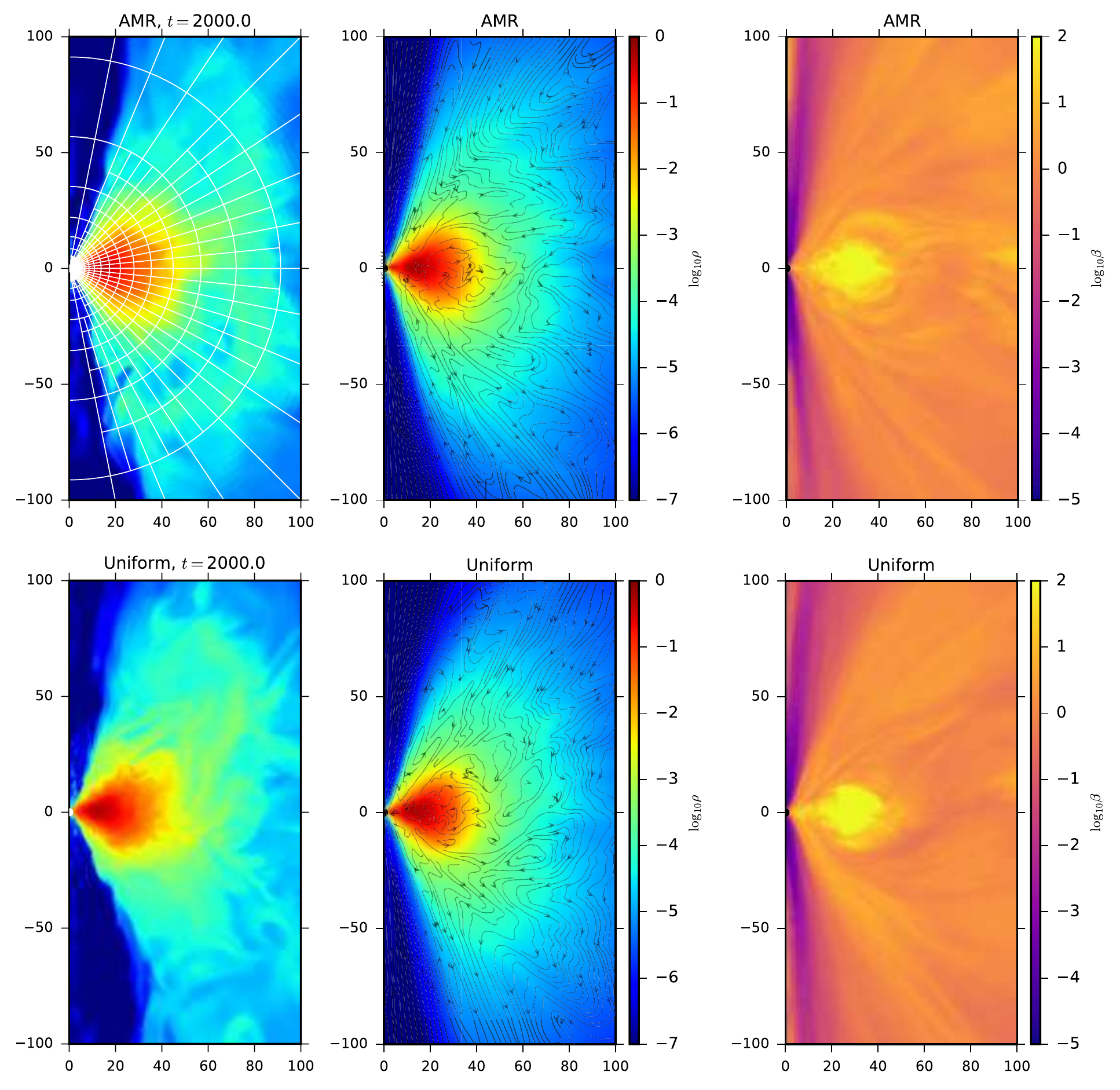}
\caption{2D logarithmic density at $t=2000\,M$ (left), averaged density
  (middle), and averaged plasma beta (right) of the 2D magnetised torus
  with three-levels AMR (top panels) and uniform resolution $512 \times
  512$ (bottom panels).  Magnetic field lines are traced out in the
  middle panels using black contour lines.  The averaged quantities are
  calculated in the time interval $t\in[1000\,M,2000\,M]$. AMR blocks
  containing $16^2$ cells are indicated in the upper left panel.}
\label{fig:AMR-2D}
\end{center}
\end{figure}

\begin{figure}[htbp]
\begin{center}
\includegraphics[width=.7\textwidth]{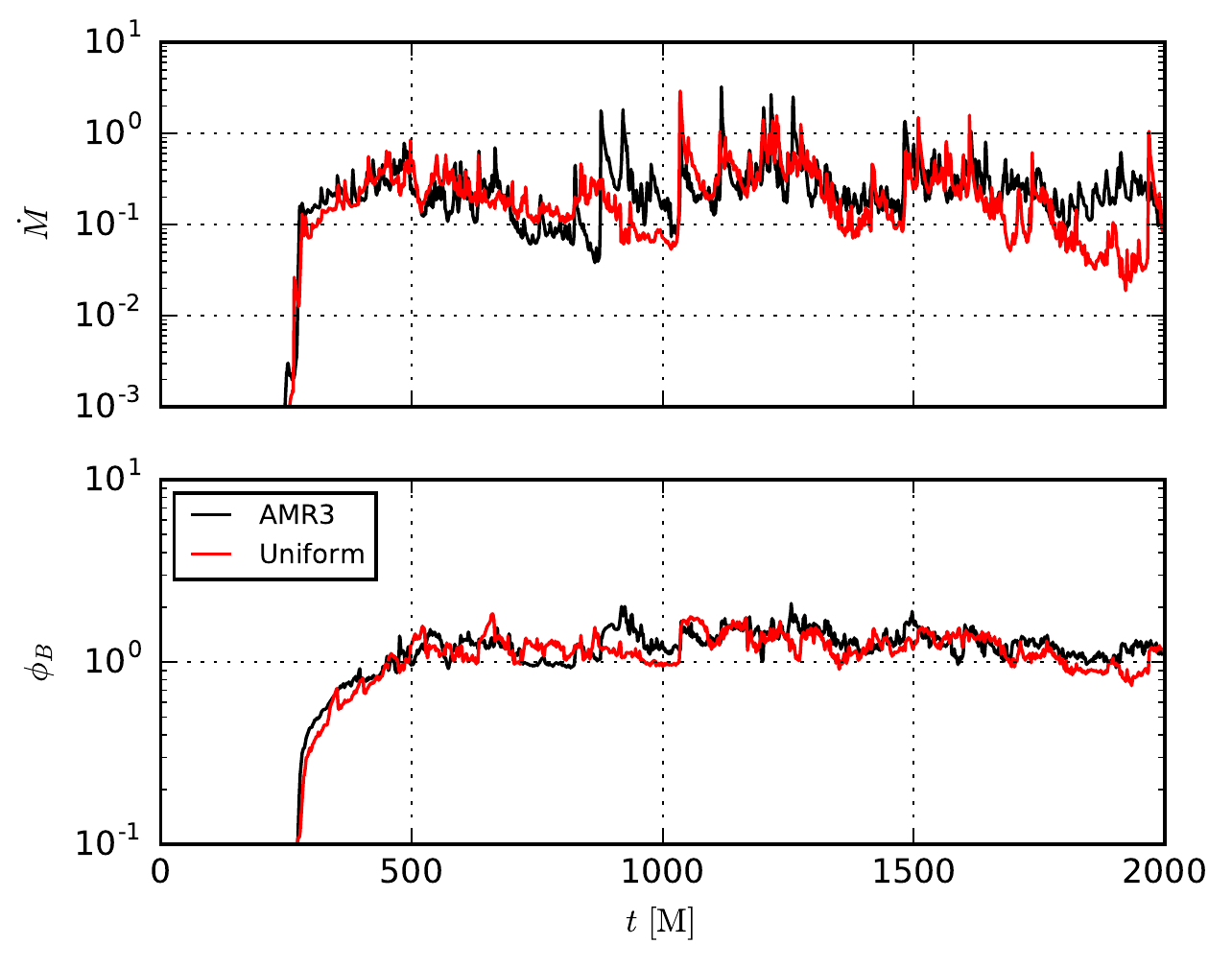}
\caption{Accretion rates and horizon penetrating magnetic flux of the 2D
  magnetised torus with three levels of AMR (black) and uniform
  resolution $512 \times 512$ (red). }
\label{fig:AMR-Mdot}
\end{center}
\end{figure}

Figure~\ref{fig:AMR-2D} shows the densities at $t=2000\,M$ as well as the
time-averaged density and plasma beta for the AMR and uniform cases. The
averaged quantities are calculated in the time interval of
$t\in[1000\,M,2000\,M]$. The overall behaviour is quite similar in both
cases. Naturally, differences are seen in the turbulent structure in the
torus and wind region for a single snapshot. However, in terms of
averaged quantities, the difference becomes marginal. In order to better
quantify the difference between the AMR and uniform runs, the mass
accretion rate and horizon penetrating magnetic flux are shown in
Fig.~\ref{fig:AMR-Mdot}. These quantities exhibit a similar behaviour in
both cases. In particular, the difference between the AMR run and the
uniform run is smaller than the one from different resolution uniform
runs and compatible with the run-to-run variation due to a different
random number seed (cf. Sect.~\ref{sec:2dtori}). This is unsurprising
since the error estimator triggers refinement of the innermost torus
region to the highest level of AMR during most of the simulation
time. The development of small scale turbulence by the MRI is clearly
captured and it leads to similar mass accretion onto the BH.

\begin{table}[h!]
\begin{center}
\caption{CPU hours (CPUH) spent by the simulations of the 2D magnetised
  torus at uniform resolution and fraction of that time spent by the
  equivalent AMR runs up to $t=2000\,M$.}\label{tab:AMRrun}
\begin{tabular}{ccc}
\hline
Grid size & CPU time & Equiv. AMR \\
$(N_r \times N_\theta)$ & uniform & time fraction \\
   & [CPUH] & [$\epsilon_{\mathrm{t}}=0.1$] \\
\hline
$512 \times 512$ & 674.0 & 0.643 \\
\hline
\end{tabular}
\end{center}
\end{table}
\begin{figure}[htbp]
\begin{center}
\includegraphics[width=.8\textwidth]{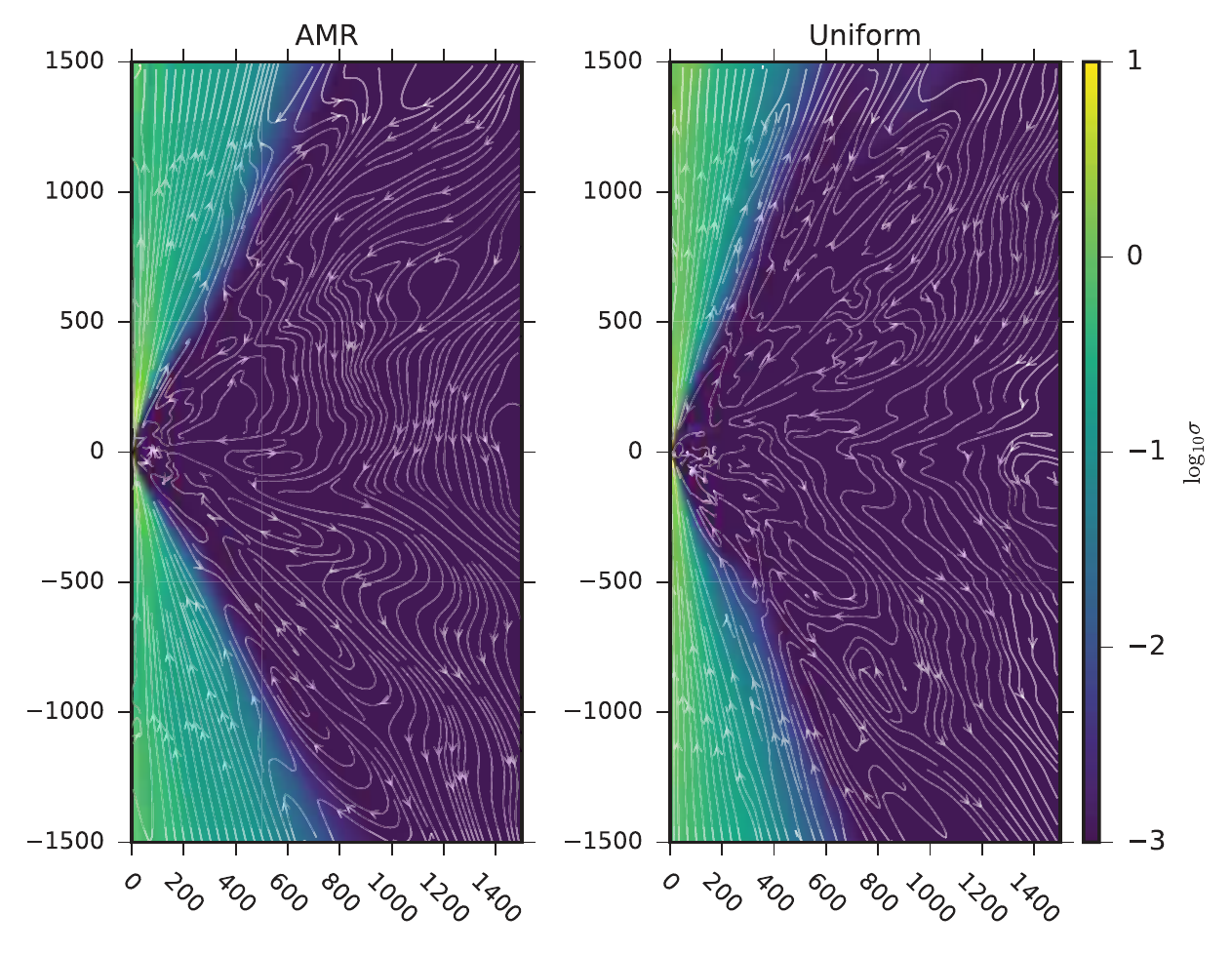}
\caption{2D logarithmic averaged magnetisation of the magnetised torus with three levels  of AMR (left) and uniform resolution $512 \times 512$ (right). 
Magnetic field lines are traced out by white contour-lines. The averaged quantities are calculated in the time interval of $t\in[6000\,M,10000\,M]$. }
\label{fig:AMR-2D-large}
\end{center}
\end{figure}

\begin{figure}[htbp]
\begin{center}
\includegraphics[width=.7\textwidth]{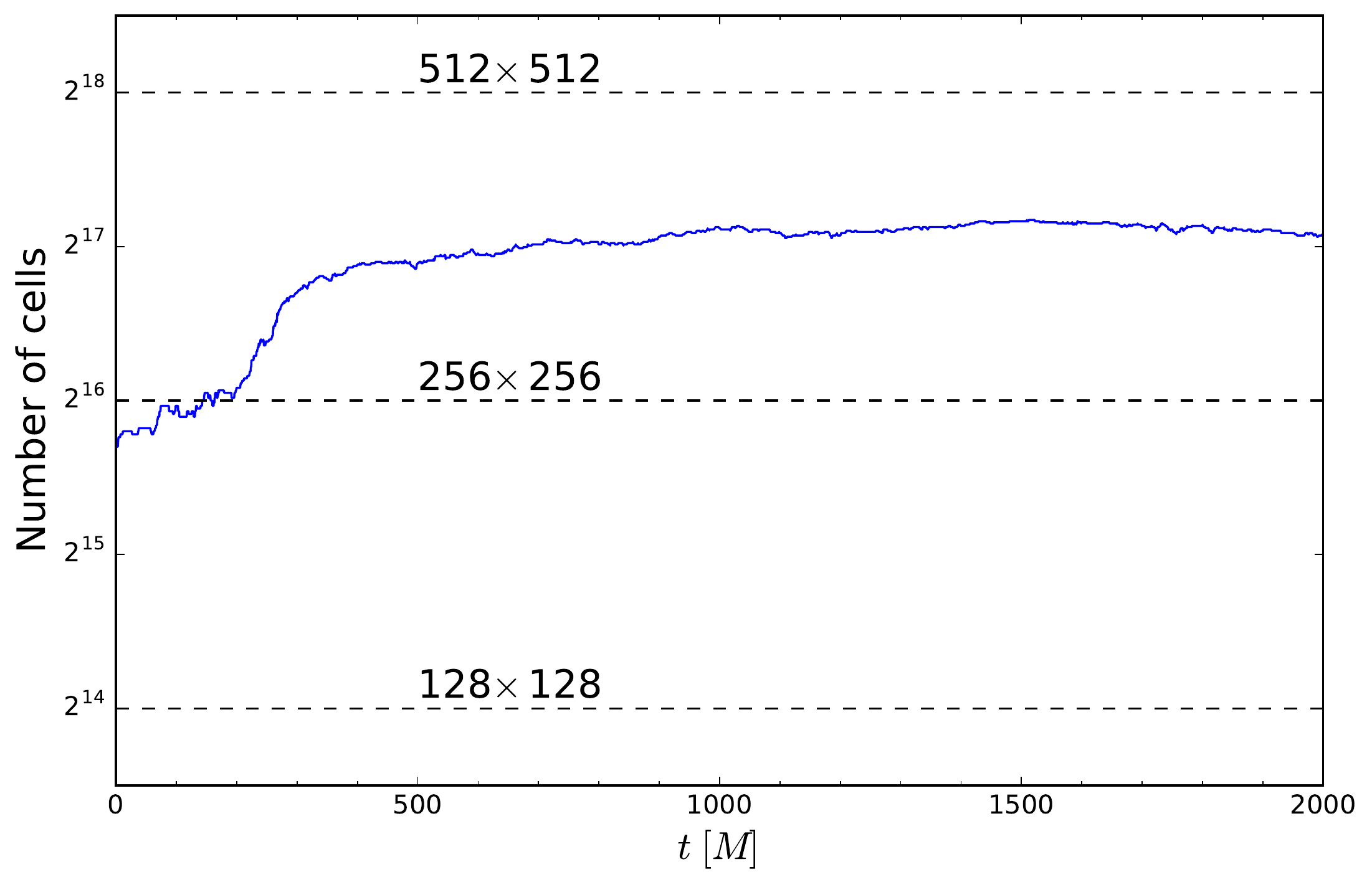}
\caption{Number of cells as a function of time for the AMR simulation. The dotted lines show the resolution of uniform grids equivalent to each of the three AMR levels.}
\label{fig:AMR-CellNo}
\end{center}
\end{figure}

One of the important merits of using AMR is the possibility to
  resolve small and large scale dynamics simultaneously with lower
  computational cost than uniform grids.  Figure~\ref{fig:AMR-2D-large}
shows the large scale structure of the averaged magnetisation after
$10000\,M$ of simulation time. The averaged quantities are calculated in
the time interval $t\in[6000\,M,10000\,M]$.  In order to allow the
  large-scale magnetic field structure to settle down, we average over a
  later simulation time compared to the previous non-AMR cases.
From the figure the collimation angle and magnetisation of the highly
  magnetised funnel in the AMR case are slightly wider than those in uniform
  case but the large-scale global structure is very similar in both cases.

A comparison of the computational time for a uniform resolution with
$512^2$ and the equivalent AMR run (three-level AMR) is shown in
Table~\ref{tab:AMRrun}.  It is encouraging that even in the naive
three-level AMR simulation we obtain qualitatively similar results
comparable to the high resolution uniform run, but with having spent only
$64\%$ of the computational time of the uniform run.\footnote{Since
    we use the same Courant limited timestep for all grid-levels, the
    speedup is entirely due to saving in computational cells.  The
    additional speedup that would be gained from \cite{Berger84}-type
    hierarchical timesteps can be estimated from the level population of
    the simulation: the expected additional gain is only $\sim8\%$ for
    this setup.}  Figure~\ref{fig:AMR-CellNo} shows the evolution of the
total number of cells during the simulations of AMR cases. Initially less
than $2^{16}$ cells are used even when we use three AMR levels,
which is a similar number of cells as the uniform grid case with $256
\times 256$. When the simulation starts, the total cell number increases
rapidly due to development of turbulence in the torus which is triggering
higher refinement. We note that the total number of cells is still half
of the total number of cells in the corresponding high-resolution uniform
grid simulation ($512 \times 512$), thus resulting in a direct reduction
of computational cost. With increasing dynamic range, we expect the
advantages of AMR to increase significantly, rendering it a useful tool
for simulations involving structures spanning multiple scales. We leave a
more detailed discussion on the effect of the AMR refinement strategy and
various divergence-control methods to a future paper.

\section{Radiation post-processing} \label{sec:radiation}

In order to compute synthetic observable images of the BH shadow and
surrounding accretion flow it is necessary to perform
general-relativistic ray-tracing and GRRT post-processing \citep[see,
  \eg][]{Fuerst2004, Vincent2011, Younsi2012, Younsi2015, Chan2015,
  Dexter2016, Pu2016, Younsi2016}. In this article the GRRT code
\bhoss~(Black Hole Observations in Stationary Spacetimes)
\cite{Younsi2017} is used to perform these calculations. From \bhac,
GRMHD simulation data are produced which are subsequently used as input
for \bhoss. Although \bhac~has full AMR capabilities, for the GRRT it is
most expedient to output GRMHD data that has been re-gridded to a uniform
grid.

Since these calculations are performed in post-processing, the effects of
radiation forces acting on the plasma during its magnetohydrodynamical
evolution are not included. Additionally, the fast-light approximation
has also been adopted in this study, \ie it is assumed that the
light-crossing timescale is shorter than the dynamical timescale of the
GRMHD simulation and the dynamical evolution of the GRMHD simulation as
light rays propagate through it is not considered. Such calculations are
considered in an upcoming article \cite{Younsi2017}.

Several different coordinate representations of the Kerr metric are
implemented in \bhoss, including Boyer-Lindquist (BL), Logarithmic BL,
Cartesian BL, Kerr-Schild (KS), Logarithmic KS, Modified KS and Cartesian
KS. All GRMHD simulation data used in this study are specified in
Logarithmic KS coordinates. Although \bhoss~can switch between all
coordinate systems on the fly, it is most straightforward to perform the
GRRT calculations in the same coordinate system as the GRMHD data, only
adaptively switching to \eg Cartesian KS when near the polar region.
This avoids the need to transform between coordinate systems at every
point along every ray in the GRMHD data interpolation, saving
computational time.

\subsection{Radiative transfer equation}
Electromagnetic radiation is described by null geodesics of the
background spacetime (in this case Kerr), and these are calculated in
\bhoss~using a Runge-Kutta-Fehlberg integrator with fourth order adaptive
step sizing and 5th order error control. Any spacetime metric may be
considered in \bhoss, as long as the contravariant or covariant metric
tensor components are specified, even if they are only tabulated on a
grid. For the calculations presented in this article the Kerr spacetime
is written algebraically and in closed-form.

The observer is calculated by constructing a local orthonormal tetrad
using trial basis vectors. These basis vectors are then orthonormalized
using the metric tensor through a modified Gram-Schmidt procedure. The
initial conditions of each ray for the coordinate system under
consideration are then calculated and the geodesics are integrated
backwards in time from the observer, until they either: (i) escape to
infinity (exit the computational domain), (ii) are captured by the BH, or
(iii) are effectively absorbed by the accretion flow.

In order to perform these calculations the GRRT equation is integrated in
parallel with the geodesic equations of motion of each ray. Written in
covariant form, the (unpolarized) GRRT equation, in the absence of
scattering, may be written \citep[][]{Younsi2012} as
\begin{equation}
\frac{d\mathcal{I}}{d\lambda} = -k_{\mu}u^{\mu} \left( -\alpha_{\nu,0} \, \mathcal{I}+\frac{j_{\nu,0}}{\nu_{0}^{3}} \right) \,, \label{GRRT_eqn}
\end{equation}
where $\mathcal{I}:= I_{\nu}/\nu^{3}$ is the Lorentz-invariant intensity,
$I_{\nu}$ is the specific intensity, $\nu$ is the frequency of radiation,
$\alpha_{\nu,0}$ is the specific absorption coefficient and $j_{\nu,0}$
is the specific emission coefficient. The subscript ``$\nu$" denotes
evaluation of a quantity at a specific frequency, $\nu$, and a subscript
``$0$" denotes evaluation of a quantity in the local fluid rest frame.
The terms $k_{\mu}$ and $u^{\mu}$ are the photon $4$-momentum and the
fluid $4$-velocity of the emitting medium, respectively. The former is
calculated from the geodesic integration and the latter is determined
from the GRMHD simulation data. The affine parameter is denoted by
$\lambda$.

By introducing the optical depth along the ray
\begin{equation}
\tau_{\nu}\left(\lambda \right) = -\int_{\lambda_{0}}^{\lambda}\ \! \alpha_{\nu,0} \left(\lambda' \right) k_{\mu}u^{\mu} \, d\lambda'  \,,
\end{equation}
together with the Lorentz-invariant emission coefficient $\left(\eta =
j_{\nu}/\nu^{2}\right)$ and Lorentz-invariant absorption coefficient
$\left( \chi = \nu\alpha_{\nu} \right)$, the GRRT equation
(\ref{GRRT_eqn}) may be rewritten as
\begin{equation}
\frac{d\mathcal{I}}{d\tau_{\nu}} = -\mathcal{I} +
\frac{\eta}{\chi} \,. \label{GRRT_eqn_tau}
\end{equation}
Following \cite{Younsi2012}, equation (\ref{GRRT_eqn_tau}) may be reduced to two differential equations
\begin{eqnarray}
\gamma\frac{d\tau_{\nu}}{d\lambda} &=&
\alpha_{\nu,0} \,, \label{GRRT_1} \\
\gamma\frac{d\mathcal{I}}{d\lambda} &=&
\frac{j_{\nu,0}}{\nu_{0}^{3}} \ \!
\mathrm{exp}\left(-\tau_{\nu}\right) \label{GRRT_2} \,,
\end{eqnarray}
where
\begin{equation}
\gamma = \frac{\nu}{\nu_{0}} = \frac{\left(k_{\alpha}u^{\alpha}\right)_{\mathrm{obs}}}{\left(k_{\beta}u^{\beta}\right)_{0}} \,,
\end{equation}
is the relative energy shift between the observer (``obs") and the
emitting fluid element. Integrating the GRRT equation in terms of the
optical depth in the manner presented provides two major advantages.
Firstly, the calculation of the geodesic and of the radiative transfer
equation may be performed simultaneously, rather than having to calculate
the entire geodesic, store it in memory, and then perform the radiative
transfer afterwards. Secondly, by integrating in terms of the optical
depth we may specify a threshold value (typically of order unity) whereby
the geodesic integration is terminated when encountering optically thick
media exceeding this threshold. The combination of these two methods
saves significant computational time and expense.

\subsection{\bhoss-simulated emission from Sgr A*}

Having in mind the upcoming radio observations of the BH candidate Sgr A*
at the Galactic Centre, the following discussion presents synthetic
images of Sgr A*. The GRMHD simulations evolve a single fluid (of ions)
and are scale-free in length and mass. Consequently a scaling must be
applied before performing GRRT calculations. Within \bhoss~this means
specifying the BH mass, which sets the length and time scales, and
specifying either the mass accretion rate or an electron density scale,
which scales the gas density, temperature and magnetic field strength to
that of a radiating electron.

Since the GRMHD simulation is of a single fluid, it is necessary to adopt
a prescription for the local electron temperature and rest-mass density.
Several such prescriptions exist, some which scale using the mass
accretion rate
\citep[see, \eg][]{Moscibrodzka2009,Moscibrodzka2014,Dexter2010}, scale using
density to determine the electron number density and physical accretion
rate \citep[see, \eg][]{Chan2015,Chan2015b}, and some by employing a
time-dependent smoothing model of the mass accretion rate
\citep[see, \eg][]{Shiokawa2012}.

The dimensionless proton temperature, $\Theta_{\mathrm{p}}$, is defined as
\begin{equation}
\Theta_{\mathrm{p}} := \frac{k_{\mathrm{B}}T_{\mathrm{p}}}{m_{\mathrm{p}}c^{2}} \,,
\end{equation}
where $k_{\mathrm{B}}$ is the Boltzmann constant, $T_{\mathrm{p}}$ is the
geometrical (\ie in physical units) proton temperature and
$m_{\mathrm{p}}$ is the proton mass. This is then calculated from the
GRMHD simulation density ($\rho$) and pressure ($p$) as
\begin{equation}
\Theta_{\mathrm{p}} = \frac{3p}{\rho} \,,
\end{equation}
where the fact that the equation of state is ideal and that
$\hat{\gamma}=4/3$ has been assumed. The magnetic field strength in
geometrical units, $B_{\mathrm{geo}}$, is readily obtained from the code
magnetic field strength $B=\sqrt{b_{\mu}b^{\mu}}$ as
\begin{equation}
B_{\mathrm{geo}} = c\left( \frac{\rho_{\mathrm{geo}}}{\rho} \right)^{1/2}B \,.
\end{equation}
What remains is to specify $T_{\mathrm{e}}$ (or $\Theta_{\mathrm{e}}:=
k_{\mathrm{B}}T_{\mathrm{e}}/m_{\mathrm{e}}c^{2}$) and
$\rho_{\mathrm{geo}}$. For simplicity we adopt the prescription of
\cite{Moscibrodzka2009}, wherein $T_{\mathrm{p}}/T_{\mathrm{e}}$ is
assumed to be a fixed ratio. Whilst such an approximation is rather
crude, to zeroth order the protons and electrons may be assumed to be
coupled in this way. To scale the electron number density we adopt the
method of \cite{Chan2015}, assuming a density scale typically of order
$10^{7}~\mathrm{cm}^{-3}$. A somewhat more sophisticated approach is to
employ a thresholding of the fluid plasma beta where, when the local
plasma beta exceeds some threshold the electrons and protons are coupled
as previously mentioned (disk region), but when not exceeded (typically
in the funnel region) the electron temperature is assumed to be constant
\citep[][]{Moscibrodzka2014,Moscibrodzka2016,Chan2015}. Since plasma beta
is found to decrease with resolution \citep[][]{Shiokawa2012} and in this
paper we seek only to demonstrate the convergence of our simulated shadow
images obtained from the GRMHD data in regions where the density is
non-negligible, we adopt the former model.

For the plasma emissivity we use the approximate formula for thermal
magnetobremsstrahlung \citep[][]{Leung2011}, which is given by
\begin{equation}
j_{\nu} = \left( \frac{\sqrt{2}\pi e^{2}}{3c} \right)n_{\mathrm{e}} \frac{\nu_{\mathrm{s}}}{K_{2}\left( \Theta_{\mathrm{e}}^{-1} \right)}\left( X^{1/2} + 2^{11/12} X^{1/6} \right)^{2}\exp \left( -X^{1/3} \right) \,,
\end{equation}
where $e$ is the electron charge, $n_{\mathrm{e}}$ the electron number density, and
\begin{eqnarray}
X &:=& \frac{\nu}{\nu_{\mathrm{s}}} \,, \\
\nu_{\mathrm{s}} &=& \left( \frac{e}{9\pi m_{\mathrm{e}}c} \right)B \Theta_{\mathrm{e}}^{2}\sin\vartheta \,,
\end{eqnarray}
and $\vartheta$ is the pitch angle of the photon with respect to the
magnetic field. The absorption coefficient is readily obtained from
Kirchoff's law.

Each image is generated using a uniform grid of $1000\times1000$ rays,
sampling 60 uniformly logarithmically spaced frequency bins between
$10^{9}$~Hz and $10^{15}$~Hz. All panels depict the observed image as
  seen at an observational frequency of $230~$GHz, i.e.~the frequency at
  which the EHT will image Sgr A*.  This resolution is chosen because the
  integrated flux over the entire ray-traced image is convergent:
  doubling the resolution from $500\times500$ to $1000\times1000$ yields
  an increase of $0.17\%$, and from $1000\times1000$ to $2000\times2000$
  an increase of $0.09\%$.

In practical GRRT calculations only simulation data which has already
reached a quasi-steady state, typically $t>2000\,M$, is used. In this
study we focus on the observational appearance of the accretion flow and
BH shadow image. The detailed discussion of the spectrum, variability and
plasma models warrants a separate study.

\subsection{Comparison of images}
A natural and important question arises from GRRT calculations of BH
shadows: do ray-traced images of GRMHD simulation data converge as the
resolution of the GRMHD simulation is increased?  The existence of an
optimal resolution, beyond which differences in images are small, implies
that one can save additional computational time and expense by running
the simulation at this optimal resolution. It would also imply that the
GRMHD data satisfactorily capture the small-scale structure, turbulence
and variations of the accretion flow. As such, it is informative to
investigate the convergence of BH shadow images obtained from GRMHD
simulation data of differing resolutions, both quantitatively and
qualitatively.

\begin{figure}[htbp]
\begin{center}
\includegraphics[width=0.9\textwidth]{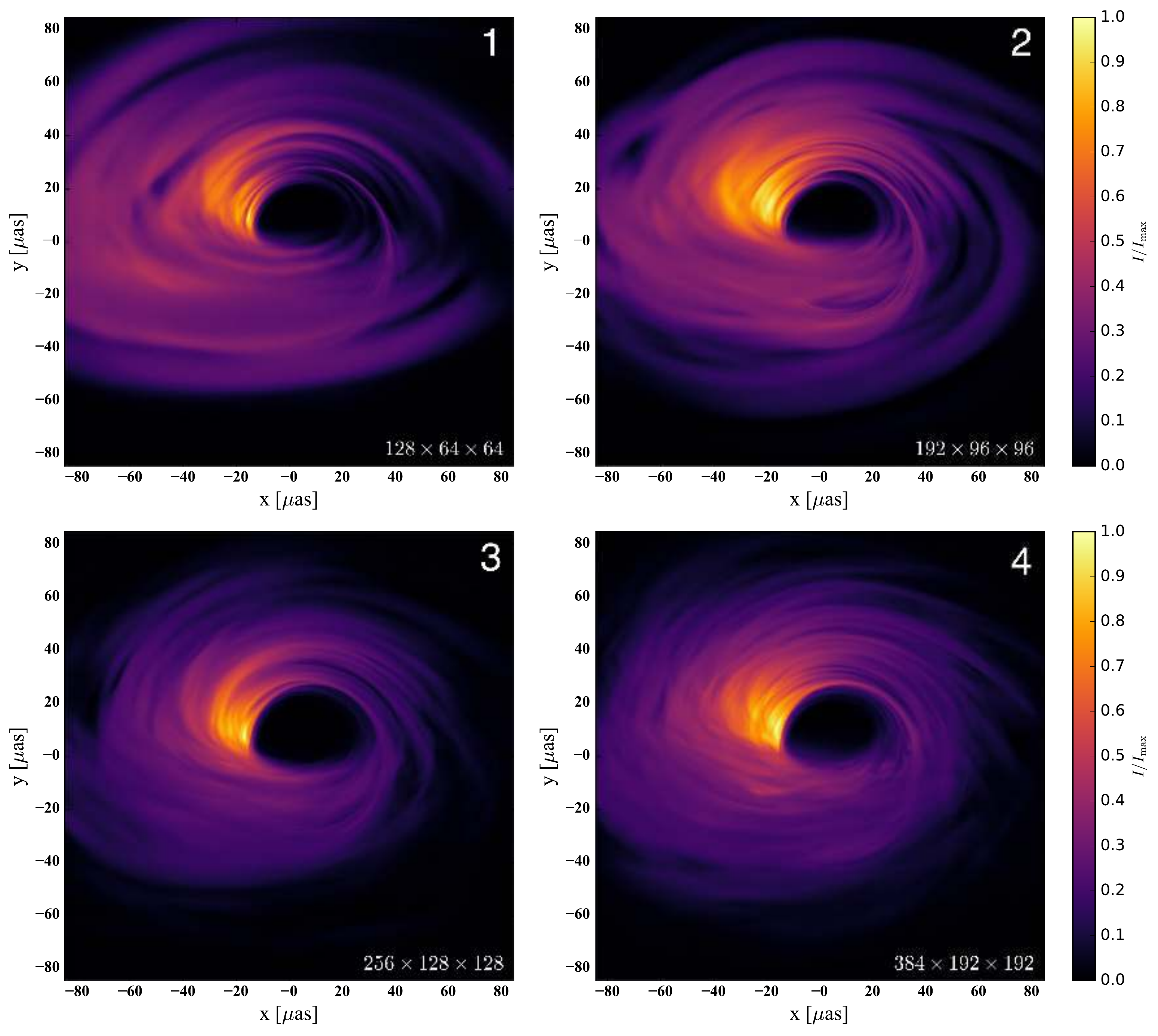}
\caption{Snapshot images of 3D GRMHD simulation data with parameters
  chosen to mimic the emission from Sgr A*. The resolution of the
  simulation data is indicated in the bottom-right corner of each panel
  and discussed in the text. }
\label{fig:Torus_4_resolutions}
\end{center}
\end{figure}
To address this question we first generate a series of four snapshot
images at $t=2500\,M$ of the the accretion flow and BH shadow from GRMHD
simulation data. The resolution of these data are
$2\mathcal{N}\times\mathcal{N}\times\mathcal{N}$ in $(r,\theta,\phi)$,
\ie twice as much resolution in the radial direction compared to the
zenith and azimuthal directions. The images depicted in
Fig.~\ref{fig:Torus_4_resolutions} correspond to $\mathcal{N}=64$, $96$,
$128$ and $192$ respectively. Here, the proton to electron temperature
ratio was chosen as $T_{\mathrm{p}}/T_{\mathrm{e}}=3$ (similar to
\cite{Moscibrodzka2009,Moscibrodzka2014}), the electron number density
scaling as $5\times 10^{7}~\mathrm{cm}^{-3}$, the BH mass is set to
$4.5\times 10^{6}~M_{\odot}$, the source distance is $8.4\times
10^{3}~\mathrm{pc}$, the dimensionless BH spin parameter $0.9375$ and the
observer inclination angle with respect to the BH spin axis is
$60^{\circ}$.

A direct consequence of increasing the resolution of the GRMHD data
  is resolving the fine-scale turbulent structure of the accretion flow.
  The characteristic dark shadow delineating the BH shadow can be clearly
  seen in all images.  As the resolution of the GRMHD data is increased,
  the images become less diffuse.  It is difficult with the naked eye to
  draw firm physical conclusions, and so in the following we perform a
  quantitative pixel-by-pixel analysis.  

With these snapshot images we may perform a quantitative measure of the
difference between any two images through introducing the (normalised)
cross-correlation. For two given two-dimensional arrays $f(x,y)$ and
$g(x,y)$ (\ie 2D images), a measure of similarity or difference may be
calculated through the cross-correlation $\mathcal{C}$, where
$\mathcal{C}\in[-1,1]$. The normalised cross-correlation is defined as
\begin{equation}
\mathcal{C} := \mathcal{C}_{_{i,j}} := \frac{1}{N \sigma_{f}\sigma_{g}}\sum\limits_{x,y}\left\{\left[f(x,y)-\mu_{f}\right]\left[g(x,y)-\mu_{g}\right]\right\} \,, \label{correlation}
\end{equation}
where $\mu_{f}, \ \sigma_{f}$ and $\mu_{g}, \ \sigma_{g}$ correspond to
the mean and standard deviation of $f$ and $g$ respectively, and $N$ is
equal to the size of either $f$ or $g$. In the examples considered here
the images are all of equal size and dimension, so $N=N_{f}=N_{g}$.
Equation (\ref{correlation}) may be interpreted as the inner product
between two data arrays, with the value of $\mathcal{C}$ expressing the
degree to which the data are aligned with respect to each other. When
$\mathcal{C}=1$ the data are identical, save for a multiplicative
constant, when $\mathcal{C}=0$ the data are completely uncorrelated, and
when $\mathcal{C}<0$ the data are negatively correlated.

Each image pixel has an intensity value represented as a single
  greyscale value between zero and one.  Given the relative intensity
  data of two different images, Equation (\ref{correlation}) is then
  employed to determine the normalised cross-correlation between the two
  images.  This procedure applied to the panels in
Fig.~(\ref{fig:Torus_4_resolutions}) yields the following symmetric
matrix of cross-correlation values between the images
\begin{equation}
\mathcal{C}_{i,j} = 
\left(
\begin{array}{cccc}
 1 & 0.839495 & 0.809205 & 0.856958 \\
 -  & 1 & 0.867578 & 0.917560 \\
 -  & -  & 1 & 0.948544 \\
 -  & -  & -  & 1 \\
\end{array}
\right) \,. \label{C_ij}
\end{equation}
Indices $i$ and $j$, where $\left( i, j \right)=(1,4)$, denote the images
being cross-correlated.

The rightmost column of equation (\ref{C_ij}) denotes the
cross-correlation values, $\mathcal{C}_{i,4}$, in descending order
between images, \ie the cross-correlation of image 4 with images 1, 2, 3
and 4 respectively. Since $\mathcal{C}_{i+1,4}>\mathcal{C}_{i,4}$ it is
clear that the similarity between images increases as the resolution of
the GRMHD simulation is increased. Similarly, for image 3 it is found
that $\mathcal{C}_{i+1,3}>\mathcal{C}_{i,3}$. Finally, it also follows
that $\mathcal{C}_{3,4}>\mathcal{C}_{2,3}>\mathcal{C}_{1,2}$, \ie the
correlation between successive pairs of images increases with increasing
resolution, demonstrating the convergence of the GRMHD simulations with
increasing grid resolution. Whilst the lowest resolution of $128\times
64 \times 64$ is certainly insufficient, both difference images and
cross-correlation measures indicate that a resolution of $256\times 128
\times 128$ is sufficient and represents a good compromise.

\begin{figure}[htbp]
\begin{center}
\includegraphics[width=0.9\textwidth]{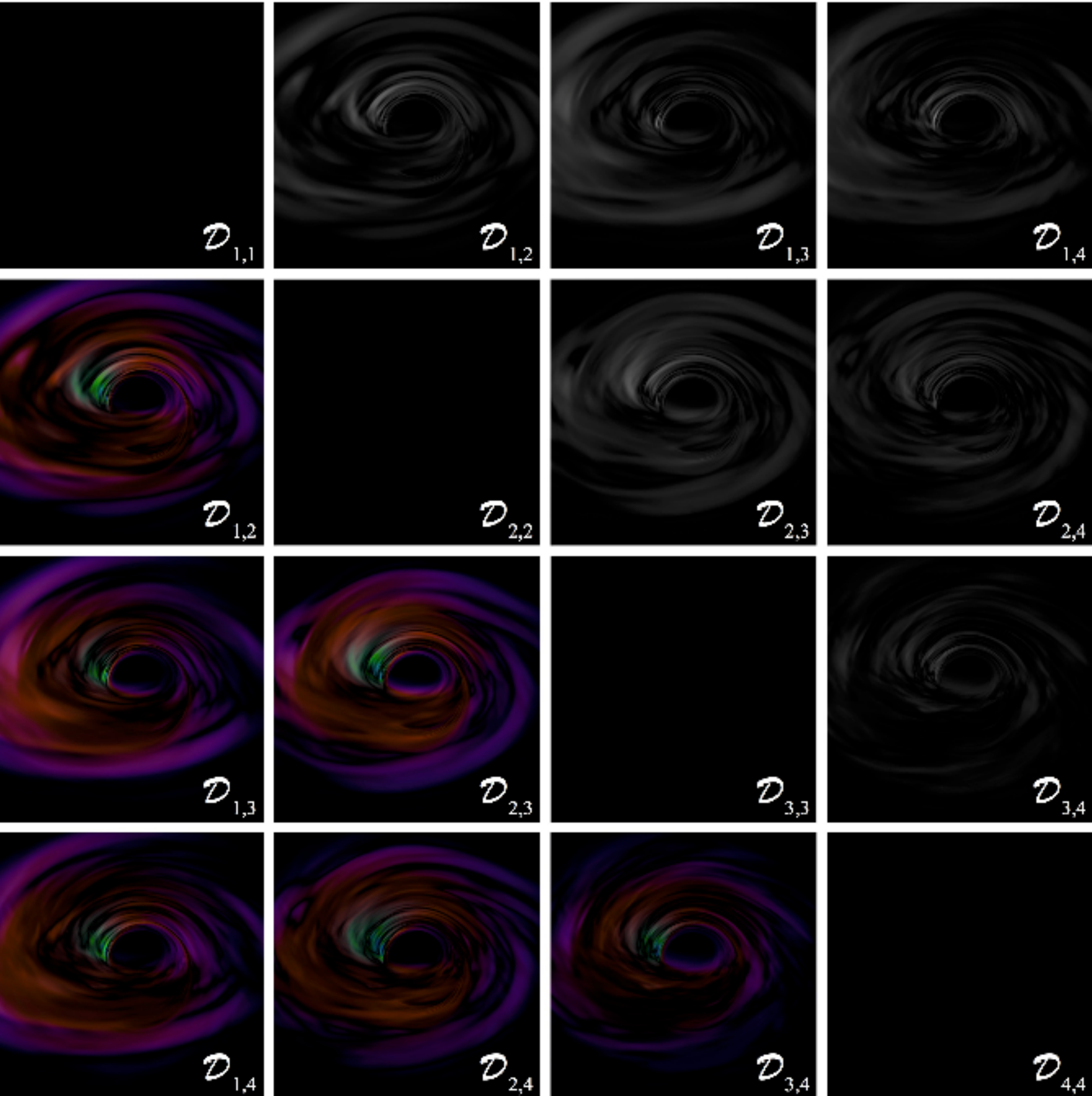}
\caption{ Matrix of image differences $\mathcal{D}_{i,j}$ of the four
  panels in Fig.~\ref{fig:Torus_4_resolutions}. Upper diagonal panels
  are greyscale differences. Lower diagonal panels are identical to
  corresponding upper diagonal panels but with differences illustrated
  with RGB pixel values. Black panels correspond to $\mathcal{D}_{i,i}$,
  \ie trivially the difference between an image and itself.}
\label{fig:Diff_Torus}
\end{center}
\end{figure}

\section{Conclusions and outlook}\label{sec:outlook}

We have described the capabilities of \bhac, a new multidimensional
general-relativistic magnetohydrodynamics code developed to perform
hydrodynamical and MHD simulations of accretion flows onto compact
objects in arbitrary stationary spacetimes exploiting the numerous
advantages of AMR techniques. The code has been tested with several one-,
two- and three- dimensional scenarios in special-relativistic and
general-relativistic MHD regimes. For validation, GRMHD simulations of
MRI unstable tori have been compared with another well-known and tested
GRMHD code, the \harmthreed code. \bhac~shows very good agreement with
the \harmthreed results, both qualitatively and quantitatively. As a
first demonstration of the AMR capabilities in multi-scale simulations,
we performed the magnetized-torus accretion test with and without
AMR. Despite the latter intrinsically implies an overhead of $\sim10\%$,
the AMR runtime amounted to $65\%$ of that relative to the uniform grid,
simply as a result of the more economical use of grid cells in the block
based AMR. At the same time, the AMR results agree very well with the
more expensive uniform-grid results. With increasing dynamic range, we
expect the advantages of AMR to increase even more significantly,
rendering it a useful tool for simulations involving structures of
multiple physical scales.

Currently, two methods controlling the divergence of the magnetic field
are available in \bhac~and we compared them in three test
problems. Although solutions obtained with the cell-centered
flux-interpolated constrained transport (FCT) algorithm and the
divergence cleaning scheme (GLM) yield the same (correct) physical
behaviour in the case of weak magnetic fields, FCT performs considerably
better in the presence of strong magnetic fields. In particular, FCT is
less diffusive than GLM, is able to preserve a stationary solution, and
it creates less spurious structures in the magnetic field. For example,
the use of GLM in the case of accretion scenarios with strong magnetic
fields leads to worrisome artefacts in the highly magnetised funnel
region. The development of a constrained transport scheme
compatible with AMR is ongoing and will be presented in a separate work
\cite{Olivares2017}.

The EHTC and its European contribution, the BlackHoleCam project
\citep{Goddi2016}, aim at obtaining horizon-scale radio images of the BH
candidate at the Galactic Center. In anticipation of these results, we
have used the 3D GRMHD simulations as input for GRRT calculations with
the newly developed \bhoss~code \citep{Younsi2017}. We found that the
intensity maps resulting from different resolution GRMHD simulations
agree very well, even when comparing snapshot data that was not time
averaged. In particular, the normalised cross-correlation between images
achieves up to $94.8\%$ similarity between the two highest resolution
runs. Furthermore, the agreement between two images converges as the
resolution of the GRMHD simulation is increased. Based on this
comparison, we find that moderate grid resolutions of
$256\times128\times128$ (corresponding to physical resolutions of
  $\Delta r_{\rm KS}\times\Delta \theta_{\rm KS}\times \Delta\phi_{\rm
    KS} = 0.04 M \times 0.024 {\rm rad}\times0.05 {\rm rad}$ at the
  horizon) yield sufficiently converged intensity maps. Given the large
and likely degenerate parameter space and the uncertainty in modelling of
the electron distribution, this result is encouraging, as it demonstrates
that the predicted synthetic image is quite robust against the
ever-present time variability, but also against the impact that the grid
resolution of the GRMHD simulations might have.  In addition, independent
information on the spatial orientation and magnitude of the spin, such as
the one that could be deduced from the dynamics of a pulsar near Sgr A*
\cite{Psaltis2016}, would greatly reduce the space of degenerate
solutions and further increase the robustness of the predictions that
\bhac~will provide in terms of synthetic images.

Finally, we have demonstrated the excellent flexibility of \bhac~with a
variety of different astrophysical scenarios that are ongoing and will be published
shortly. These include: oscillating hydrodynamical equilibrium tori for
the modelling of quasi-periodic oscillations \cite{deAvellar2017},
episodic jet formation \cite{Porth2017} and magnetised tori orbiting
non-rotating dilaton BHs \cite{Mizuno2017}.

\begin{appendix}
\section{Evolution of the scalar $\phi$}\label{sec:phi}

To obtain the evolution equation for $\phi$ in the augmented Faraday's
law, we project (\ref{eq:maxglm}) onto the Eulerian observer by
contracting with $-n_{\mu}$ as
\begin{align}
-\nabla_{\nu} (^{*}\!F^{\mu\nu}n_{\mu} - \phi n^{\nu}) &=-\kappa \phi - (^{*}\!F^{\mu\nu} - \phi g^{\mu\nu})\nabla_{\nu} n_{\mu}\\
\Rightarrow
\nabla_{\nu} B^{\nu} + \nabla_{\nu} \phi n^{\nu} &= -\kappa \phi - (^{*}\!F^{\mu\nu} - \phi g^{\mu\nu})\nabla_{\nu} n_{\mu}\\
\Rightarrow
(-g)^{-1/2}\partial_{i}[\gamma^{1/2}\alpha B^{i}] + \nabla_{\nu} \phi n^{\nu} &= -\kappa \phi - (^{*}\!F^{\mu\nu} - \phi g^{\mu\nu})\nabla_{\nu} n_{\mu} \,,
\end{align}
where we used $B^{\nu}=-n_{\mu}\,^{*}\!F^{\mu\nu}$. Using the definition
of extrinsic curvature $K_{\mu\nu}:=
-\gamma^{\lambda}_{\mu}\nabla_{\lambda}n_{\nu}$, we can write
\citep[Eq. (7.62) in][]{Rezzolla_book:2013}
\begin{align}
\nabla_{\nu}n_{\mu} = -n_{\nu} a_{\mu} - K_{\mu\nu} \,, \label{eq:Kmunu}
\end{align}
where we used the ``acceleration'' of the Eulerian observer $a_{\mu}:=
n^{\lambda}\nabla_{\lambda}n_{\mu}$ which satisfies $n^{\mu} a_{\mu}=0$.
With the identity $a_{i}=\alpha^{-1} \partial_{i} \alpha$ \citep{York79}
and exploiting the symmetries of $^{*}\!F^{\mu\nu}$ and $K_{\mu\nu}$, is
straightforward to show that
\begin{align}
(\gamma)^{-1/2}\partial_{i}[(\gamma)^{1/2}\alpha B^{i}] + \alpha F^{*\mu\nu} \nabla_{\nu}n_{\mu} = (\gamma)^{-1/2}\partial_{i}[(\gamma)^{1/2}\alpha B^{i}] - B^i\partial_i\alpha \,.
\end{align}
Hence it follows that
\begin{align}
\begin{split}
\partial_{t} (\sqrt{\gamma}\phi) + \partial_{i}[\sqrt{\gamma}(\alpha B^{i}-\phi \beta^{i})] = &-\alpha \gamma^{1/2} \kappa \phi + \alpha \gamma^{1/2}\phi g^{\mu\nu}\nabla_{\nu}n_{\mu} \\
&+ \gamma^{1/2} B^i\partial_i\alpha\,.
\end{split}
\end{align}
Using again Eq. (\ref{eq:Kmunu}), the source term
$\mathcal{S}:=\sqrt{\gamma} \alpha \phi g^{\mu\nu} \nabla_{\nu} n_{\mu}$
can be rewritten as
\begin{align}
\mathcal{S}= - \sqrt{\gamma} \alpha \phi g^{\mu\nu} K_{\mu\nu} \,,
\end{align}
where the first term drops out due to the orthogonality $n^{\mu} a_{\mu}=0$. 
For a symmetric tensor $S^{\mu\nu}$, we have
\begin{align}
\alpha S^{\mu\nu} K_{\mu\nu} = \alpha S^{ij} K_{ij} = S^{i}_{j} \partial_{i} \beta^{j} + \frac{1}{2} S^{ij} \beta^{k}\partial_{k}\gamma_{ij} \,.
\end{align}
This follows from the relation $\Gamma^{0}_{ij}=-K_{ij}\alpha^{-1}$ where
$\Gamma^{0}_{ij}$ are elements of the 4-Christoffel symbols \citep[see
  \eg (B.9) of][]{Alcubierre:2008}. Thus
\begin{align}
\mathcal{S} &= -\sqrt{\gamma} \phi(\partial_{i}\beta^{i}+\frac{1}{2}\gamma^{ij}\beta^{k}\partial_{k}\gamma_{ij}) \\
&= -\sqrt{\gamma} \phi\partial_{i}\beta^{i} 
- \sqrt{\gamma} \phi \frac{1}{2}\gamma^{ij}\beta^{k}\partial_{k}\gamma_{ij}
\,.
\end{align}

\section{Modified Faraday's law}\label{sec:modfaraday}

The augmented Faraday's law follows from the $j$-component of (\ref{eq:maxglm}) as
\begin{align}
&\nabla_{\nu} (^{*}\!F^{j\nu} - \phi g^{j\nu}) = \kappa \phi \beta	^{j}/\alpha \\
\Rightarrow & (-g)^{-1/2} \left\{\del{t}[\sqrt{\gamma} (-B^{j})] + \del{i}\left[\sqrt{\gamma} \left(\vt^{j}B^{i}-\vt^{i}B^{j}\right)\right]\right\} + g^{j\lambda}\del{\lambda}(-\phi) = \kappa \phi \beta^{j}/\alpha \\
\Rightarrow & \del{t}\left(\sqrt{\gamma} B^{j}\right) + \del{i}\left[\sqrt{\gamma} \left(\vt^{i}B^{j}-\vt^{j}B^{i}\right)\right] 
+ \sqrt{\gamma}\alpha g^{j\lambda}\del{\lambda}\phi = - \kappa \phi \sqrt{\gamma} \beta^{j} \\
\Rightarrow & \del{t}\left(\sqrt{\gamma} B^{j}\right) + \del{i}\left[\sqrt{\gamma} \left(\vt^{i}B^{j}-\vt^{j}B^{i}\right) \right] 
+ \frac{\beta^{j}}{\alpha}\del{t}(\sqrt{\gamma} \phi) + \sqrt{\gamma}\alpha \gamma^{ji}\del{i}\phi
 \nonumber \\
\phantom{\Rightarrow} &\hskip 6.0cm - \sqrt{\gamma} \frac{\beta^{i}\beta^{j}}{\alpha}\del{i}\phi
= - \sqrt{\gamma} \kappa \phi \beta^{j} \,. \label{eq:fglm1}
\end{align}
We see that apart from the gradient $\phi$-term, we obtain another term
that involves the time-derivative of $(\sqrt{\gamma}\phi)$. Hence we need to
plug in Equation (\ref{eq:phievol}). We rewrite the term
${\beta^{j}}\del{t}(\sqrt{\gamma} \phi)/{\alpha}$ simplifying the
lengthy expression
\begin{align}
\hskip -1.0cm \frac{\beta^{j}}{\alpha} \partial_{t} (\sqrt{\gamma} \phi) 
& = - \frac{\beta^{j}}{\alpha}\partial_{i} \left[ \sqrt{\gamma} \left( \alpha B^{i}-\phi \beta^{i} \right) \right] \nonumber \\
& \phantom{=}- \sqrt{\gamma} \kappa \phi \beta^{j} 
- \frac{\beta^{j}}{\alpha}\sqrt{\gamma} \phi \partial_{i}\beta^{i} 
- \frac{1}{2} \frac{\beta^{j}}{\alpha}\sqrt{\gamma} \phi \gamma^{il}\beta^{k}\partial_{k}\gamma_{il}
+ \frac{\beta^j}{\alpha} \sqrt{\gamma} B^i\partial_i \alpha
\\
& = -\del{i}\left[\sqrt{\gamma}B^{i}\beta^{j}\right] 
+ \sqrt{\gamma} B^{i} \del{i}\beta^{j} 
+ \sqrt{\gamma} \frac{\beta^{i}\beta^{j}}{\alpha} \del{i}\phi 
- \sqrt{\gamma} \kappa \phi \beta^{j} 
\end{align}
Substituting this into (\ref{eq:fglm1}) yields the modified Faraday's law (\ref{eq:fglm2}).

\section{Derivation of cell centred formulas for FCT}\label{sec:fvolumeFCT}

In the 3+1 decomposition, for the case of a stationary spacetime the
induction equation can be written in component form as
\begin{equation}
\label{eq:faraday_cons}
\partial_t \sqrt{\gamma} B^a + \partial_b (-\eta^{abc}E_c) = 0 \,.
\end{equation}
\noindent
Integrating this on each of the surfaces bounding a cell with vertexes at
$x^1_{i+l_1}$, $x^2_{i+l_2}$, $x^3_{i+l_3}$ with $l=\pm 1/2$, and using
the Stokes theorem, we obtain the evolution equations for the magnetic
flux in CT, for instance
\begin{equation}
\label{eq:CT_update}
\frac{d\Phi_{i+1/2,j,k}}{dt} = G_{i+1/2,j+1/2,k} - G_{i+1/2,j-1/2,k} - G_{i+1/2,j,k+1/2} + G_{i+1/2,j,k-1/2} \,,
\end{equation}
\noindent
where
\begin{equation}
\Phi_{i+1/2,j,k}=\int_{\partial V (x^1_{i+1/2})} \gamma^{1/2} B^1\, dx^2 \, dx^3 \,,
\end{equation}
\noindent
with each $G$ representing a line integral of the form
\begin{equation}
\label{eq:integralG}
G_{i+1/2,j+1/2,k} = - \int_{x^3_{k-1/2}}^{x^3_{k+1/2}} \left. E_3\right|_{x^1_{i+1/2},x^2_{j+1/2}} \, dx^3 \,.
\end{equation}

\noindent
The fact that each of these integrals appear in the evolution equation of two magnetic fluxes guarantees the conservation of divergence,
as will be explained in the next Section.

On the other hand, the numerical fluxes corresponding to the magnetic field components that are returned by the Riemann solver
are surface integrals of the electric field, for example, the flux in the $x^2$-direction for $B^1$ is
\begin{equation}
\label{eq:fluxEfield_integrals}
\left. \Delta S^2 \bar{F}^2\right|_{i,j+1/2,k} = \int_{x^1_{i-1/2}}^{x^1_{i+1/2}}\int_{x^3_{k-1/2}}^{x^3_{k+1/2}}\left. E_{x^3}\right|_{j+1/2} \, dx^3 \, dx^1 \,.
\end{equation}

\noindent
The innermost integral is the same as that of Eq.~(\ref{eq:integralG}), so the average flux can be interpreted as
\begin{equation}
\left. \Delta S^2 \bar{F}^2 \right|_{i,j+1/2,k} = -  \Delta x_i \tilde{G}_{i,j+1/2,k} \,,
\end{equation}

\noindent
where $\tilde{G}_{i,j+1/2,k}$ is the mean value of the integral from Eq.~(\ref{eq:integralG}).
To second-order accuracy, this integral takes the value $\tilde{G}_{i,j+1/2,k}$ at the middle of the cell,
therefore $G_{i+1/2,j+1/2,k}$ can be found by interpolating the averaged fluxes from the four adjacent cell faces as
\begin{align}
\begin{split}
G_{i+1/2,j+1/2,k} =  \frac{1}{4}
\Biggl(
&\frac{\Delta S^2 \bar{F}^2  \bigr|_{i,j+1/2,k}}{\Delta x}
+\frac{\Delta S^2 \bar{F}^2 \bigr|_{i+1,j+1/2,k}}{\Delta x_{i+1}} - \\
&\frac{\Delta S^1 \bar{F}^1 \bigr|_{i+1/2,j,k}}{\Delta y_j}
-\frac{\Delta S^1 \bar{F}^1 \bigr|_{i+1/2,j+1,k}}{\Delta y_{j+1}}\Biggr) \,.
\end{split}
\label{eq:interpolationG_fluxes}
\end{align}
Since we implemented a cell-centred version of FCT, we are interested in the evolution of the average
magnetic field at the cell centres. To second order accuracy, the rate of change of the average value of the
$x^1-$component of the magnetic field is

\begin{equation}
\label{eq:B_i_integration_approx}
\int_{x_{i-1/2}}^{x_{x+1/2}} \frac{d\Phi}{dt} dx = \Delta V_{ijk} \frac{d \bar{B}^x}{dt}
 \approx \frac{\Delta x_i}{2} \left( \frac{d\Phi}{dt}\Bigr|_{x_{i+1/2}} + \frac{d\Phi}{dt}\Bigr|_{x_{i-1/2}}\right).
\end{equation}
\noindent
Now we substitute Eq.~(\ref{eq:interpolationG_fluxes}) into Eq.~(\ref{eq:CT_update}) and Eq.~(\ref{eq:CT_update}) into Eq.~(\ref{eq:B_i_integration_approx}).
After some algebra, we finally obtain eqs. \ref{eq:Bbar_evol} and \ref{eq:average_fluxes}.

\section{Discretisation of $\boldsymbol{\nabla\cdot B}$ and zero-divergence initial conditions}\label{sec:initial_div}

CT schemes aim to maintain to zero at machine precision the discretisation of the divergence given by
\begin{align}
\begin{split}
(\boldsymbol{\nabla \cdot B})_{i,j,k} = \frac{1}{\Delta V_{i,j,k}}
\Biggl(&\Phi_{i+1/2,j,k} - \Phi_{i-1/2,j,k} + \Phi_{i,j+1/2,k} - \\
       &\Phi_{i,j-1/2,k} + \Phi_{i,j,k+1/2} - \Phi_{i,j,k-1/2}\Biggr) \,,
\end{split}
\label{eq:div1}
\end{align}
\noindent
which can be thought of as the volume average of the quantity $\partial_a (\gamma^{1/2} B^a)$ in the given cell. 

When calculating the evolution equation for $(\boldsymbol{\nabla \cdot
  B})_{i,j,k}$, each of the integrals $G$ appear with opposite signs in
the expression for $d\Phi/dt$ \eqref{eq:CT_update} and cancel to machine
precision.  Therefore, if this discretisation of the divergence was
originally zero, it will be zero to machine precision during the rest of
the simulation.

However, in the cell-centred version of FCT employed here, we lack
information concerning the magnetic flux at cell faces, so Equation
\eqref{eq:div1} cannot be used to monitor the creation of divergence.  We
will therefore find a derived quantity that we can monitor based on the
other available quantities.

We calculate the average value of the divergence of eight cells sharing a vertex as
\begin{equation}
\label{eq:div_average1}
 (\boldsymbol{\nabla \cdot B})_{i+1/2,j+1/2,k+1/2} = \frac{1}{\Delta V^*}\sum_{l_1,l_2,l_3=0,1}\left. \Delta V (\boldsymbol{\nabla \cdot B})\right|_{i+l_1,j+l_2,k+l_3} \,. \\
\end{equation}
\noindent
When substituting Eq.~\eqref{eq:div1}, the right hand side of Eq.~\eqref{eq:div_average1} consists of a sum of terms of the form
\begin{equation*}
\sum_{l_2,l_3=0,1} \left(\Phi_{i+3/2,j+l_2,k+l_3} - \Phi_{i+1/2,j+l_2,k+l_3} + \Phi_{i+1/2,j+l_2,k+l_3} - \Phi_{i-1/2,j+l_2,k+l_3} \right) \,,
\end{equation*} 

\noindent
for each direction. Using the same second-order approximation as for the time-update,
\begin{equation}
\label{eq:B2nd_order}
\Delta V_{i,j,k} \bar{B}^x_{i,j,k} \approx \frac{\Delta x_i}{2} (\Phi_{i+1/2,j,k} + \Phi_{i-1/2,j,k}) \,,
\end{equation}
\noindent
this becomes
\begin{equation*}
\sum_{l_1,l_2,l_3=0,1} \left[(-1)^{1+l_1}\frac{\bar{B}^1\Delta V}{\Delta x^1} \right]_{i+l_1,j+l_2,k+l_3} \,.
\end{equation*} 

\noindent
Finally, summing over the three directions, we recover
Eq.~\eqref{eq:div}. Since the same second-order approximation is used
both for the definition and for the time update of $\bar{B}^a$, the
definition of divergence given by Equation \eqref{eq:div} is conserved to
machine precision during each evolution step.

To obtain a divergence-free initial condition, we calculate the magnetic field as the curl of a vector potential.
First, we calculate the magnetic flux at each of the cell faces as
\begin{equation}
\label{eq:curlA}
\Phi_{i+1/2,j,k} = \mathcal{A}_{i+1/2,j+1/2,k} - \mathcal{A}_{i+1/2,j-1/2,k} - \mathcal{A}_{i+1/2,j,k+1/2} + \mathcal{A}_{i+1/2,j,k-1/2} \,,
\end{equation}
\noindent
where $\mathcal{A}$ are line integrals of the vector potential along the cell edges
\begin{equation}
\label{eq:integralA}
\mathcal{A}_{i+1/2,j+1/2,k} = \int_{x^3_{k-1/2}}^{x^3_{k+1/2}} \left. A_3\right|_{x^1_{i+1/2},x^2_{j+1/2}} dx^3 \,.
\end{equation}
\noindent
Then we use again the second order approximation from Equation
\eqref{eq:B2nd_order} to find the average magnetic field components at
the cell center. By construction, in this way we obtain a divergence-free
initial condition using either of the discretization of divergence in
Eqs.~\eqref{eq:div1} or \eqref{eq:div}.

\end{appendix}

\begin{backmatter}

\section*{Competing interests}
The authors declare that they have no competing interests.

\section*{Funding}
This research is supported by the ERC synergy
grant "BlackHoleCam: Imaging the Event Horizon of Black Holes" (Grant
No. 610058), by ``NewCompStar'', COST Action MP1304, by the LOEWE-Program
in HIC for FAIR, and by the European Union's Horizon 2020 Research and
Innovation Programme (Grant 671698) (call FETHPC-1-2014, project
ExaHyPE). ZY acknowledges support from an Alexander von Humboldt
Fellowship. HO is supported in part by a CONACYT-DAAD scholarship.

\section*{Author's contributions}
The implementation of the GRMHD equations was performed by OP. The FCT
algorithm was implemented and tested by HO. YM contributed with code
tests and ZY performed the radiative-transfer calculations. MM provided
\harmthreed validation data. The project was initiated by LR, HF and MK and was closely
supervised by LR. 

\section*{Acknowledgements}
It is a pleasure to thank Christian Fromm, Mariafelicia de Laurentis,
Thomas Bronzwaer, Jordy Davelaar, Elias Most and Federico Guercilena for discussions.  
We are grateful to Scott Noble for the ability to use the \harmthreed code for comparison
and to Zakaria Meliani for input on the construction of \bhac. The
initial setup for the toroidal-field equilibrium torus was kindly
provided by Chris Fragile.  The simulations were performed on LOEWE 
at the CSC-Frankfurt and Iboga at ITP Frankfurt. We acknowledge 
technical support from Thomas Coelho.

\bibliographystyle{bmc-mathphys}

\clearpage

\end{backmatter}
\end{document}